\documentclass[aps,prc,showpacs,preprint,superscriptaddress]{revtex4}

\usepackage{graphicx}
\usepackage{color}
\usepackage{amsmath}

\newcommand{\beq}[1]{\begin{equation}\label{#1}}
\newcommand{\eeq}{\end{equation}}
\newcommand{\bea}[1]{\begin{eqnarray}\label{#1}}
\newcommand{\eea}{\end{eqnarray}}

\newcommand{\Eq}[1]{Eq.~(\ref{#1})}

\newcommand{\Fi}[1]{Fig.~\ref{#1}}
\newcommand{\Ta}[1]{Table~\ref{#1}}

\newcommand{\agev}{\mbox{$A$~GeV}}               %PRL notation
\newcommand{\gevc}{\mbox{GeV$/c$}}

\newcommand{\rb}[1]{\mbox{\textrm{\scriptsize #1}}}

\newcommand{\dedx}{\ensuremath{\textrm{d}E/\textrm{d}x}}

\newcommand{\enwee}{\ensuremath{N_{\rb{w}}}}

\begin{document}

%\preprint{}


%============================================================================

\title{$K^{\ast}(892)^0$ and $\overline{K}^{\ast}(892)^0$ production 
in central Pb+Pb, Si+Si, C+C and inelastic p+p collisions at 158\agev\\
%\begin{center}
%( version 3.9 )    
%\end{center}
}

%============================================================================


%\author{}
%\thanks{}
%\affiliation{}

%============================================================================

%============================================================================
\affiliation{NIKHEF, 
             Amsterdam, Netherlands.}
\affiliation{Department of Physics, University of Athens, 
             Athens, Greece.}
%\affiliation{Comenius University,
%            Bratislava, Slovakia.}
\affiliation{E\"otv\"os Lor\'ant University, 
             Budapest, Hungary.}
\affiliation{KFKI Research Institute for Particle and Nuclear Physics,
             Budapest, Hungary.} 
\affiliation{MIT, Cambridge, USA.}
\affiliation{Henryk Niewodniczanski Institute of Nuclear Physics, 
             Polish Academy of Sciences, 
             Cracow, Poland.}
\affiliation{GSI Helmholtzzentrum f\"{u}r Schwerionenforschung,
             Darmstadt, Germany.} 
\affiliation{Joint Institute for Nuclear Research, 
             Dubna, Russia.}
\affiliation{Fachbereich Physik der Universit\"{a}t, 
             Frankfurt, Germany.}
\affiliation{CERN, 
             Geneva, Switzerland.}
\affiliation{Institute of Physics, Jan Kochanowski University, 
             Kielce, Poland.}
\affiliation{Fachbereich Physik der Universit\"{a}t, 
             Marburg, Germany.}
\affiliation{Max-Planck-Institut f\"{u}r Physik, 
             Munich, Germany.}
\affiliation{Charles University, Faculty of Mathematics and Physics,
             Institute of Particle and Nuclear Physics, 
             Prague, Czech Republic.} 
\affiliation{Nuclear Physics Laboratory, University of Washington,
             Seattle, WA, USA.} 
\affiliation{Atomic Physics Department, Sofia University St.~Kliment Ohridski, 
             Sofia, Bulgaria.} 
\affiliation{Institute for Nuclear Research and Nuclear Energy, 
             Sofia, Bulgaria.}
\affiliation{Department of Chemistry, Stony Brook Univ. (SUNYSB), 
             Stony Brook, USA.}
\affiliation{Institute for Nuclear Studies, 
             Warsaw, Poland.}
\affiliation{Institute for Experimental Physics, University of Warsaw,
             Warsaw, Poland.} 
\affiliation{Faculty of Physics, Warsaw University of Technology, 
             Warsaw, Poland.}
\affiliation{Rudjer Boskovic Institute, 
             Zagreb, Croatia.}

%============================================================================

\author{T.~Anticic} 
\affiliation{Rudjer Boskovic Institute, 
             Zagreb, Croatia.}
\author{B.~Baatar}
\affiliation{Joint Institute for Nuclear Research, 
             Dubna, Russia.}
\author{D.~Barna}
\affiliation{KFKI Research Institute for Particle and Nuclear Physics,
             Budapest, Hungary.} 
\author{J.~Bartke}
\affiliation{Henryk Niewodniczanski Institute of Nuclear Physics, 
             Polish Academy of Sciences, 
             Cracow, Poland.}
\author{H.~Beck}
\affiliation{Fachbereich Physik der Universit\"{a}t, 
             Frankfurt, Germany.}
\author{L.~Betev}
\affiliation{CERN, 
             Geneva, Switzerland.}
\author{H.~Bia{\l}\-kowska} 
\affiliation{Institute for Nuclear Studies, 
             Warsaw, Poland.}
\author{C.~Blume}
\affiliation{Fachbereich Physik der Universit\"{a}t, 
             Frankfurt, Germany.}
\author{M.~Bogusz} 
\affiliation{Faculty of Physics, Warsaw University of Technology, 
             Warsaw, Poland.}
\author{B.~Boimska}
\affiliation{Institute for Nuclear Studies, 
             Warsaw, Poland.}
\author{J.~Book}
\affiliation{Fachbereich Physik der Universit\"{a}t, 
             Frankfurt, Germany.}
\author{M.~Botje}
\affiliation{NIKHEF, 
             Amsterdam, Netherlands.}
%\author{J.~Bracinik}
%\affiliation{Comenius University, 
%             Bratislava, Slovakia.}
\author{P.~Bun\v{c}i\'{c}}
\affiliation{CERN, 
             Geneva, Switzerland.}
%\author{V.~Cerny}
%\affiliation{Comenius University, 
%             Bratislava, Slovakia.}
\author{T.~Cetner} 
\affiliation{Faculty of Physics, Warsaw University of Technology, 
             Warsaw, Poland.}
\author{P.~Christakoglou}
\affiliation{NIKHEF, 
             Amsterdam, Netherlands.}
\author{P.~Chung}
\affiliation{Department of Chemistry, Stony Brook Univ. (SUNYSB), 
             Stony Brook, USA.}
\author{O.~Chvala}
\affiliation{Charles University, Faculty of Mathematics and Physics,
             Institute of Particle and Nuclear Physics, 
             Prague, Czech Republic.} 
\author{J.G.~Cramer}
\affiliation{Nuclear Physics Laboratory, University of Washington,
             Seattle, WA, USA.} 
\author{V.~Eckardt}
\affiliation{Max-Planck-Institut f\"{u}r Physik, 
             Munich, Germany.}
\author{Z.~Fodor}
\affiliation{KFKI Research Institute for Particle and Nuclear Physics,
             Budapest, Hungary.} 
\author{P.~Foka}
\affiliation{GSI Helmholtzzentrum f\"{u}r Schwerionenforschung,
             Darmstadt, Germany.} 
\author{V.~Friese}
\affiliation{GSI Helmholtzzentrum f\"{u}r Schwerionenforschung,
             Darmstadt, Germany.} 
\author{M.~Ga\'zdzicki}
\affiliation{Fachbereich Physik der Universit\"{a}t, 
             Frankfurt, Germany.}
\affiliation{Institute of Physics, Jan Kochanowski University, 
             Kielce, Poland.}
\author{K.~Grebieszkow}
\affiliation{Faculty of Physics, Warsaw University of Technology, 
             Warsaw, Poland.}
\author{C.~H\"{o}hne}
\affiliation{GSI Helmholtzzentrum f\"{u}r Schwerionenforschung,
             Darmstadt, Germany.} 
\author{K.~Kadija}
\affiliation{Rudjer Boskovic Institute, 
             Zagreb, Croatia.}
\author{A.~Karev}
\affiliation{CERN,
             Geneva, Switzerland.}
\author{V.I.~Kolesnikov}
\affiliation{Joint Institute for Nuclear Research, 
             Dubna, Russia.}
\author{M.~Kowalski}
\affiliation{Henryk Niewodniczanski Institute of Nuclear Physics, 
             Polish Academy of Sciences, 
             Cracow, Poland.}
\author{D.~Kresan}
\affiliation{GSI Helmholtzzentrum f\"{u}r Schwerionenforschung,
             Darmstadt, Germany.} 
%\author{M.~Kreps}
%\affiliation{Comenius University, 
%             Bratislava, Slovakia.}
\author{A.~Laszlo}
\affiliation{KFKI Research Institute for Particle and Nuclear Physics,
             Budapest, Hungary.} 
\author{R.~Lacey}
\affiliation{Department of Chemistry, Stony Brook Univ. (SUNYSB), 
             Stony Brook, USA.}
\author{M.~van~Leeuwen}
\affiliation{NIKHEF, 
             Amsterdam, Netherlands.}
\author{M.~Mackowiak} 
\affiliation{Faculty of Physics, Warsaw University of Technology, 
             Warsaw, Poland.}
\author{M.~Makariev}
\affiliation{Institute for Nuclear Research and Nuclear Energy, 
             Sofia, Bulgaria.} 
\author{A.I.~Malakhov}
\affiliation{Joint Institute for Nuclear Research, 
             Dubna, Russia.}
\author{M.~Mateev}
\affiliation{Atomic Physics Department, Sofia University St.~Kliment Ohridski, 
             Sofia, Bulgaria.} 
\author{G.L.~Melkumov}
\affiliation{Joint Institute for Nuclear Research, 
             Dubna, Russia.}
\author{M.~Mitrovski}
\affiliation{Fachbereich Physik der Universit\"{a}t, 
             Frankfurt, Germany.}
\author{S.~Mr\'owczy\'nski}
\affiliation{Institute of Physics, Jan Kochanowski University, 
             Kielce, Poland.}
\author{V.~Nicolic}
\affiliation{Rudjer Boskovic Institute, 
             Zagreb, Croatia.}
\author{G.~P\'{a}lla}
\affiliation{KFKI Research Institute for Particle and Nuclear Physics, Budapest, Hungary.} 
\author{A.D.~Panagiotou}
\affiliation{Department of Physics, University of Athens, 
             Athens, Greece.}
\author{W.~Peryt}
\affiliation{Faculty of Physics, Warsaw University of Technology, 
             Warsaw, Poland.}
%\author{M.~Pikna}
%\affiliation{Comenius University, 
%             Bratislava, Slovakia.}
\author{J.~Pluta}
\affiliation{Faculty of Physics, Warsaw University of Technology, 
             Warsaw, Poland.}
\author{D.~Prindle}
\affiliation{Nuclear Physics Laboratory, University of Washington,
             Seattle, WA, USA.} 
\author{F.~P\"{u}hlhofer}
\affiliation{Fachbereich Physik der Universit\"{a}t, 
             Marburg, Germany.}
\author{R.~Renfordt}
\affiliation{Fachbereich Physik der Universit\"{a}t, 
             Frankfurt, Germany.}
\author{C.~Roland}
\affiliation{MIT, 
             Cambridge, USA.}
\author{G.~Roland}
\affiliation{MIT, 
             Cambridge, USA.}
\author{M.~Rybczy\'nski}
\affiliation{Institute of Physics, Jan Kochanowski University, 
             Kielce, Poland.}
\author{A.~Rybicki}
\affiliation{Henryk Niewodniczanski Institute of Nuclear Physics, 
             Polish Academy of Sciences, 
             Cracow, Poland.}
\author{A.~Sandoval}
\affiliation{GSI Helmholtzzentrum f\"{u}r Schwerionenforschung,
             Darmstadt, Germany.} 
\author{N.~Schmitz}
\affiliation{Max-Planck-Institut f\"{u}r Physik, 
             Munich, Germany.}
\author{T.~Schuster}
\affiliation{Fachbereich Physik der Universit\"{a}t, 
             Frankfurt, Germany.}
\author{P.~Seyboth}
\affiliation{Max-Planck-Institut f\"{u}r Physik, 
             Munich, Germany.}
\author{F.~Sikl\'{e}r}
\affiliation{KFKI Research Institute for Particle and Nuclear Physics, Budapest, Hungary.} 
%\author{B.~Sitar}
%\affiliation{Comenius University, 
%             Bratislava, Slovakia.}
\author{E.~Skrzypczak}
\affiliation{Institute for Experimental Physics, University of Warsaw,
             Warsaw, Poland.} 
\author{M.~S{\l}\-odkowski}
\affiliation{Faculty of Physics, Warsaw University of Technology, 
             Warsaw, Poland.}
\author{G.~Stefanek}
\affiliation{Institute of Physics, Jan Kochanowski University, 
             Kielce, Poland.}
\author{R.~Stock}
\affiliation{Fachbereich Physik der Universit\"{a}t, 
             Frankfurt, Germany.}
\author{H.~Str\"{o}bele}
\affiliation{Fachbereich Physik der Universit\"{a}t, 
             Frankfurt, Germany.}
\author{T.~Susa}
\affiliation{Rudjer Boskovic Institute, 
             Zagreb, Croatia.}
\author{M.~Szuba}
\affiliation{Faculty of Physics, Warsaw University of Technology, 
             Warsaw, Poland.}
\author{M.~Utvi\'{c}}
\affiliation{Fachbereich Physik der Universit\"{a}t, 
             Frankfurt, Germany.}
\author{D.~Varga}
\affiliation{KFKI Research Institute for Particle and Nuclear Physics, Budapest, Hungary.} 
\affiliation{CERN, 
             Geneva, Switzerland.}
\author{M.~Vassiliou}
\affiliation{Department of Physics, University of Athens, 
             Athens, Greece.}
\author{G.I.~Veres}
\affiliation{KFKI Research Institute for Particle and Nuclear Physics, Budapest, Hungary.} 
\affiliation{MIT, Cambridge, USA.}
\author{G.~Vesztergombi}
\affiliation{KFKI Research Institute for Particle and Nuclear Physics, Budapest, Hungary.}
\author{D.~Vrani\'{c}}
\affiliation{GSI Helmholtzzentrum f\"{u}r Schwerionenforschung,
             Darmstadt, Germany.}
\author{Z.~W{\l}odarczyk}
\affiliation{Institute of Physics, Jan Kochanowski University, 
             Kielce, Poland.}
\author{A.~Wojtaszek-Szwarc}
\affiliation{Institute of Physics, Jan Kochanowski University, 
             Kielce, Poland.}

%============================================================================

\collaboration{The NA49 collaboration}
\noaffiliation

%============================================================================

\date{\today }

%============================================================================

\begin{abstract}

Production of the $K^{\ast}(892)^0$ and $\overline{K}^{\ast}(892)^0$ resonances was studied via their
$K^+ \pi^-$ and $K^- \pi^+$ decay modes in central
Pb+Pb, Si+Si, C+C and inelastic p+p collisions at 158\agev~($\sqrt{s_{NN}}$ = 17.3~GeV) 
with the NA49 detector at the CERN SPS. 
Transverse momentum and rapidity distributions were measured and total yields were estimated.
The yield of $K^{\ast}$ exceeds that of $\overline{K}^{\ast}$ by about a factor of two in nucleus-nucleus reactions.
The total yield ratios $\langle K^{\ast} \rangle$/$\langle K^+ \rangle$ and
$\langle \overline{K}^{\ast} \rangle$/$\langle K^-\rangle$ are strongly suppressed in central Pb+Pb
compared to p+p, C+C and Si+Si collisions in agreement with the expected attenuation 
of these short-lived resonance states in the hadronic phase of the expanding
fireball. The UrQMD model, although incorporating such a scenario, does not provide 
a quantitative description of the experimental results. The statistical hadron gas
model assuming the same freeze-out parameters for stable hadrons and resonances overestimates
the $\langle K^{\ast} \rangle$/$\langle K \rangle$ ratios in central Pb+Pb collisions by about a factor of 2.5.

\end{abstract}

%============================================================================

%\keywords{}

\maketitle

\section{Introduction}

High-energy collisions of heavy nuclei produce a transient state of extreme 
energy and matter density in which quarks and gluons are probably briefly 
deconfined \cite{Hz_Jac_2000,deconf,Ga_Go_Se_2011}. 
Production of entropy and of $s$, $\overline{s}$ quarks is believed to occur at the
early stage of the collision and this process is expected to be sensitive to the phase of the
created matter \cite{stenh,gazdz}. The high-density state evolves
into a hadron-resonance gas which finally decouples into the observed
hadrons. The $K^{\ast}(892)$ and $\overline{K}^{\ast}(892)$ resonance states contain
an $\overline{s}$ and $s$ valence quark, respectively, and are therefore sensitive to the
level of strangeness production. However, resonance states have lifetimes similar to that of the 
fireball and may interact in the dense medium in which they are produced. Their mass and width
could be affected \cite{rapp00} and scattering processes might destroy or regenerate 
them. Furthermore, daughters of those $K^{\ast}$ that decay inside the fireball may rescatter 
resulting in a changed invariant-mass spectrum. Thus the yields contained in the $K^{\ast}$
mass peak were conjectured to be sensitive to the duration and properties of
the hadronic fireball stage \cite{torr01}. 

Studies of $K^{\ast}(892)$ production at mid-rapidity in Au+Au, Cu+Cu and p+p collisions 
at RHIC energies were performed by the STAR collaboration \cite{adler02,adams05}.
This paper reports measurements of $K^{\ast}(892)^0$ and 
$\overline{K}^{\ast}(892)^0$ resonance production via their $K^+ \pi^-$ and $K^- \pi^+$ decay modes
at the CERN SPS in central Pb+Pb, Si+Si, C+C and inelastic p+p collisions at 158\agev~
($\sqrt{s_{NN}}$~=~17.3~GeV). Preliminary results
were presented in \cite{qm08}. Section~II briefly describes the NA49 detector. Section~III discusses 
the analysis procedure. Distributions of transverse momentum $p_T$ and center-of-mass 
rapidity $y$ as well as total yields are presented in Section~IV. 
These results are compared to predictions of the 
ultrarelativistic quantum molecular dynamics (UrQMD) model \cite{urqmd_bass} and 
a statistical hadron gas model (HGM) \cite{Be:05} in Section~V. 
The paper ends with the summary Section~VI.

%-----------------------------------------------------------------

\section{Detector}

The NA49 experimental apparatus~\cite{na49_nim} consists of four large-volume
time projection chambers (TPC).  Two of these (VTPC) are placed
in the fields of two super-conducting dipole magnets.  The other two
(MTPC) are positioned downstream of the magnets and are optimized
for high-precision measurements of the ionization energy loss \dedx\ 
with a resolution of about 4\%. The particle identification provided by 
the \dedx\ measurement is complemented in the mid-rapidity domain 
by a measurement of the
time-of-flight (TOF) with a resolution of about 60~ps in two TOF
detector arrays positioned downstream of the MTPCs. The magnetic fields
were set to about 1.5~T (upstream magnet) and 1.1~T (downstream magnet).
With the lower momentum cut employed for \dedx\ identification the 
detector acceptance covers the forward rapidity region for $K^{\ast}(892)$.

The precise transverse position of each beam particle at the target was 
measured by three pairs of small proportional wire chambers (BPD) upstream of the target
with a precision of better than 200~$\mu$m.
Lead ions of 158\agev~impinged on a thin Pb-foil target 
of 337~mg/cm$^2$ (approximately 1.5 \% interaction probability for Pb ions)
which was positioned 80~cm upstream from the first VTPC. 
For the study of C+C and Si+Si collisions a 3~mm thick C
(2.4\% interaction length) and 5~mm thick Si target (4.4\%) were used,
respectively. The incident C and Si nuclei were produced by
fragmentation of a Pb beam of 158\agev~beam energy~\cite{na49_nim}
and were selected by magnetic rigidity ($Z/A = 0.5$) and 
by specific energy loss in the BPDs.
The "C-beam" as defined by the online trigger
and offline selection was a mixture of ions with $Z=6$ and 7
(intensity ratio 69:31); the "Si-Beam" of ions with $Z=13, 14$ and
15 (intensity ratio 35:41:24).
The trigger selected the centrality of the collisions based on a
measurement of the energy deposited by projectile spectator nucleons in
a downstream calorimeter.

For the study of p+p collisions the beam line was set to
select secondary protons of 158~GeV/$c$ momentum which were produced in a Be
target by the 400~GeV/$c$ SPS proton beam. The secondary protons were identified by
Cherenkov counters in the H2 beamline resulting in a contamination by pions and kaons
of less then $10^{-3}$. Liquid hydrogen targets of 14~cm (year
1996) and 20~cm (later years, 2.8~\% interaction length) and 3~cm diameter were used. A
scintillation counter S4 of 2~cm diameter was positioned about 5~m downstream
on the deflected beam line between the two VTPCs. It was used in
anticoincidence with the beam in order to select p+p interactions.
For a detailed description of detector aspects for p+p collisions see~\cite{na49_pions_in_pp}.

\section{Data analysis}

The analysis of Pb+Pb reactions is based on a high-statistics data run which 
recorded about $3 \cdot 10^6$ collisions. The trigger selected the 23.5\% 
most central Pb+Pb collisions.
The corresponding mean number of wounded nucleons $\enwee$~\cite{bialas}
was calculated using the VENUS simulation code~\cite{venus} following 
the Glauber model approach, and found to be 
$\langle \enwee \rangle = 262 $ with a systematic uncertainty of $\pm5$.
More details on the procedure can be found in Ref.\cite{Laszlo_cent}.

The C+C and Si+Si collision data are more limited in statistics. For both
systems about $45\cdot 10^{3}$ events were recorded for the
$(15.3\pm2.4)$\% and $(12.2\pm1.8)$\% most central C+C and Si+Si
collisions, respectively. The corresponding mean numbers of wounded
nucleons based on VENUS simulations are $14\pm2$ and
$37\pm3$~\cite{syssz_2005}.

Results on p+p reactions are based on $1.125\cdot 10^{6}$ ($4.18\cdot 10^{5}$)
events collected with the 20~(14)~cm long liquid hydrogen target. The trigger cross section
was 28.3$\pm$0.1 mb which contained about 86~\% of the inelastic 
cross section and excludes most of the elastic collisions
(about 1~mb remaining contamination)~\cite{na49_pions_in_pp}.

Charged particle tracks were reconstructed from the charge
deposited along the particle trajectories in
the TPCs using a global tracking scheme which combines track segments that
belong to the same physical particle but were detected in different TPCs.
A vertex fit was then performed using the reconstructed tracks.
Particle identification is based on measurements of the 50~\% truncated mean of the specific energy
loss $dE/dx$ in the TPCs which provide up to 234 charge samples on a track. The uncertainty
of the $dE/dx$ measurement for a specific track depends on its visible length and the
number of associated charge clusters. The average value of $dE/dx$ is a universal 
function of the velocity of a charged particle (Bethe-Bloch curve) and thus 
depends on its mass for a given momentum.

\subsection{Pb+Pb collisions}

For Pb+Pb collisions the event vertex was determined using the tracks 
reconstructed in the TPCs. The resulting
vertex distribution had widths $\sigma(x) = 0.21$~cm and $\sigma(y) = 0.15$~cm
in the coordinates transverse to the beam, and $\sigma(z) = 1.3$~cm in the
longitudinal direction. Events were accepted if the vertex $z$ coordinate
was within 1 cm of the nominal target position. With this requirement the background
from non-target interactions was negligible. Further cuts were applied at the
track level. The distance of the back-extrapolated track from the fitted
vertex position had to be below 5~cm in the (horizontal) bending plane and
below 3~cm in the vertical direction. Moreover, the number of measured
points on the track had to exceed 25 and constitute more than 50 \% of
the geometrically possible maximum in order to eliminate split tracks. 
Finally, the track momentum fit was
repeated including the vertex position resulting in a typical momentum resolution of 
$\sigma(p)/p^2 \approx (0.3 - 7) \cdot 10^{-4}$~(\gevc)$^{-1}$ depending on
track length.
 
\begin{figure}[htb]
\includegraphics[width=0.45\linewidth]{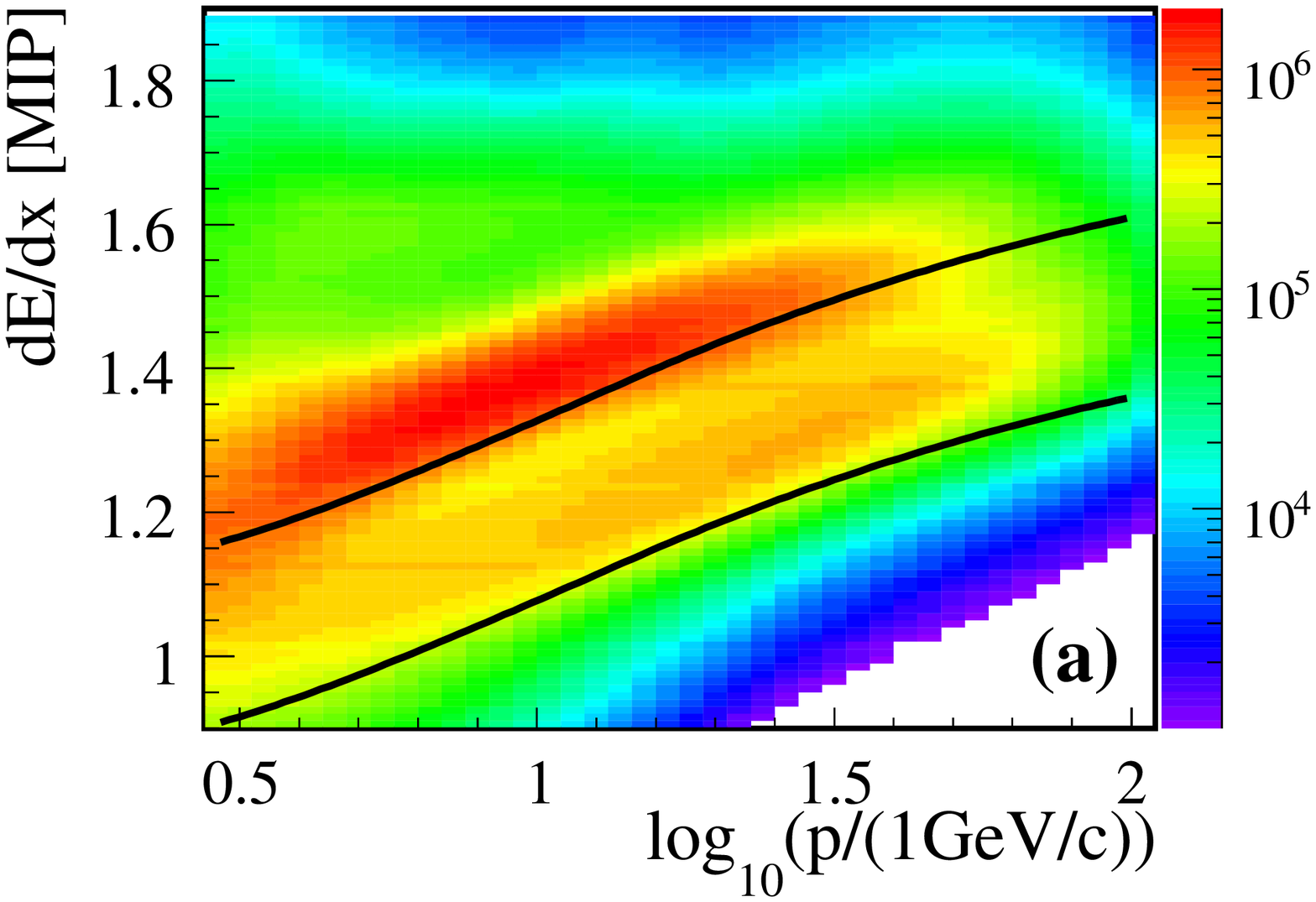}
\includegraphics[width=0.45\linewidth]{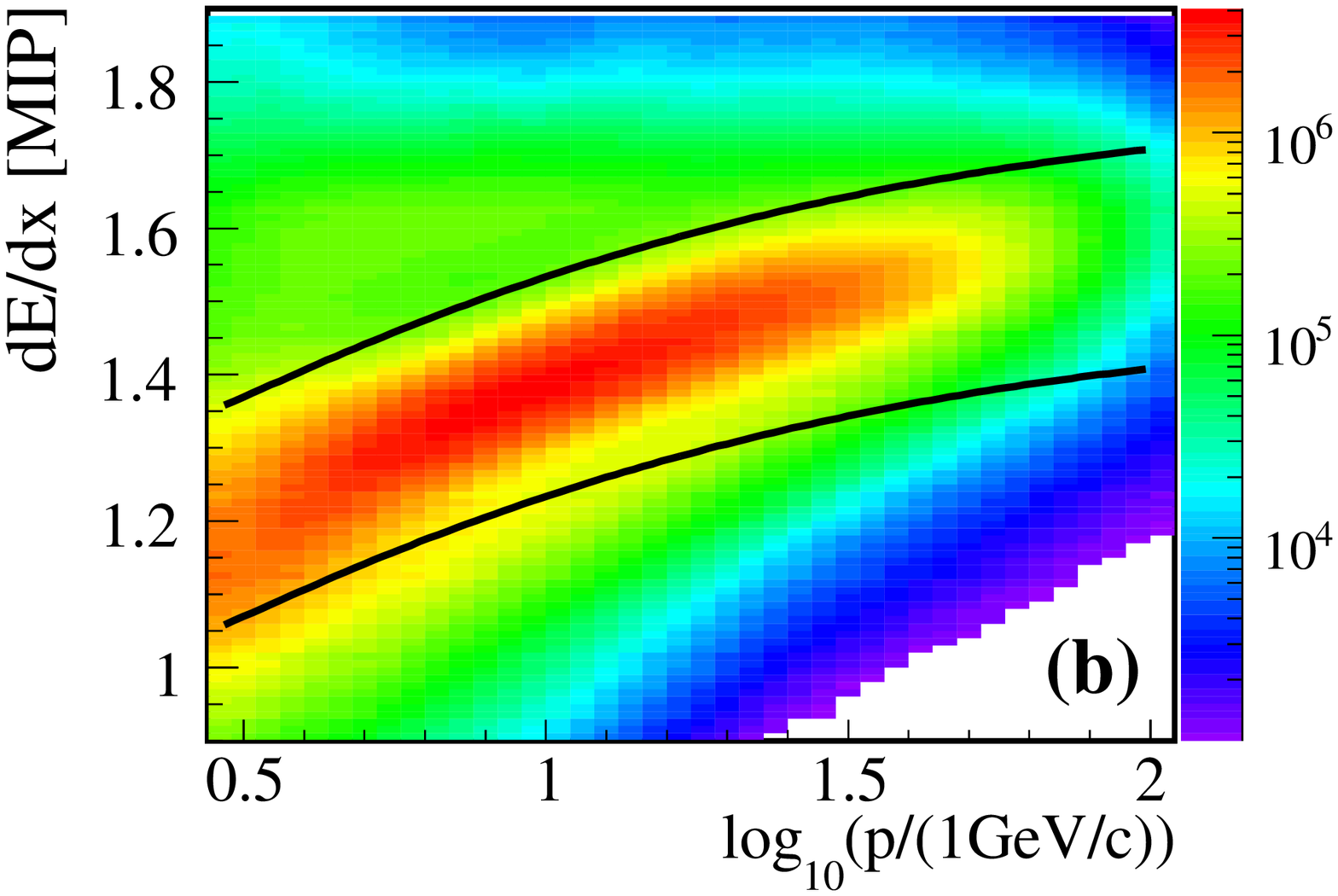}
\caption{\label{fig1} (Color online) 
  Specific energy loss $dE/dx$ measured in the NA49 TPCs versus momentum $p$ for positively (a)
  and negatively (b) charged particles in central Pb+Pb collisions. 
  Curves in (a) and (b) show the acceptance limits for $K^+$ and $\pi^-$, respectively.
}
\end{figure}

Figure~\ref{fig1} shows a density plot of $dE/dx$ as a function of momentum $p$ 
for accepted positively (a) and negatively (b) charged particles 
showing bands for various particle species. $K$ and $\pi$ meson candidates were
selected by requiring a momentum in the range 3$<p<$100~GeV/$c$  and
a measured $dE/dx$ value in a band of 2.5 and 3 standard deviations,
respectively, around the expected mean values. Expected losses due to the cuts
are small ($<$~2~\%) and no correction was applied. 
Systematic biases of the $K^*$ yields from uncertainties in the fit of the Bethe-Bloch 
function are estimated to be below $<$~5~\%.

% 
%-----------------------------------------------------------------
%------------------------------------------------------------------
%

Raw yields of the $K^{\ast}(892)^0$ and $\overline{K}^{\ast}(892)^0$ resonance states were
extracted from the invariant-mass distributions calculated for $K^+\pi^-$ and $K^-\pi^+$
pair candidates, respectively. First, invariant-mass  
\begin{eqnarray}\label{equEM}
m_{inv}(K\pi) = \sqrt{ (E_K + E_{\pi})^2 - (\overrightarrow{p}_K + \overrightarrow{p}_{\pi})^2 }
\end{eqnarray}
distributions were computed for all selected $K^+\pi^-$ and $K^-\pi^+$ candidate pairs in the events.
Next, similarly obtained distributions from pairs taken from different events of
the same multiplicity class and normalised to the number of real pairs
were subtracted in order to reduce the dominant contribution from
combinatorial background. The resulting invariant-mass distributions are plotted in \Fi{fig2}
for transverse momenta $p_T < 2.0$~GeV/$c$ and the rapidity range $0.6 < y < 0.9$. The peaks due to the 
$K^{\ast}(892)^0$ and $\overline{K}^{\ast}(892)^0$ resonance states are clearly seen above a strongly
mass-dependent residual background. The mixing procedure preserves the inclusive single-particle
phase space distributions, but destroys all correlations between particles. It therefore cannot 
fully describe the combinatorial background in real events which is presumably partly shaped by effects
such as energy and momentum conservation as well as reflections from other resonance states.
%
%--------------------------------------------------------------
\begin{figure}[htb]
\includegraphics[width=0.45\linewidth]{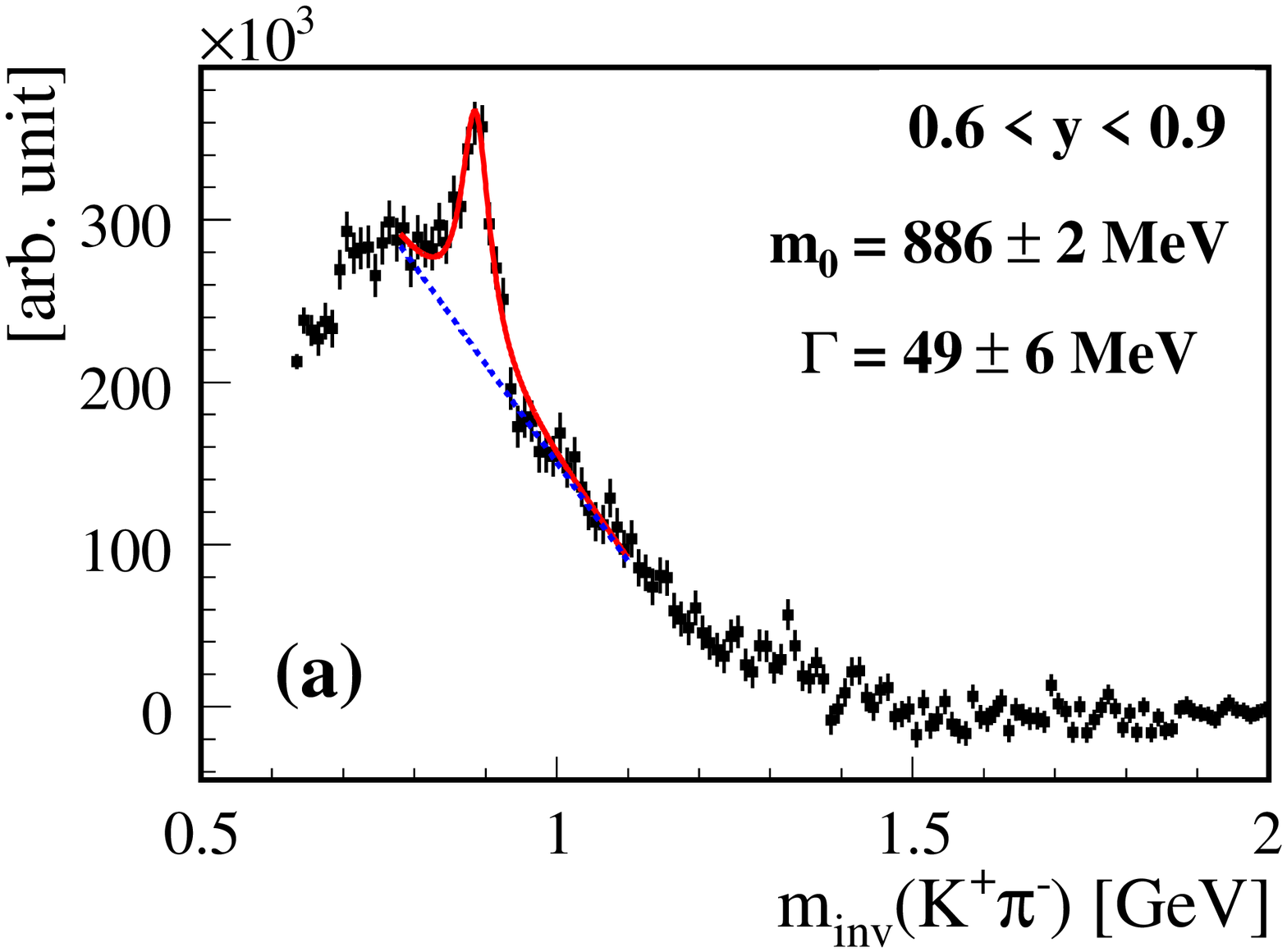}
\includegraphics[width=0.45\linewidth]{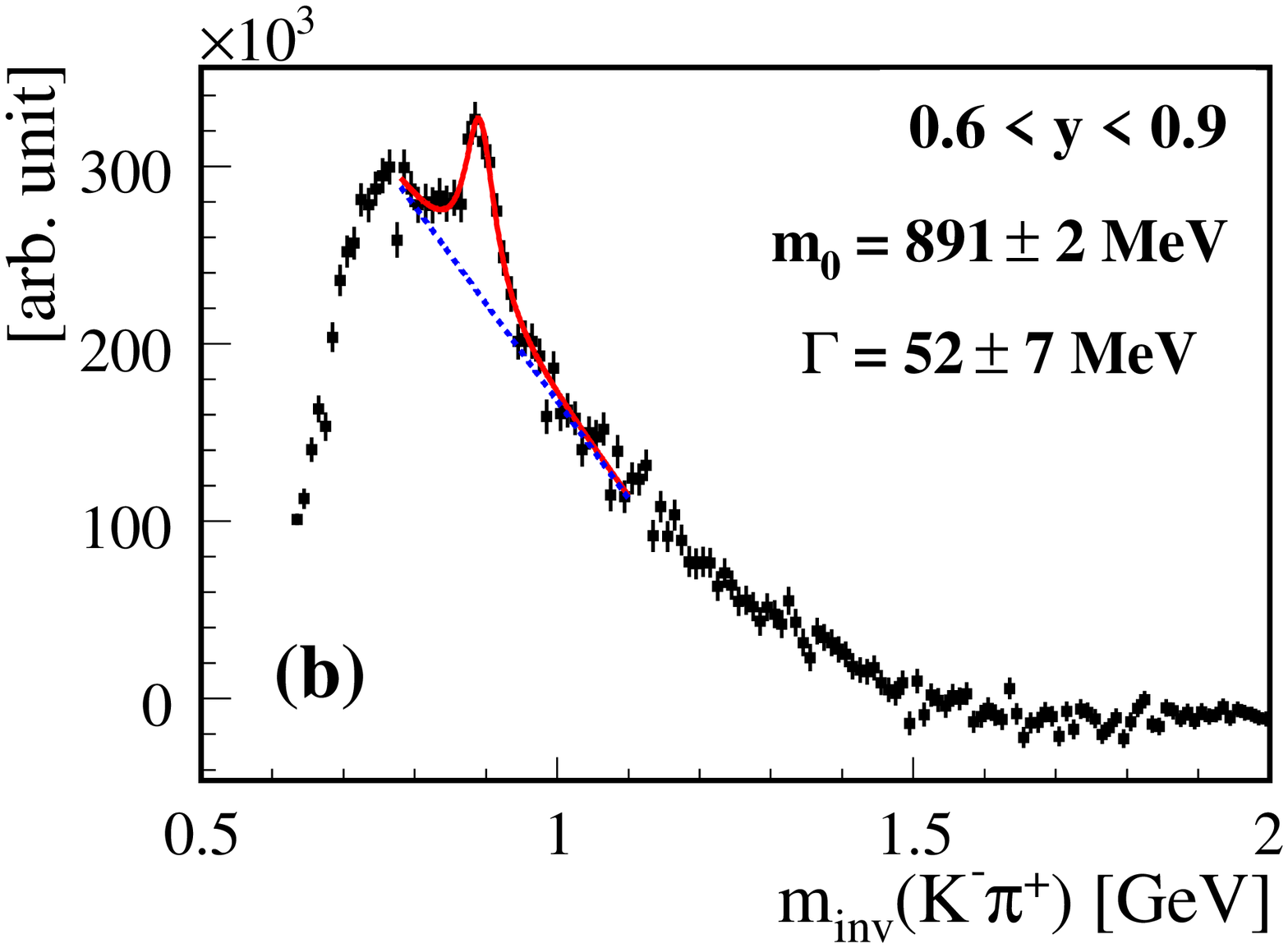}
\caption{\label{fig2} (Color online)
   Invariant mass distribution of $K^+\pi^-$ (a) and $K^-\pi^+$ (b)
   for $p_T < 2.0$~GeV/$c$ in the rapidity region $0.6 < y < 0.9$ after subtraction
   of mixed-pair background in central Pb+Pb collisions. 
   The fitted polynomial background is shown by the dashed
   curves, the sum of fitted polynomial background and signal Breit-Wigner function
   by the solid curves.
  }
\end{figure}
%---------------------------------------------------------------
%
The raw number of $K^{\ast}(892)^0$ and $\overline{K}^{\ast}(892)^0$ were derived from fits to the 
mass distributions after the mixed-event background had been subtracted. The fit function
was chosen as a sum of a linear or a 2nd order polynomial background and a 
Breit-Wigner distribution:
\begin{eqnarray}\label{equBW}
\frac{dN}{dm_{inv}} = C \cdot  \frac{\Gamma}{\pi\left(\left(
m_{inv}-m_0\right)^2+\left(\frac{\Gamma}{2}\right)^2\right)} ,
\end{eqnarray}
where $m_0$ and $\Gamma$ are mass and width of the $K^{\ast}$,
and $C$ is a normalisation factor. Examples of such fits in the rapidity range
$0.6 < y < 0.9$ are shown in \Fi{fig2}. The fits are seen to provide
a good description of the mass distributions in the fit range 
$780 < m_{inv} < 1100$~MeV and result in $m_0 = 886 \pm 2$~MeV and 
$\Gamma = 49 \pm 6$~MeV for $K^{\ast}(892)^0$
and $m_0 = 891 \pm 2$~MeV and $\Gamma = 52 \pm 7$~MeV for $\overline{K}^{\ast}(892)^0$.
The mass values are somewhat smaller than the world average of 
$895.94 \pm 0.22$~MeV \cite{pdg2010}, and were found not to depend significantly
on rapidity or $p_T$ as demonstrated by \Fi{fig3}~(a,b) (example for the $K^{\ast}(892)^0$
which provides the better statistical accuracy). Scaling of the magnetic field
value by the upper limit of its systematic uncertainty of $\pm$~1~\% changes the fitted mass value
by about $\pm$~5~MeV. Thus the observed mass shift is at the limit of significance. The STAR experiment
at RHIC also found a similarly reduced mass (see \Fi{fig3}~(b)) but only for
$p_T$ below about 1 \gevc~\cite{adams05}. The fitted width agrees well with the 
world average of $48.7 \pm 0.8$~MeV \cite{pdg2010} (the
invariant mass resolution is about 6 MeV~\cite{minvres}). No significant variation with
rapidity or $p_T$ (see~\Fi{fig3}~c,d) was found in agreement with results from STAR~\cite{adams05}. 

%
%--------------------------------------------------------------
\begin{figure}[htb]
\mbox{
\includegraphics[width=0.45\linewidth]{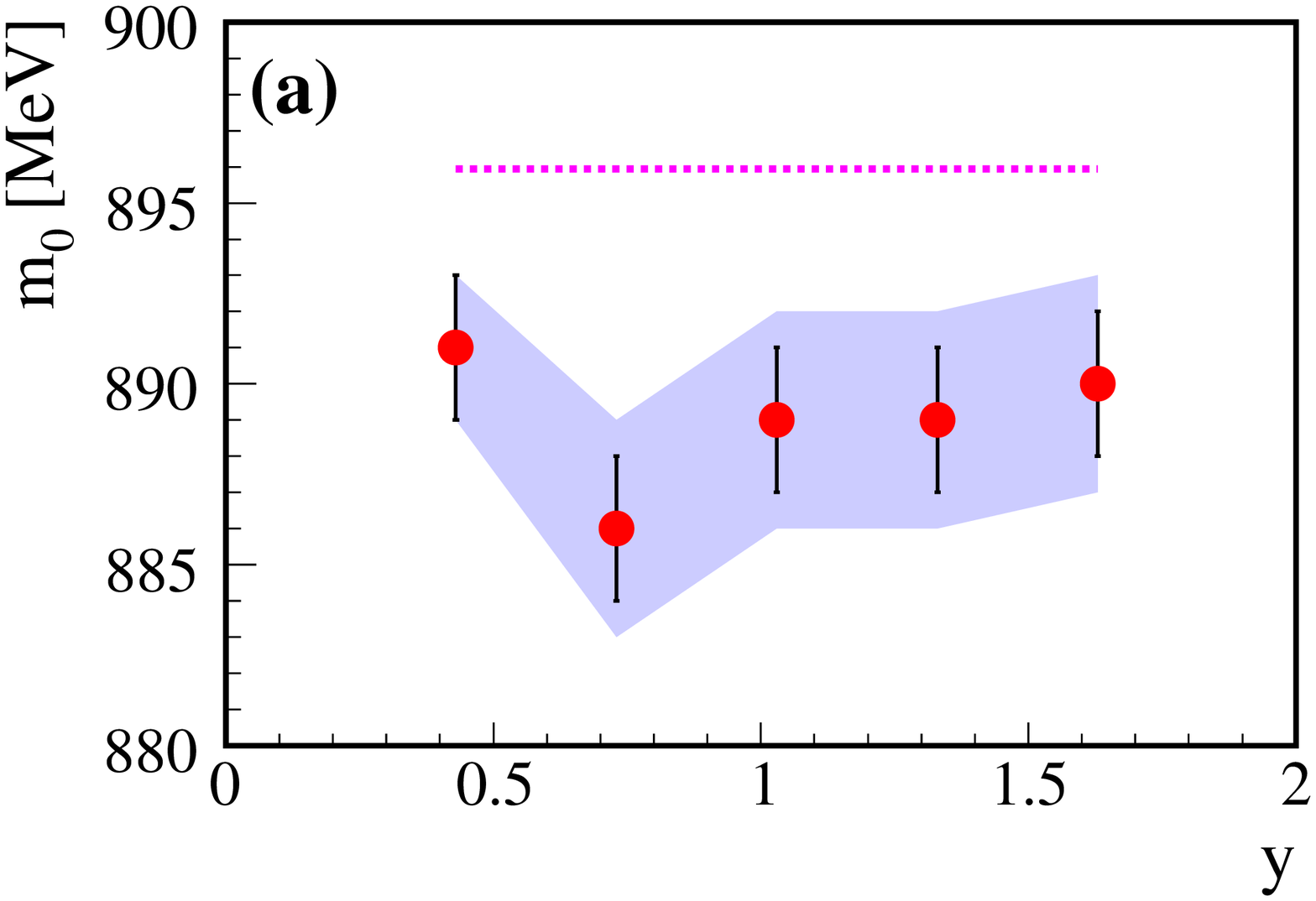}
\includegraphics[width=0.45\linewidth]{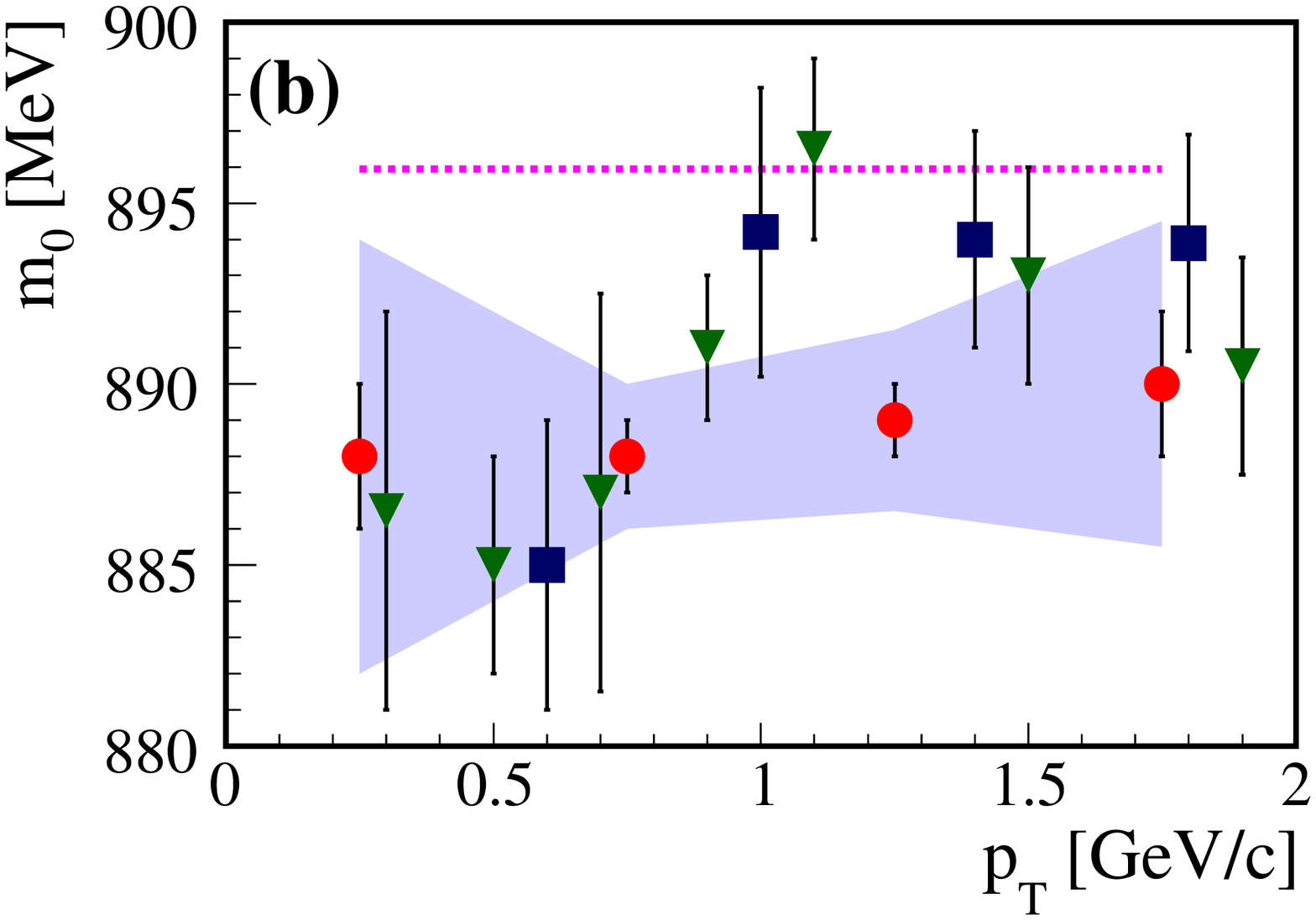}
}
\mbox{
\includegraphics[width=0.45\linewidth]{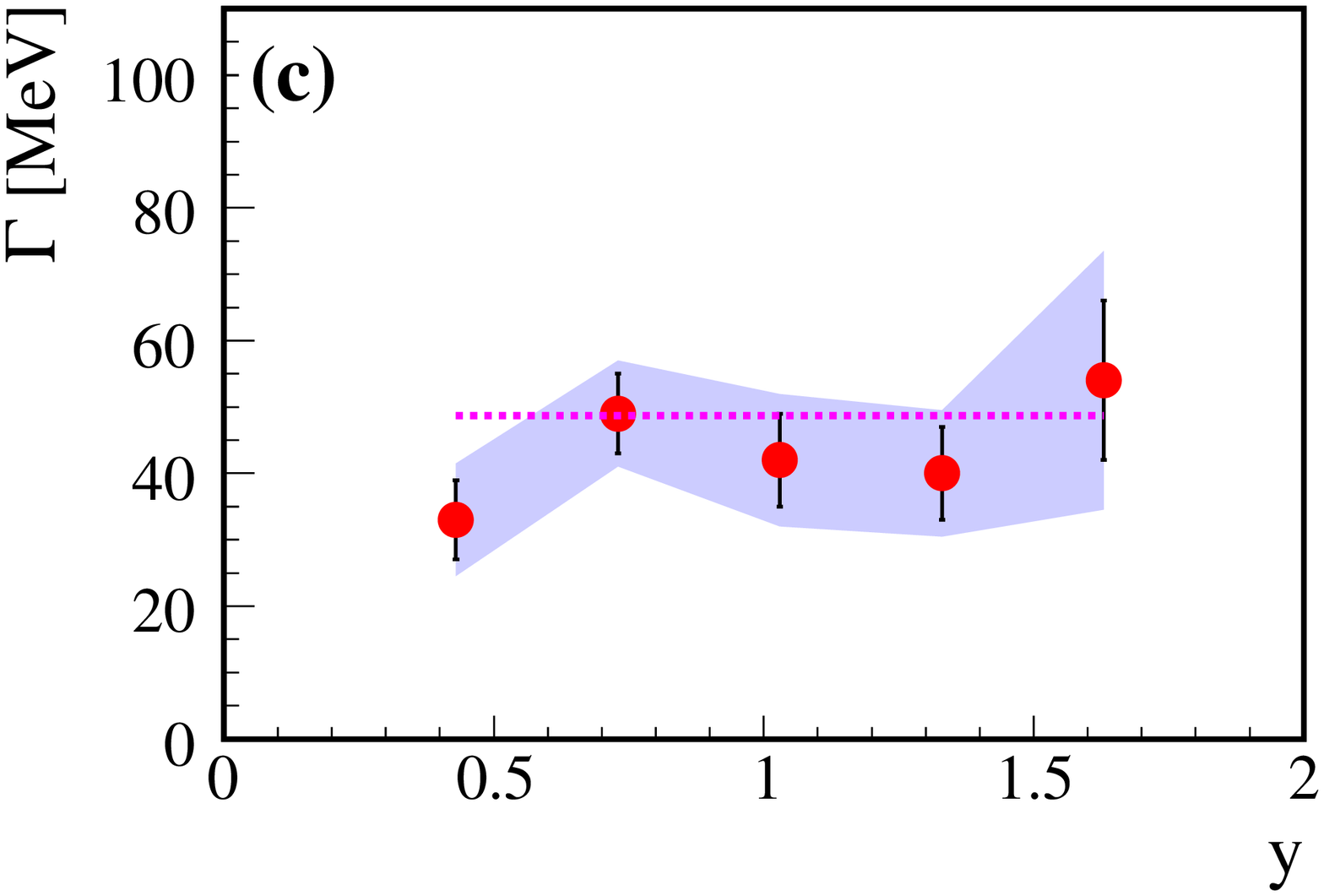}
\includegraphics[width=0.45\linewidth]{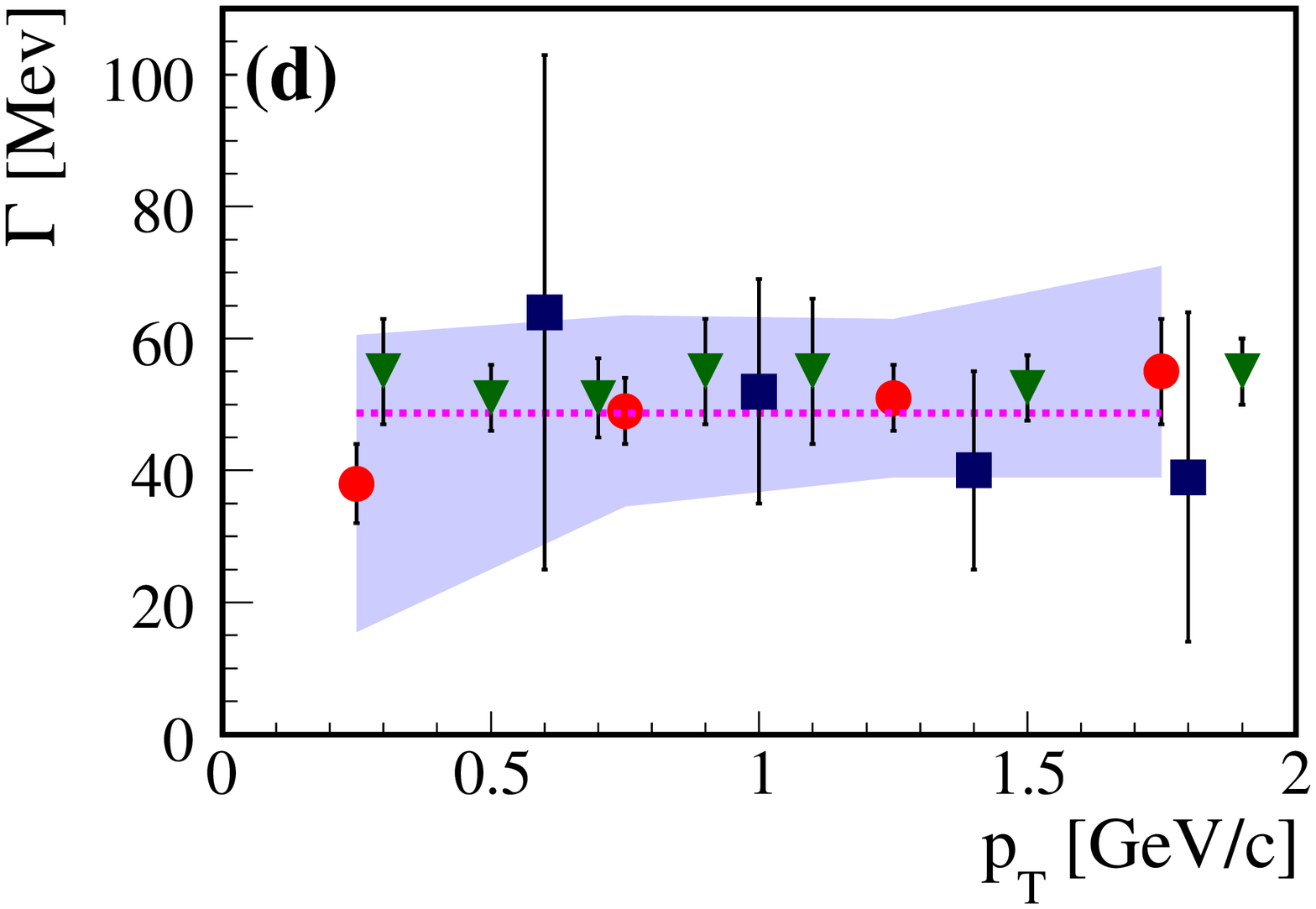}
}
\caption{\label{fig3} (Color online)
   Fitted mass values $m_0$ (a,b) and width $\Gamma$ (c,d) of the $K^{\ast}(892)^0$ 
   peak in the $K^+\pi^-$ mass distribution versus rapidity $y$ (a,c) 
   and transverse momentum $p_T$ (b,d)
   in central Pb+Pb collisions. Dots show the fitted values with statistical
   error bars, and bands indicate the systematic uncertainties. The dotted horizontal
   lines indicate the world average values for $m_0$ and $\Gamma$~\cite{pdg2010}. For comparison
   results are shown from the STAR collaboration at RHIC~\cite{adams05} for Au+Au collisions 
   at $\sqrt{s_{NN}}$= 62 GeV (triangles) and $\sqrt{s_{NN}}$= 200 GeV (squares).
  }
\end{figure}
%---------------------------------------------------------------
%

Correction factors for acceptance and reconstruction efficiency were derived from
Monte Carlo simulations. $K^{\ast}(892)^0$ and $\overline{K}^{\ast}(892)^0$ 
were generated with realistic distributions
in transverse momentum and rapidity and then passed through the NA49 simulation
chain based on GEANT~3.21~\cite{geant} and a specific TPC signal simulation software. These
signals were then added at the TPC signal level to the raw data of real 
events (embedding). Finally, the hybrid events were reconstructed and 
analysed like real events. A matching step associated the reconstructed tracks with the
originally generated tracks. Resulting invariant-mass spectra are plotted in
\Fi{fig4} and demonstrate that neither the mass peak position nor its width are 
affected by the measuring resolution.
%

%--------------------------------------------------------------
\begin{figure}[htb]
\includegraphics[width=0.45\linewidth]{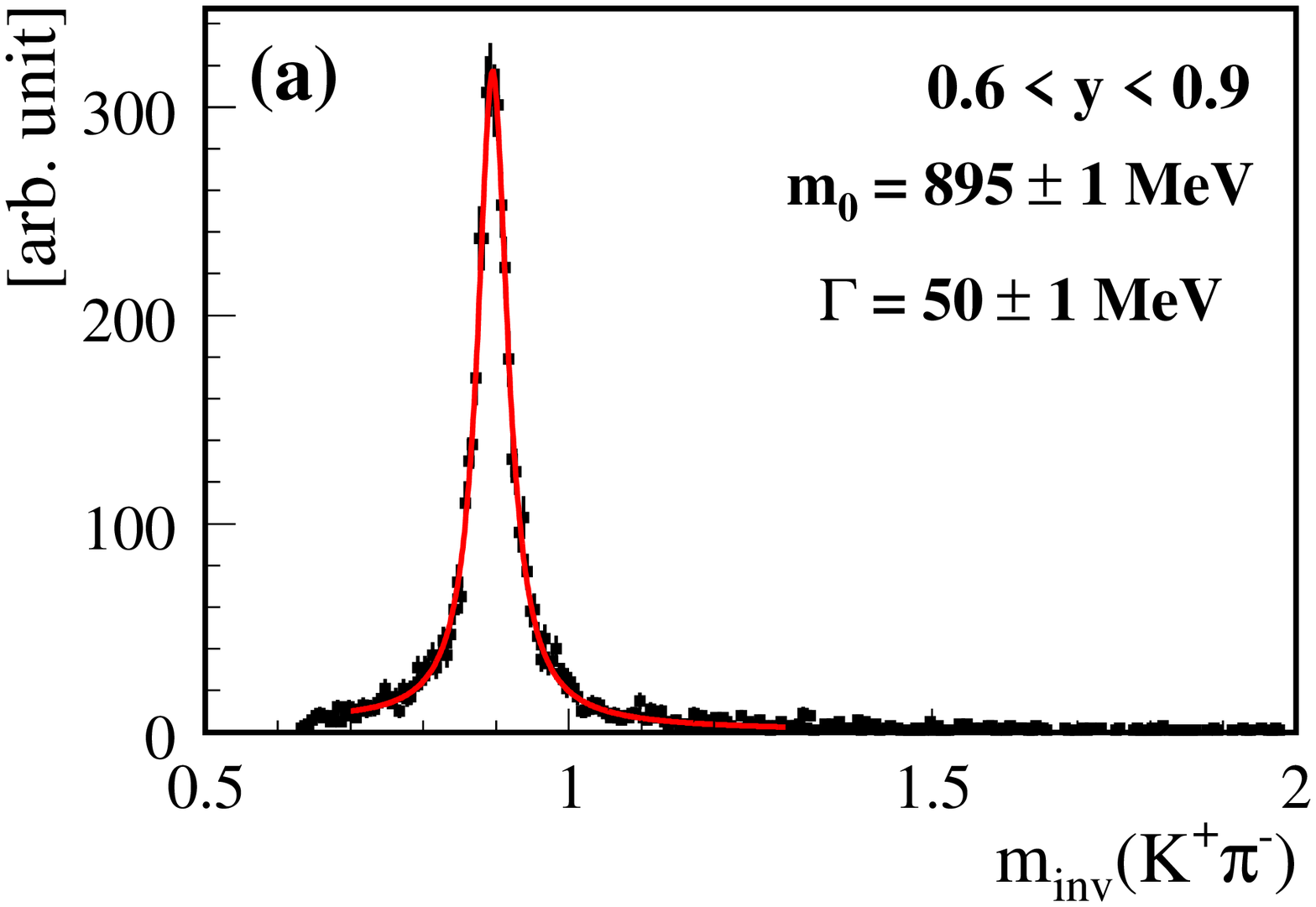}
\includegraphics[width=0.45\linewidth]{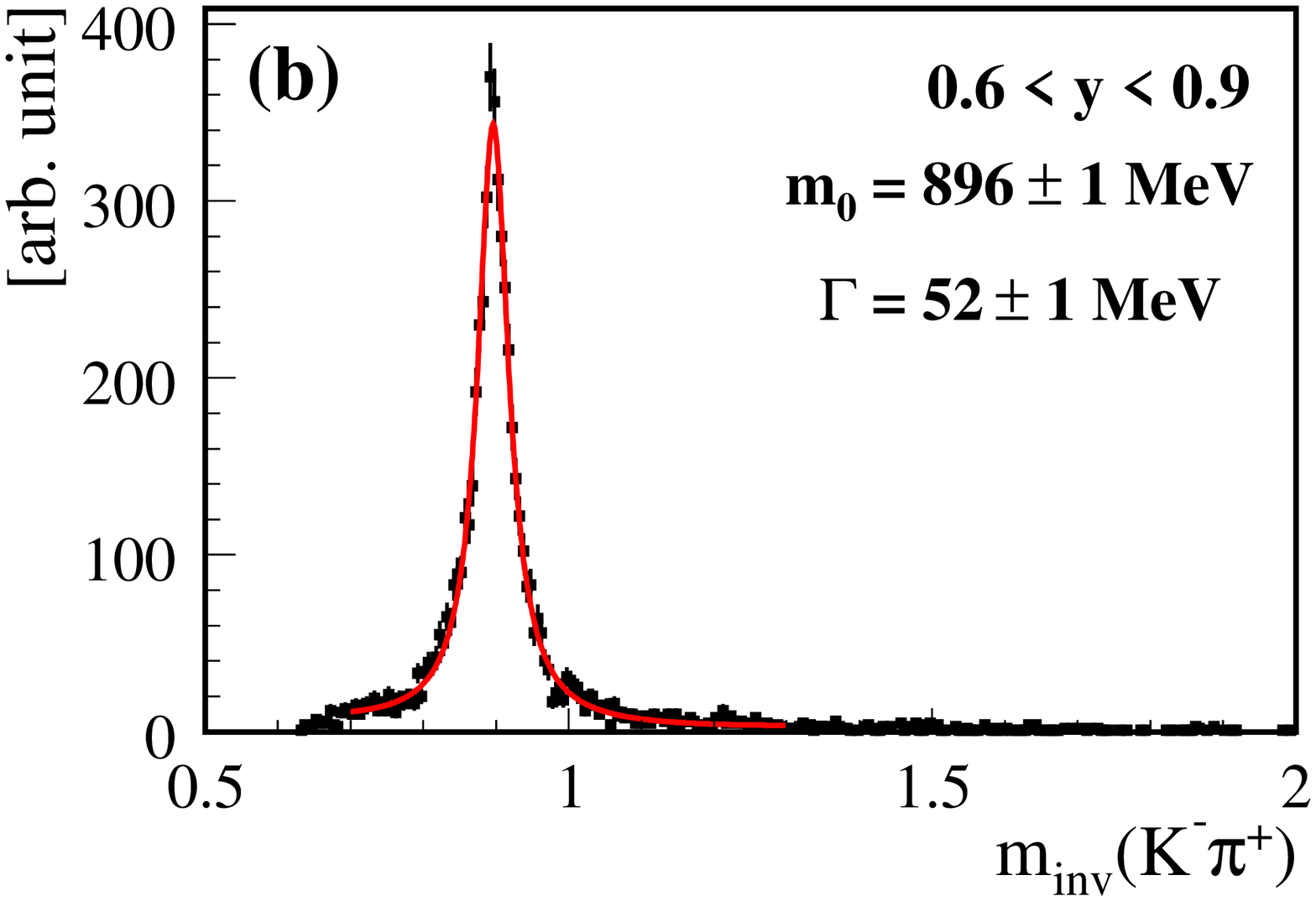}
\caption{\label{fig4} (Color online)
   Invariant mass spectra of simulated and embedded $K^{\ast}(892)^0$ (a) and 
   $\overline{K}^{\ast}(892)^0$ (b) calculated from reconstructed matched tracks.
  }
\end{figure}
%---------------------------------------------------------------
%
Correction factors for reconstruction inefficiencies, limited geometrical
acceptance and in-flight decays were obtained by comparing the
$K^{\ast}(892)^0$ and $\overline{K}^{\ast}(892)^0$ yields extracted from the reconstructed
hybrid events to the generated yield.
Resulting efficiencies for $K^{\ast}(892)^0 \rightarrow K^+ \pi^-$ 
are shown as a function of rapidity and transverse
momentum in \Fi{fig5} and range between 0.4 and 0.8. 
Values for $\overline{K}^{\ast}(892)^0 \rightarrow K^- \pi^+$
are the same within statistical errors. Efficiencies with embedding (full
symbols) are lower than those obtained by simulating only $K^{\ast}$ (open symbols)
but suggest only a modest effect of the high track density in the TPCs.
%
%--------------------------------------------------------------
\begin{figure}[htb]
\includegraphics[width=0.45\linewidth]{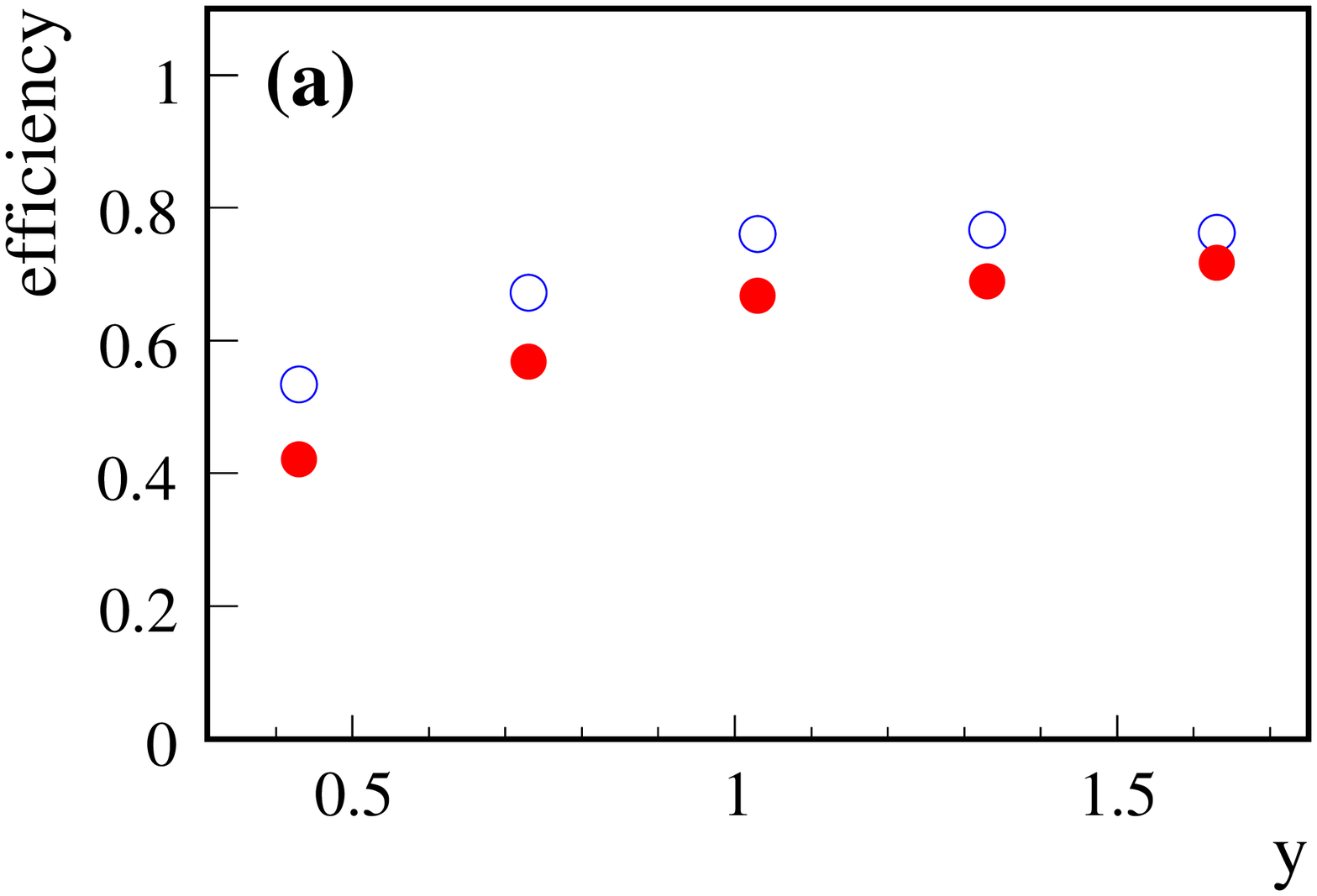}
\includegraphics[width=0.45\linewidth]{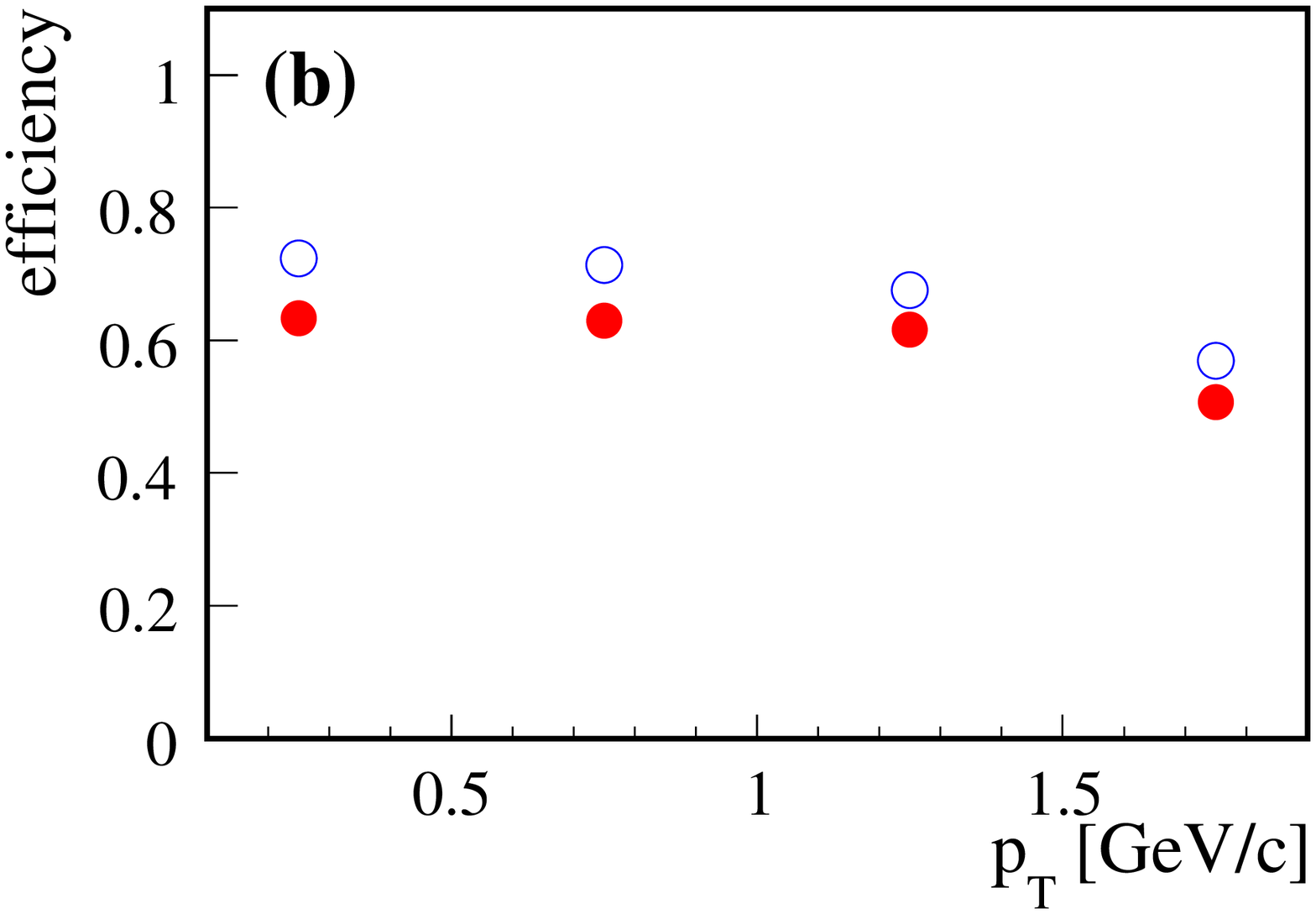}
\caption{\label{fig5} (Color online)
   Combined reconstruction efficiency and acceptance 
   of $K^{\ast}(892)^0 \rightarrow K^+ \pi^-$ as a function of rapidity 
   (transverse momentum range $0 < p_T < 2.0$~\gevc) (a)
   and transverse momentum (rapidity range $0.43 < y < 1.78$) 
   (b) in central Pb+Pb collisions. 
   Full and open symbols show results with and
   without embedding into real events (see text).
   The decay branching ratio is not included in the plotted efficiencies.
  }
\end{figure}
%---------------------------------------------------------------
%

Differential $K^{\ast}(892)^0$ yields were obtained by fitting the invariant-mass spectra
after subtraction of combinatorial background
in bins of transverse momentum and rapidity. In order
to ensure the stability of the fits the $K^{\ast}(892)^0$ mass and width 
were fixed to the world averages. The number of $K^{\ast}(892)^0$ in
each bin was calculated as the integral of the Breit-Wigner function. Correction factors 
were then applied for the reconstruction efficiency and the decay 
branching ratio (66.7~\%). 

Systematic errors were estimated by varying
the identification criteria for kaons and the details of the fit procedure applied to 
the mass distributions. Increasing or decreasing the width of the $dE/dx$ selection by 
half a standard deviation of the $dE/dx$ measurements led to changes in the yields of
around $\pm$~7\%. Extending or narrowing the mass range of the fit by 50 MeV or changing from
a linear to a 2nd order polynomial background affected the results by about $\pm$~10\%. Other
sources of uncertainty, like using a mass value different from the world average or
changing the inverse slope parameter $T$ in the efficiency
calculation, were much smaller. The total systematic error was estimated as half the range
covered by the results obtained when varying the $dE/dx$ cuts and the fitting procedure as
just described. More details of the analysis procedure for Pb+Pb collisions
can be found in \cite{thesis_slod}.

%--------------------------------------------------------------
\begin{figure}[htb]
\includegraphics[width=0.45\linewidth]{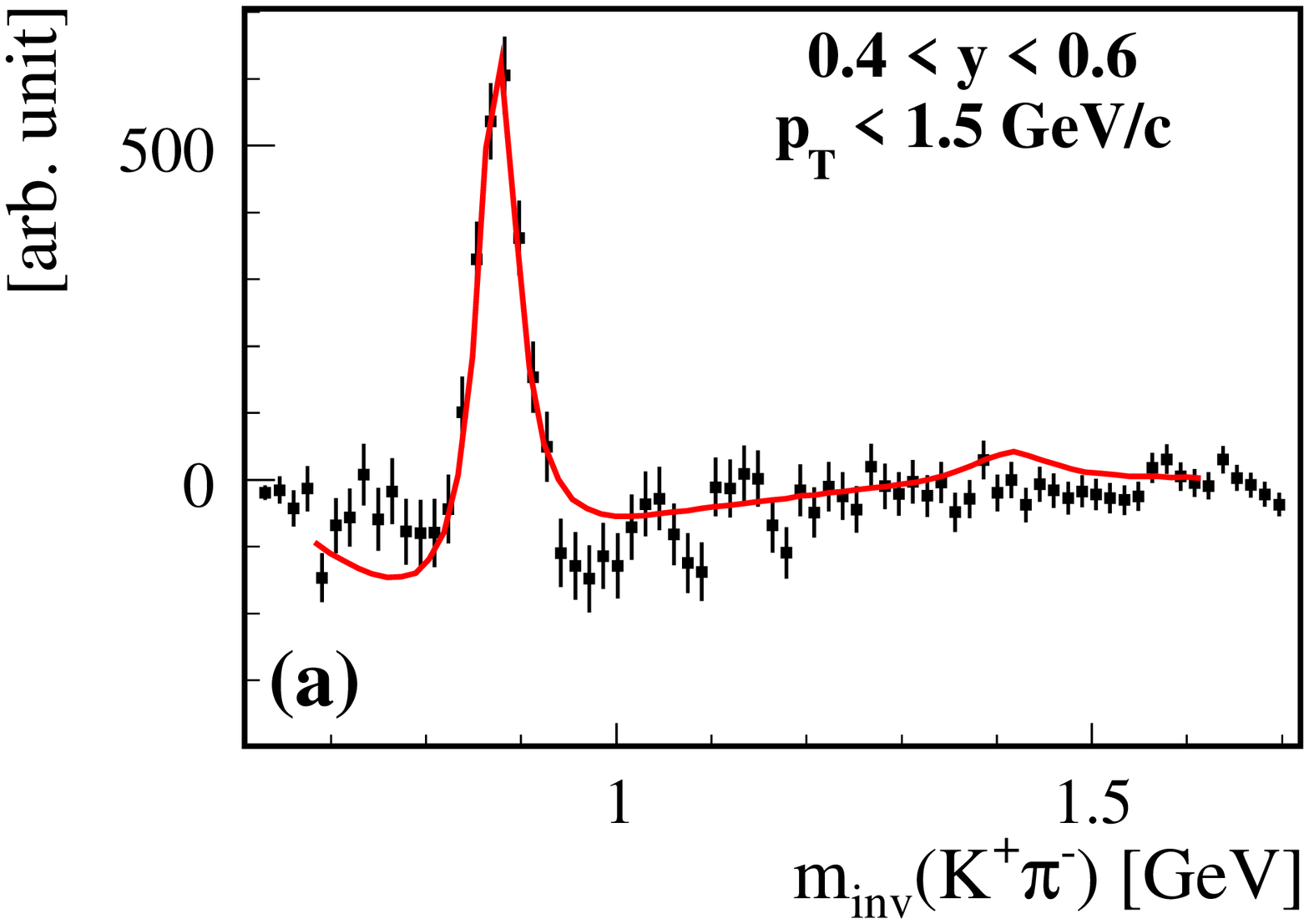}
\includegraphics[width=0.45\linewidth]{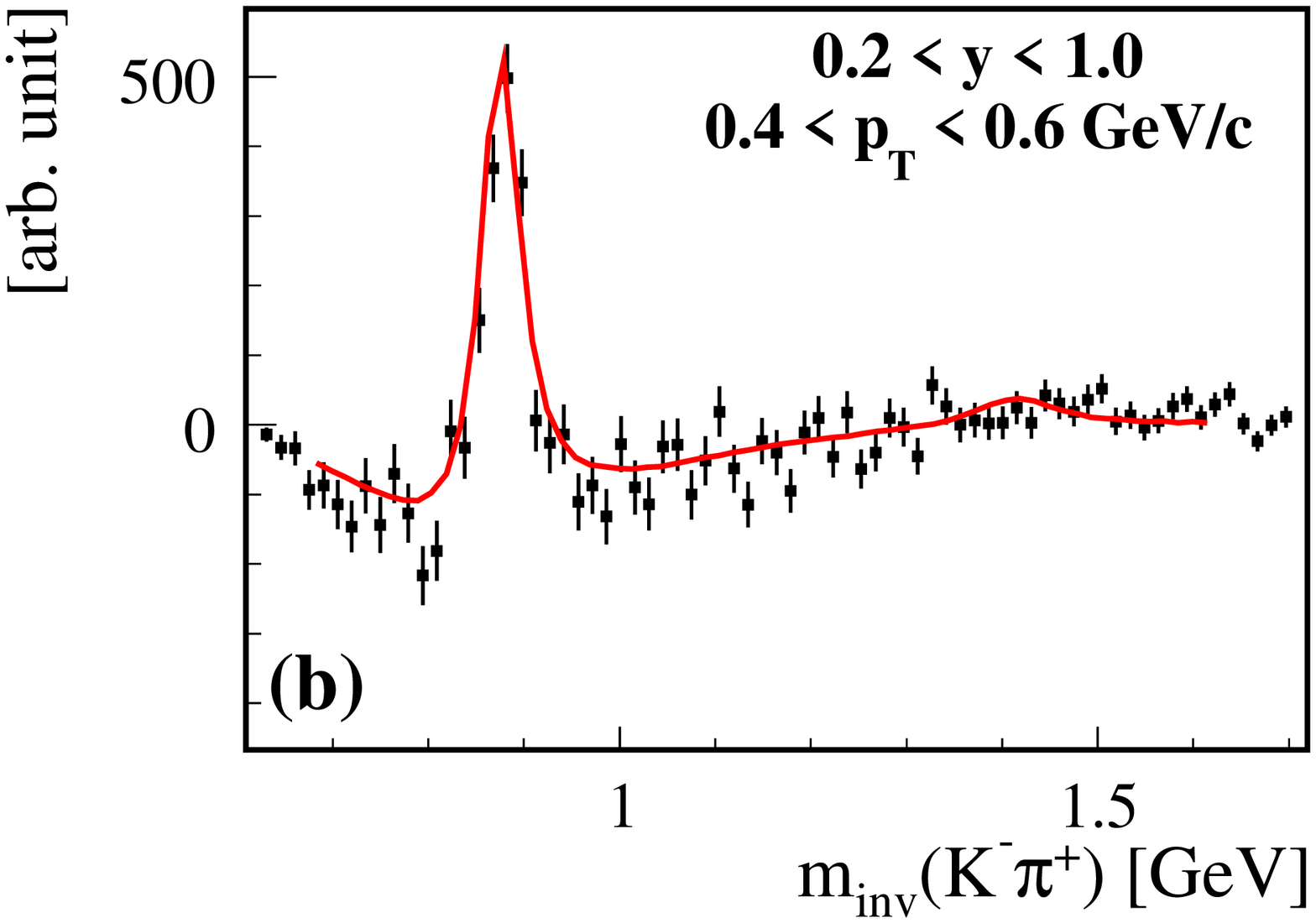}
\caption{\label{kstar_pp}
   Examples of invariant-mass distributions for $K^+\pi^-$ (a) and $K^-\pi^+$ (b)
   after subtraction of mixed-event background in inelastic p+p collisions. 
   The curves show the fits with the sum of Breit-Wigner functions to describe 
   the signals of the $K^{*}(892)^0$ and $K^{*}_{2}(1430)^0$ and the 
   contributions from their reflections (see text).
  }
\end{figure}
%---------------------------------------------------------------

\subsection{p+p collisions}

For p+p collisions the interaction vertex was determined from the back-extrapolated tracks
and the trajectory of the individual beam particle in the target which was measured
by the BPDs.
In order to obtain a clean sample of p+p collisions, only events with a successfully
fitted vertex differing in position by less than $\pm 9$~cm ($\pm 5.5$~cm for short target) in
z-(beam-)direction from the target center and having a radial distance of
less than 1~cm from the beam axis were accepted, thus minimizing
contributions from interactions in the mylar windows. 
From data taken with the liquid hydrogen removed from the
target vessel ("empty-target runs") the remaining fraction
of background events was estimated to be below 1~\% and therefore no
correction was applied.

Correction procedures were devised to obtain the yield per inelastic p+p collision.
The efficiency and accuracy of vertex
reconstruction using tracks back-extrapolated from the TPCs
vary with the charged-particle multiplicity of the event.
The efficiency was derived from the probability that a successful vertex fit is
obtained and that the fitted location fell inside the cuts. 
The corrections amount to 30\% for events
with 3 tracks, but rapidly drop to 6\% for events with 7 tracks.
Furthermore, about 7\% of the triggered events have no accepted tracks in the detector.
Half of these can be attributed to the 1 mb contribution of elastic scattering events
to the 28.3 mb trigger cross section, the other half are most likely due to singly diffractive
events. In order to obtain the yield per triggered inelastic event, the number
of events with tracks in the TPCs was scaled up by 3.5\% for the per event normalisation.

A further correction was applied for the $14.4 \pm 1$~\% of inelastic events which do not give rise 
to a trigger~\cite{na49_pions_in_pp}. Their contribution to the inclusive production cross section
was found to generally depend on $p_T$, Feynman $x$ and the type of the produced particle 
under consideration~\cite{na49_pp-kaon-paper}. However, for charged kaons only
a weak rapidity and no significant $p_T$ dependence was observed. 
For $K^*$ resonance production we assume a similar behavior and thus estimate that on
average the observed $K^*$ yield should be scaled up by $5 \pm 1$~\%. On the other hand,
the number of events used for normalisation has to be increased by 14.4~\% in order to obtain 
the yield per inelastic event. To account for the trigger loss we therefore scale down
the measured $K^{*}(892)^0$ and $\overline{K}^{\ast}(892)^0$ yields per event as defined
in the previous paragraph by $10 \pm 1$~\%. 

For p+p reactions invariant-mass spectra could be extracted with much 
lower combinatorial background than for A+A collisions. Selected kaons and pions 
were required to have more than 30 points per track, a momentum
in the interval  $4 < p < 50$~\gevc, a transverse
momentum of $ p_{T} \leq 1.5$~\gevc, and a measured $dE/dx$ value 
within $\pm 1.5$~$\sigma_{dE/dx}$ around the expected $dE/dx$ position.
Pairs were entered into the invariant-mass distributions with the
appropriate event-multiplicity dependent correction factors.
The background in the invariant-mass spectrum was
determined by mixing kaons and pions from different events. This
distribution was normalized to the same number of entries as the
real event spectrum and subtracted, resulting in a small undershoot around the $K^{*}(892)^0$
signal (see Fig.~\ref{kstar_pp}). This undershoot structure is well
described by simulations of invariant-mass distributions resulting from $K^{*}(892)^0$ decays
and the contributions of the $K^{*}(892)^0$ decay products to the 
mixed event background~\cite{event_mixing} (see curves in Fig.~\ref{kstar_pp}). 
No additional subtraction of remaining background is necessary here. 
In the simulation an expected contribution
from the $K^{*}_{2}(1430)^0$ (24~\% of the $K^{*}(892)^0$ yield~\cite{ehs86}) was accounted 
for. Its inclusion does improve the description of the invariant-mass
distribution, but does not influence the $K^{*}(892)^0$ yield. The
fitted mass of the $K^{*}(892)^0$ was 892 $\pm$ 5 MeV,
consistent with the world average \cite{pdg2010} The width was also found
to agree with the world average.

Yields of the $K^{*}(892)^0$ and $\overline{K}^{\ast}(892)^0$ were 
extracted by fitting the normalisation factor 
of the simulated invariant-mass distributions in bins of rapidity and
transverse momentum. The simulation used the world average values for masses and widths 
and took into account the effect of geometrical acceptance and kaon decay in flight.
Losses owing to reconstruction inefficiency are negligible
in p+p collisions. Global corrections were applied for $dE/dx$ particle
identification cuts and the decay branching fraction. Systematic errors were 
evaluated by changing track cuts of the selected kaons and pions and amount to 8\% for
the integrated yield. A conservative systematic error of 4\% is assigned for
the uncertainties of the vertex reconstruction efficiency and the trigger loss
corrections. The final systematic error is taken as the quadratic sum of all
these contributions and amounts to 9\%. More details on the analysis 
procedure for p+p collisions can be found in~\cite{hoehne2003}.

%
%--------------------------------------------------------------
\begin{figure}[htb]
\mbox{
\includegraphics[width=0.45\linewidth]{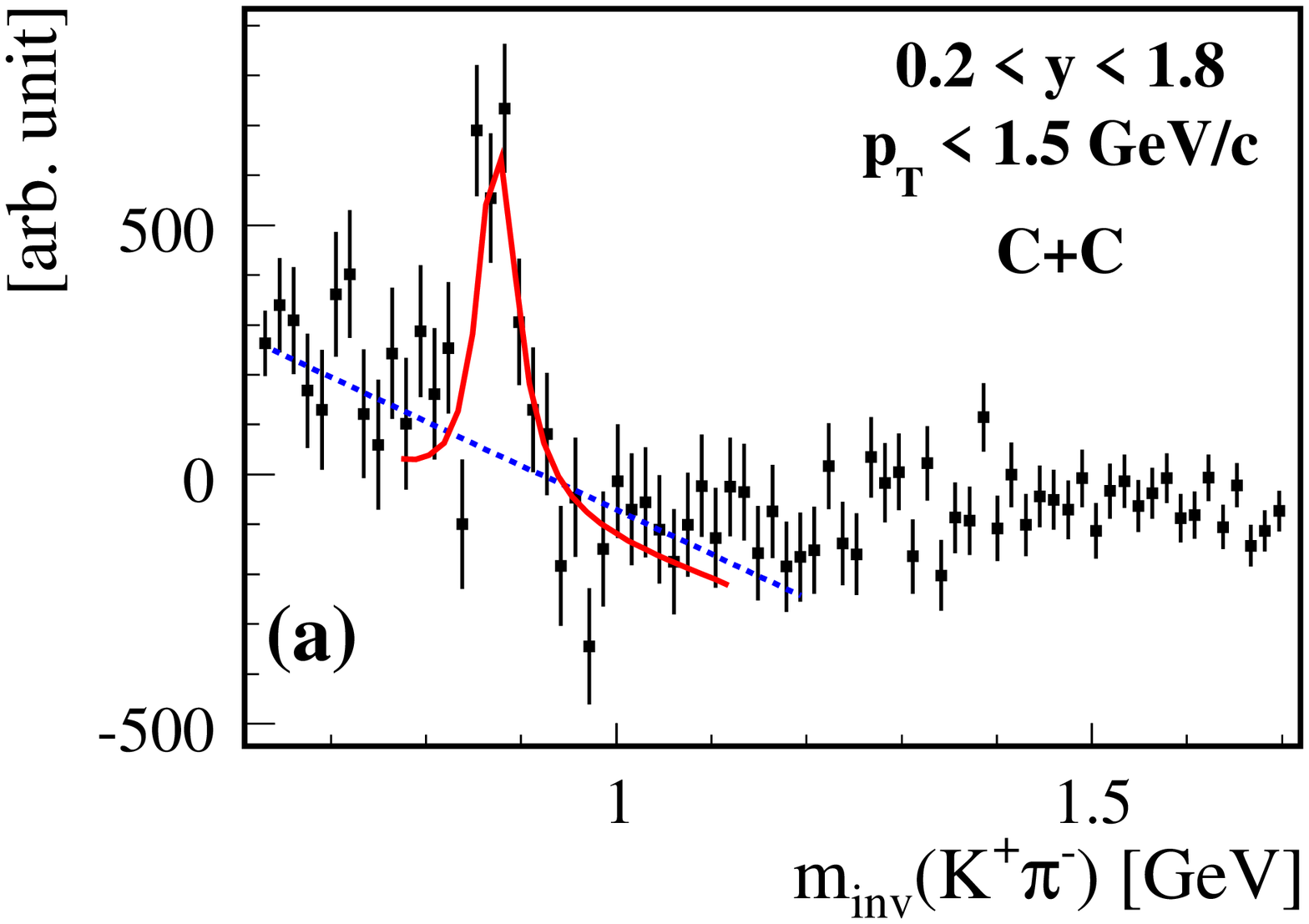}
\includegraphics[width=0.45\linewidth]{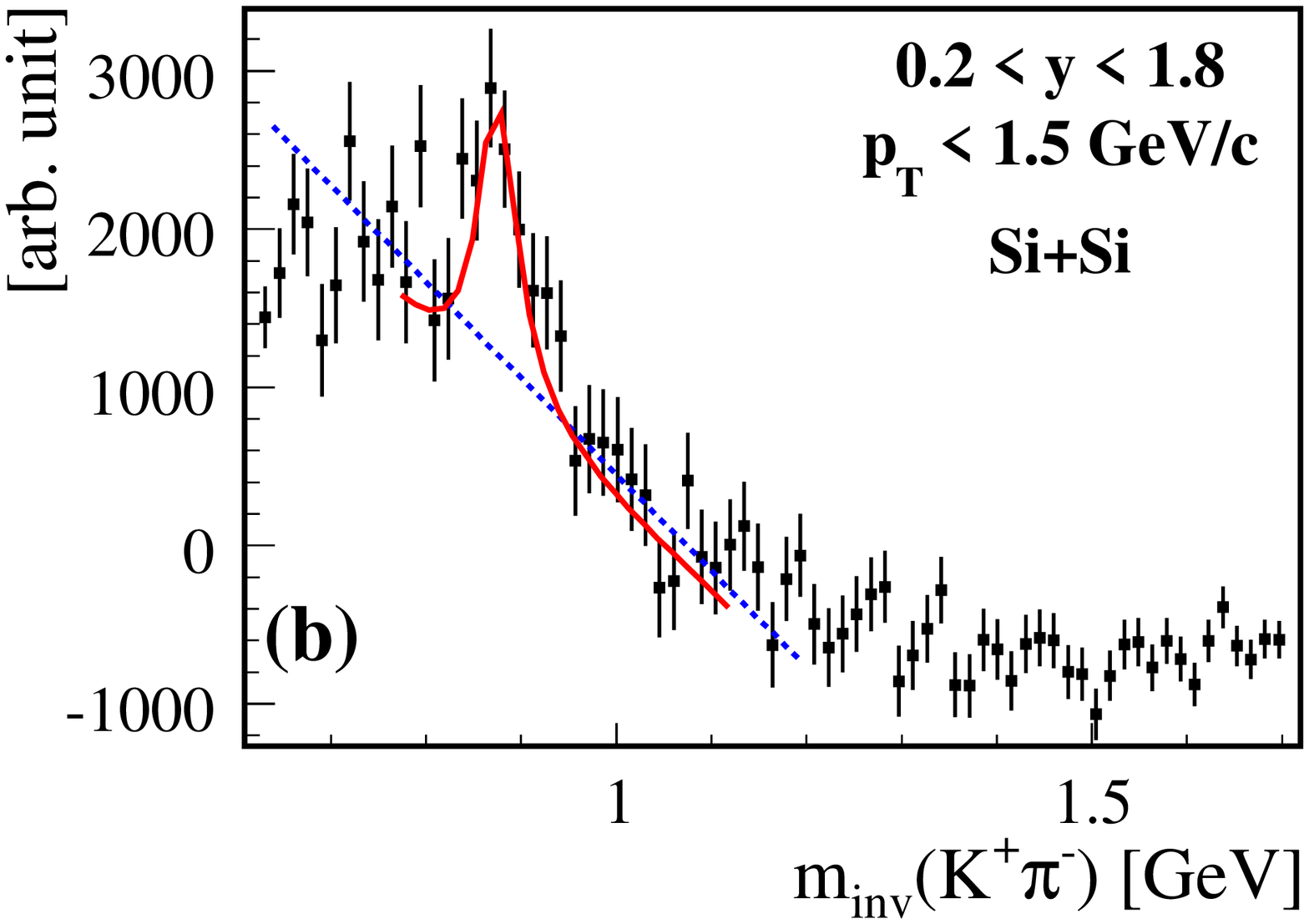}
}
\mbox{
\includegraphics[width=0.45\linewidth]{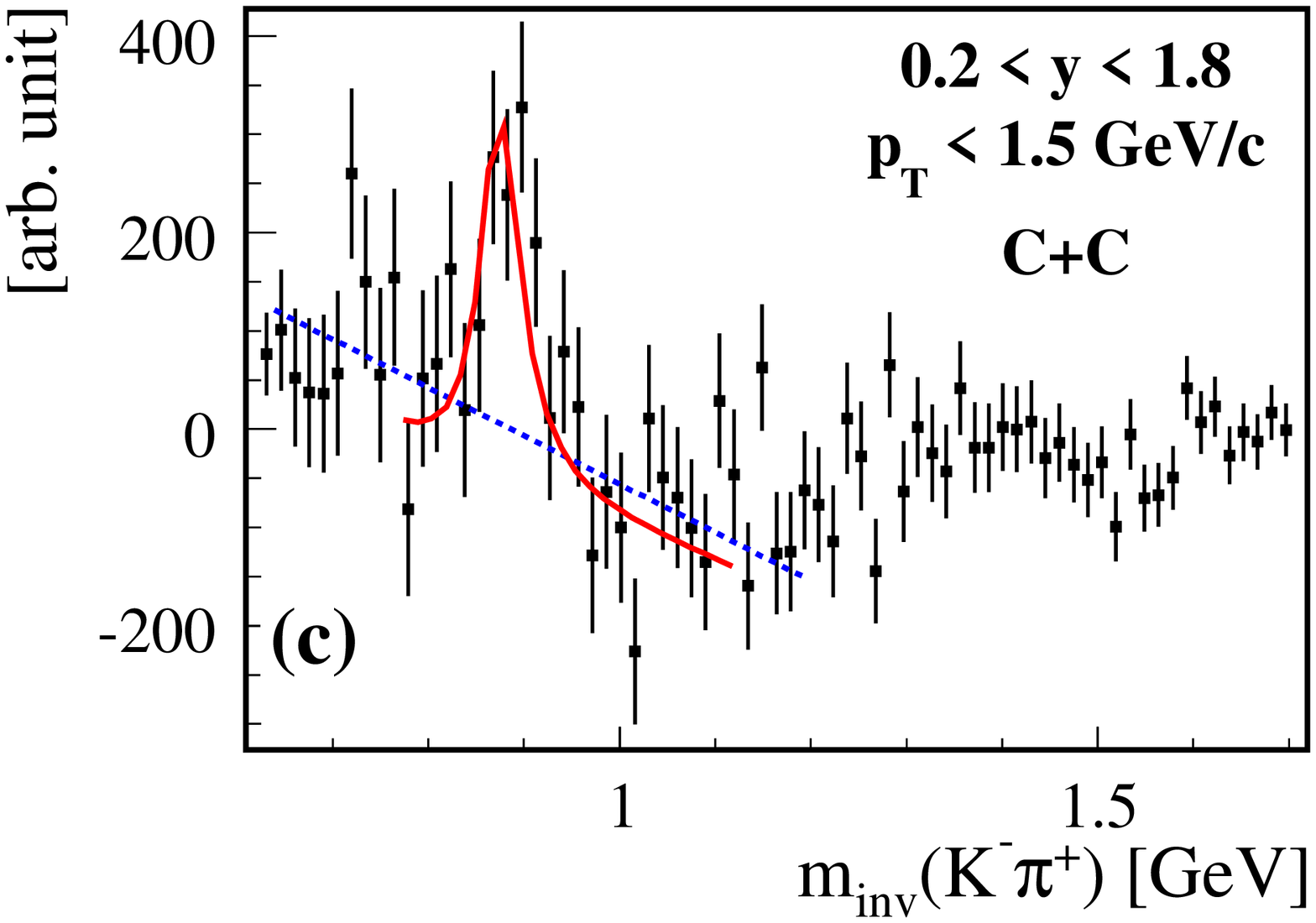}
\includegraphics[width=0.45\linewidth]{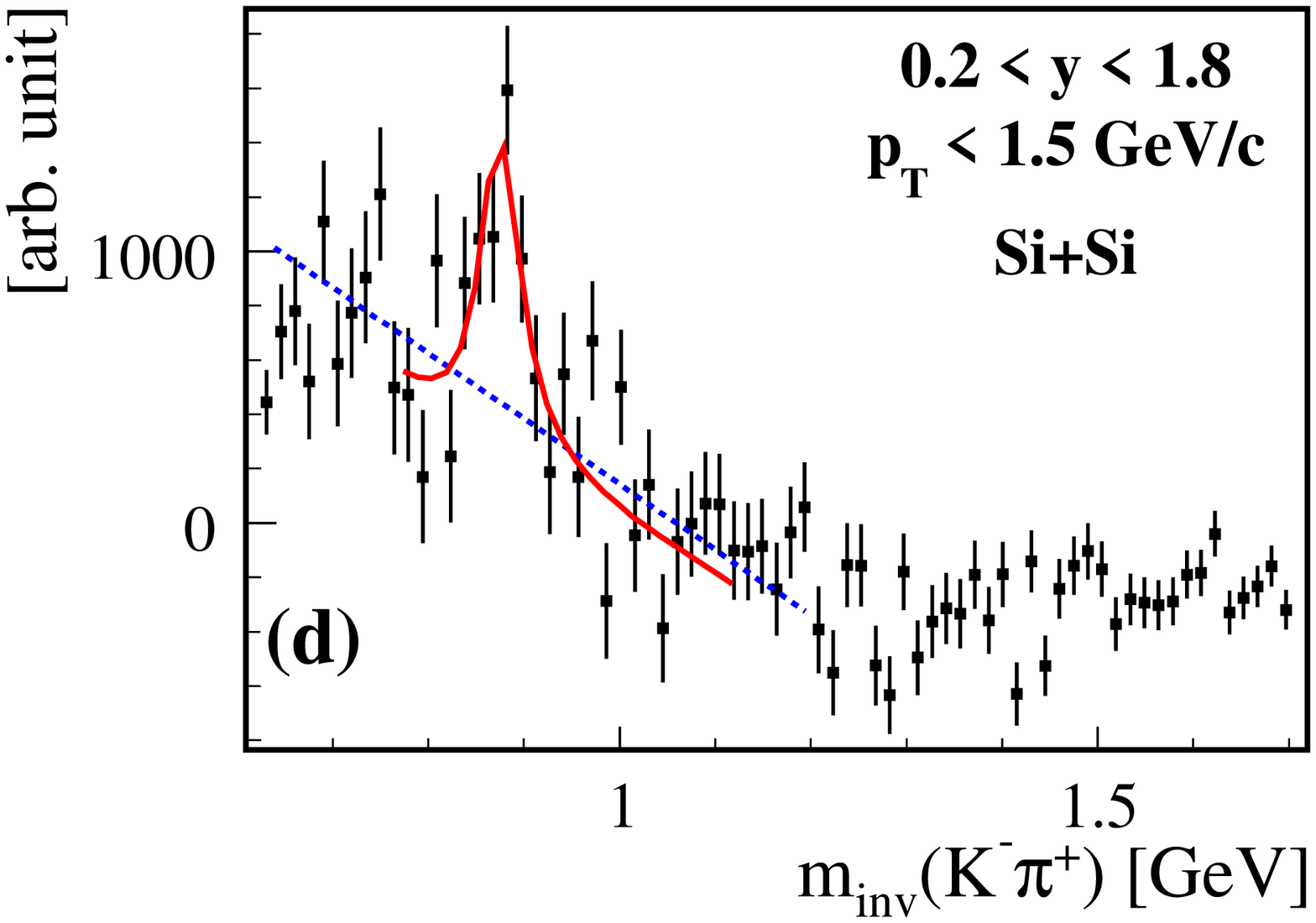}
}
\caption{\label{fig_minv_cc_sisi} (Color online)
   Invariant mass distribution of $K^+\pi^-$ (a,b) and $K^-\pi^+$ (c,d)
   in the rapidity region $0.2 < y < 1.8$ and $p_{T} < 1.5$~\gevc~after subtraction
   of mixed-pair background. Results are plotted for C+C collisions in (a,c) and 
   for Si+Si collisions in (b,d).
   The fitted residual background is shown by the dashed
   lines, the sum of fitted background and signal Breit-Wigner function
   by the solid curves.
  }
\end{figure}
%---------------------------------------------------------------
%
\subsection{C+C and Si+Si collisions}

The analysis of C+C and Si+Si collisions follows a strategy similar
to that employed for Pb+Pb and p+p collisions, however being limited by low statistics. 
The available number of events was not sufficient to
extract rapidity and transverse mass distributions, but allowed only to
estimate a total yield. $K^{\ast}(892)^0$ and $\overline{K}^{\ast}(892)^0$ invariant-mass
spectra (see \Fi{fig_minv_cc_sisi}) were extracted in a wide range of 
rapidity $0.2 < y <1.8$ and transverse momentum $p_{T} < 1.5$~\gevc.
Kaons and pions were selected in the momentum range of
$4 < p < 50$~\gevc~and  $p_{T} < 1$~\gevc. The minimum number of points required 
per track was 50. Pions and kaons were identified by $dE/dx$ within a
band of typically $\pm 1.5$ standard deviations around the Bethe-Bloch value. 
Both cuts were varied in order to estimate the systematic uncertainties. 

A similar strategy for invariant-mass spectra, background calculation, and signal
extraction was used as for p+p collisions. However, subtraction of the mass distributions
from mixed events did not completely remove the combinatorial background as can be
seen from \Fi{fig_minv_cc_sisi}. The remaining background was parameterized as a
straight line in the $K^{\ast}(892)$ mass region (dashed lines in \Fi{fig_minv_cc_sisi}). 
The signal was then obtained by fitting the sum of this linear background and the 
$K^{\ast}(892)$ line shape (using the world average values for mass and width \cite{pdg2010})
including its reflections as was done for p+p reactions.
The results of the fits are shown as the solid curves in \Fi{fig_minv_cc_sisi}.

The extracted raw yields were corrected for acceptance, the $dE/dx$ identification
cuts and the decay branching fraction. The mean acceptance
of $K^{\ast}(892)^0$ and $\overline{K}^{\ast}(892)^0$ in the selected wide phase space was 
calculated by Monte Carlo simulations assuming a width of
the rapidity distribution similar to that of charged kaons and inverse
$p_{T}$-slopes similar to that of the $\phi$-meson~\cite{syssz_2005} 
as its mass is close to that of the $K^{\ast}(892)$. A typical value of the mean
acceptance requiring more than 100 points for both, the pion and
kaon track, is 14\%. Varying the assumptions on background shape and the kinematic
distributions of the $K^{\ast}(892)$ changes the mean acceptance by 10\%
only. Extracted yields have statistical errors of 15-20\%. Varying
background assumptions and selection criteria of kaons and pions,
yields change by 15\% at most. We thus assign a combined
systematic and statistical error of 30\%. More details on the
analysis can be found in~\cite{hoehne2003}.
\section{Results}

\subsection{Pb+Pb collisions}

Yields of $K^{\ast}(892)^0$ and $\overline{K}^{\ast}(892)^0$ per event were
extracted for the region of
rapidity $0.3 < y < 1.8$ and transverse momentum $0 < p_T < 2.0$~\gevc.
Efficiency-corrected results are plotted as a function of rapidity $y$ 
in \Fi{fig6} and listed in \Ta{tab:results_dndy}.
They include a small extrapolation in $p_T$ beyond 2.0~\gevc~based on the exponential
parameterisation of the invariant $p_T$ distribution using the temperature parameter $T$
fitted in the measured $p_T$ range (see below). The rapidity distributions decrease
with increasing $y$ and suggest a maximum at mid-rapidity in view of the forward-backward
symmetry of the reaction. 

\begin{table}[tbh]
\caption{\label{tab:results_dndy}
Yields of $K^{\ast}(892)^0$ and $\overline{K}^{\ast}(892)^0$ per event 
in central Pb+Pb collisions as a function of
rapidity $y$ and integrated over transverse momentum. 
Both statistical (first) and systematic (second) errors are listed.
}
\begin{center}
\begin{tabular}{c|cc}
\hline
\hline
    $y$      & \hspace{1.5cm} $\frac{dn}{dy}(K^{\ast}(892)^0)$ \hspace{1.5cm} & \hspace{1.5cm} $\frac{dn}{dy}(\overline{K}^{\ast}(892)^0)$ \hspace{1.5cm} \\
\hline
 0.3 - 0.6   &      2.52 $\pm$ 0.22 $\pm$ 0.73   &      1.52 $\pm$ 0.19 $\pm$ 0.45   \\
 0.6 - 0.9   &      2.71 $\pm$ 0.18 $\pm$ 0.47   &      1.93 $\pm$ 0.13 $\pm$ 0.37   \\
 0.9 - 1.2   &      2.31 $\pm$ 0.16 $\pm$ 0.40   &      1.21 $\pm$ 0.11 $\pm$ 0.31   \\
 1.2 - 1.5   &      2.08 $\pm$ 0.15 $\pm$ 0.49   &      0.81 $\pm$ 0.10 $\pm$ 0.24   \\
 1.5 - 1.8   &      1.88 $\pm$ 0.13 $\pm$ 0.65   &      0.45 $\pm$ 0.08 $\pm$ 0.26   \\
\hline
 total yield (Gauss fit)     &  10.3 $\pm$ 0.4 $\pm$ 2.5     &  5.2 $\pm$ 0.3 $\pm$ 1.7     \\
\hline
\hline
\end{tabular}
\end{center}
\end{table}
%------------------------------------------------------------------
%

%
%--------------------------------------------------------------
\begin{figure}[htb]
\includegraphics[width=0.45\linewidth]{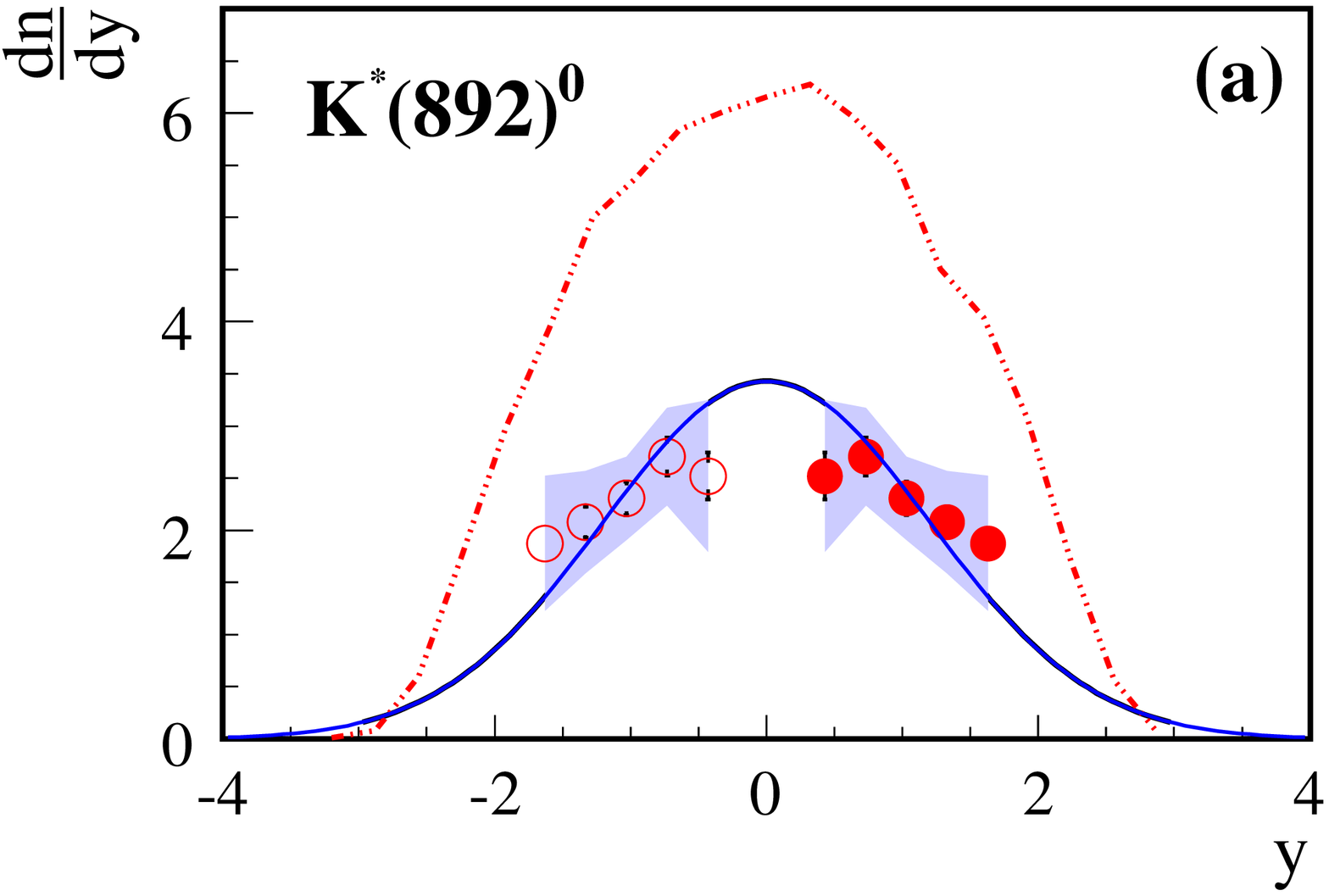}
\includegraphics[width=0.45\linewidth]{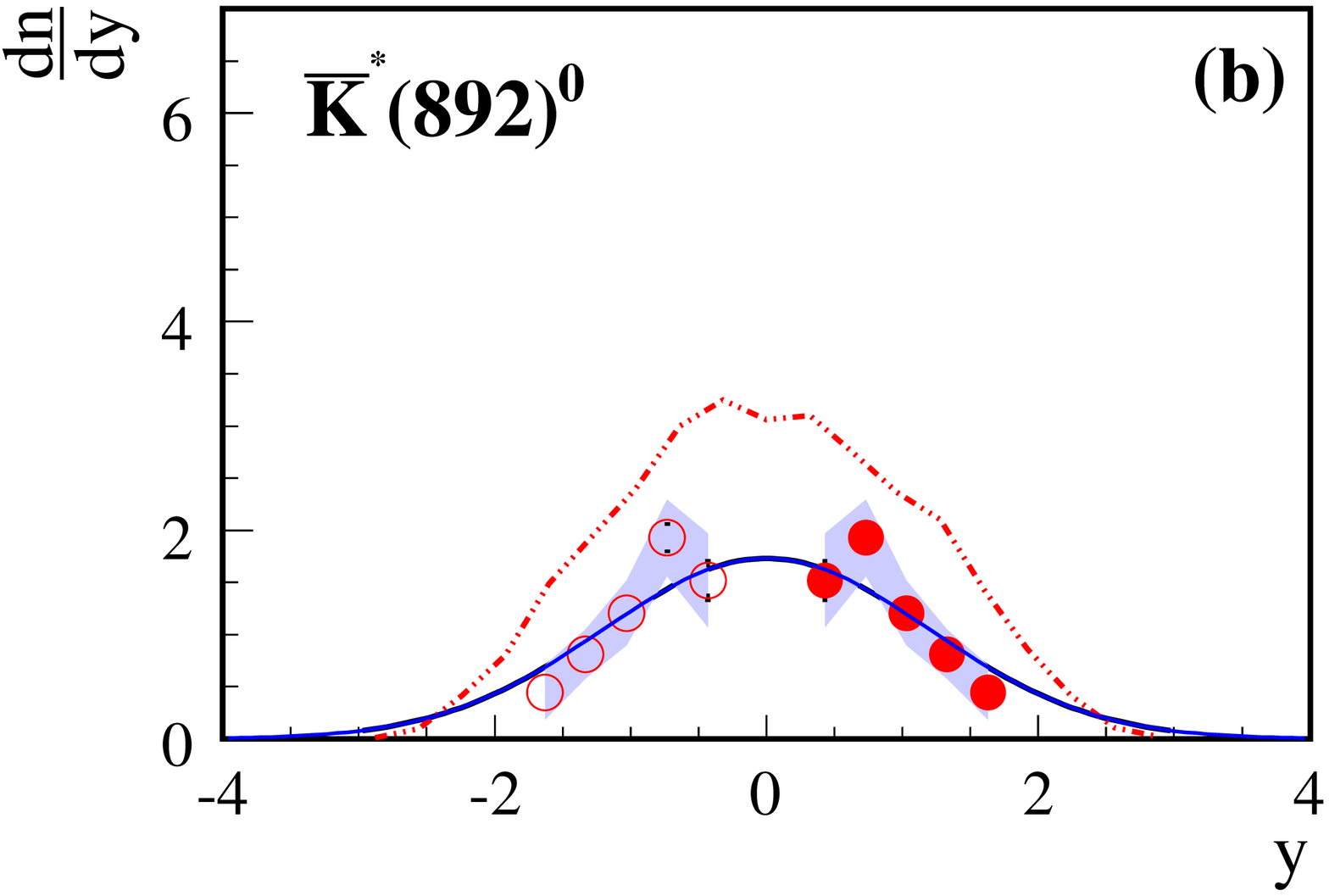}
\caption{\label{fig6} (Color online)
   Rapidity distribution of $K^{\ast}(892)^0$ (a) and $\overline{K}^{\ast}(892)^0$
   (b) integrated over $p_T$ in central Pb+Pb collisions. 
   Full symbols represent the measurements, 
   open symbols were obtained by reflection 
   around mid-rapidity. The bars show statistical errors, the bands indicate 
   systematic errors. Solid curves show fits of Gaussian functions for 
   estimating total yields, dashed-dotted curves depict predictions of the UrQMD 
   model~\cite{urqmd_vogel}.   
  }
\end{figure}
%---------------------------------------------------------------
%

Estimates of total yields were obtained by fitting Gauss functions to the
rapidity distributions centered at mid-rapidity. In order to better constrain the 
fit the width of the Gaussian was fixed at 1.2 units in rapidity. This choice was
motivated by the observed systematics of the widths of rapidity distributions in
Pb+Pb collisions \cite{phina49} and the prediction of the UrQMD 
model~\cite{urqmd_vogel}.
The results are also included in \Ta{tab:results_dndy}. 

%
%--------------------------------------------------------------
\begin{figure}[htb]
\mbox{
\includegraphics[width=0.40\linewidth]{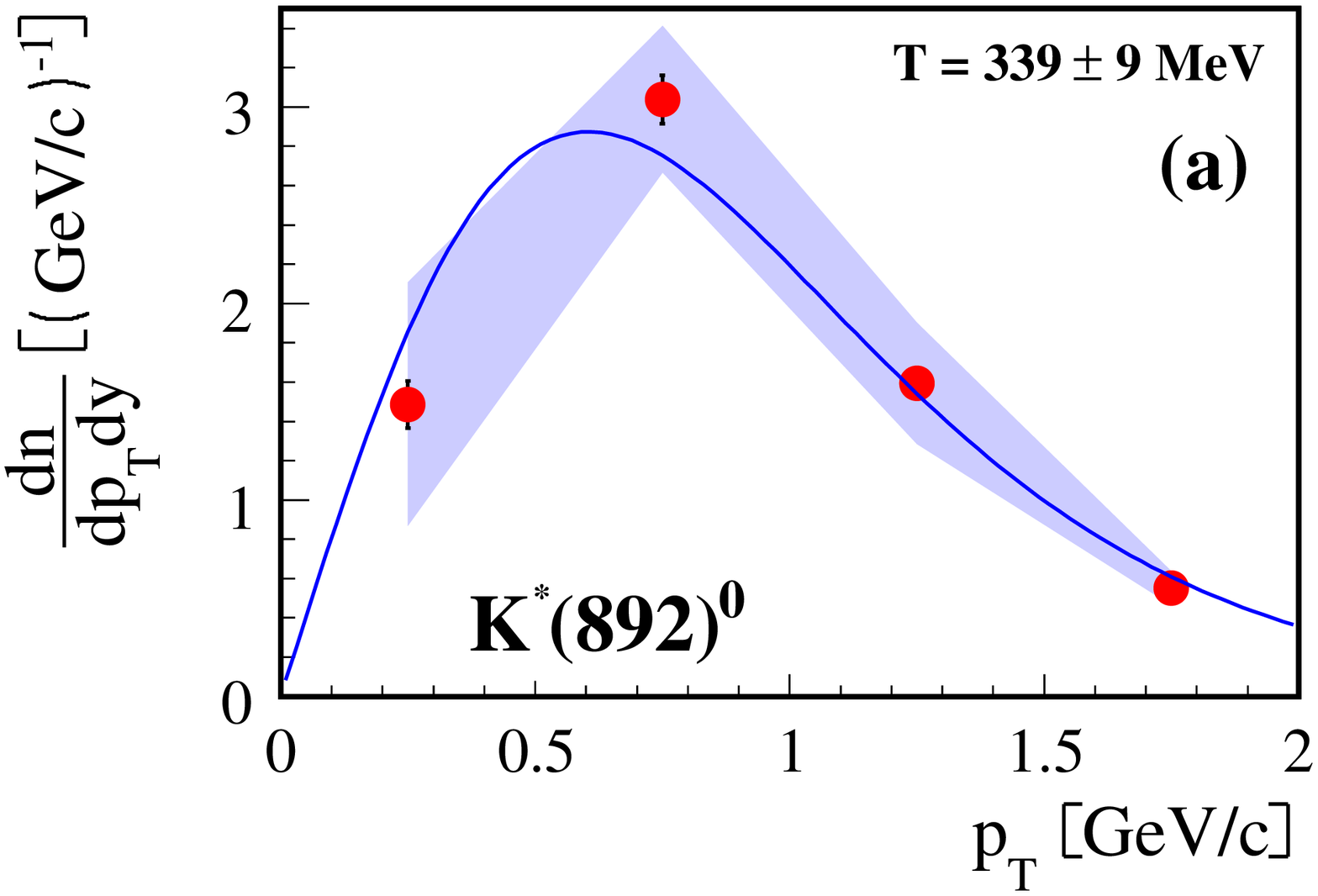}
\includegraphics[width=0.40\linewidth]{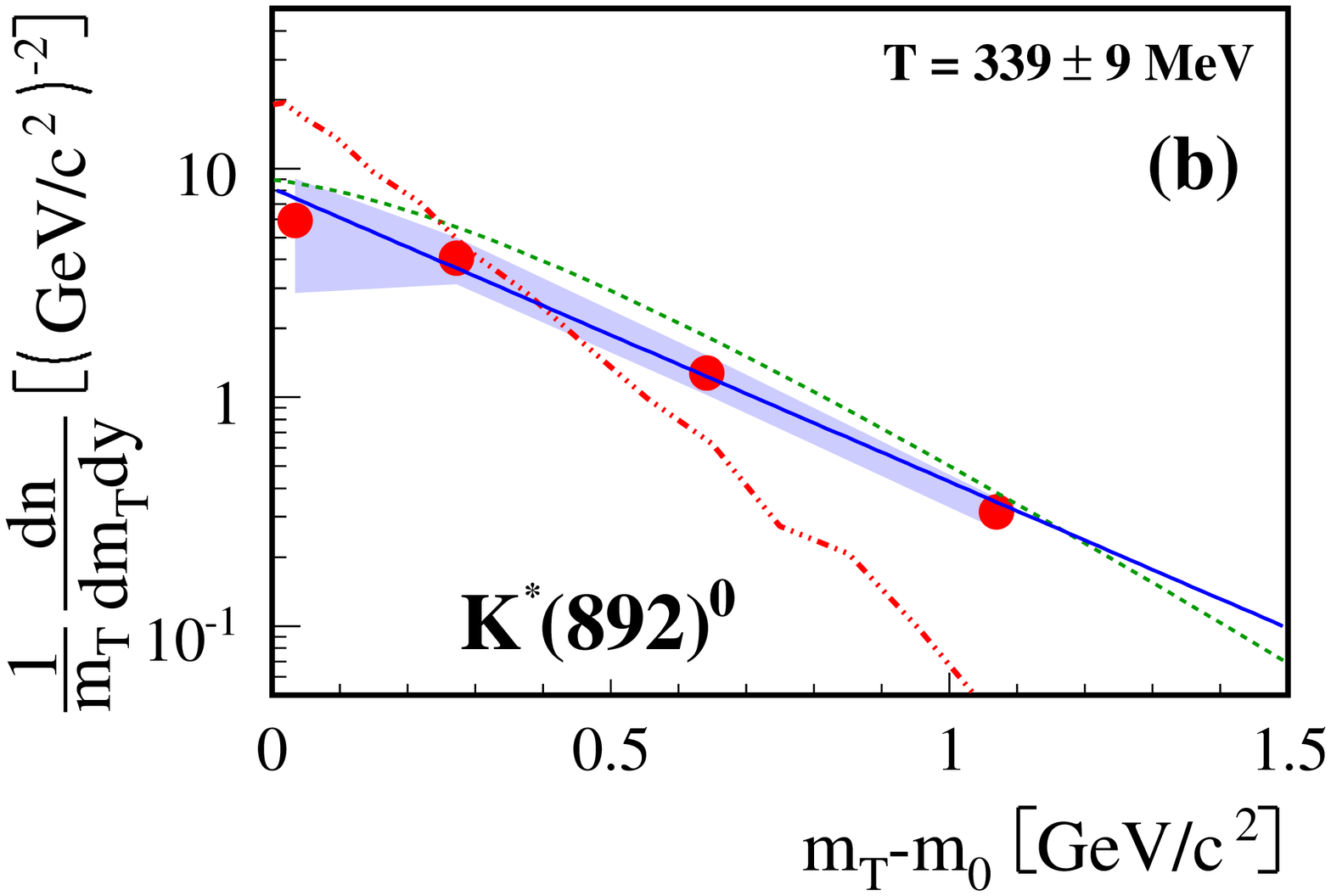}
}
\mbox{
\includegraphics[width=0.40\linewidth]{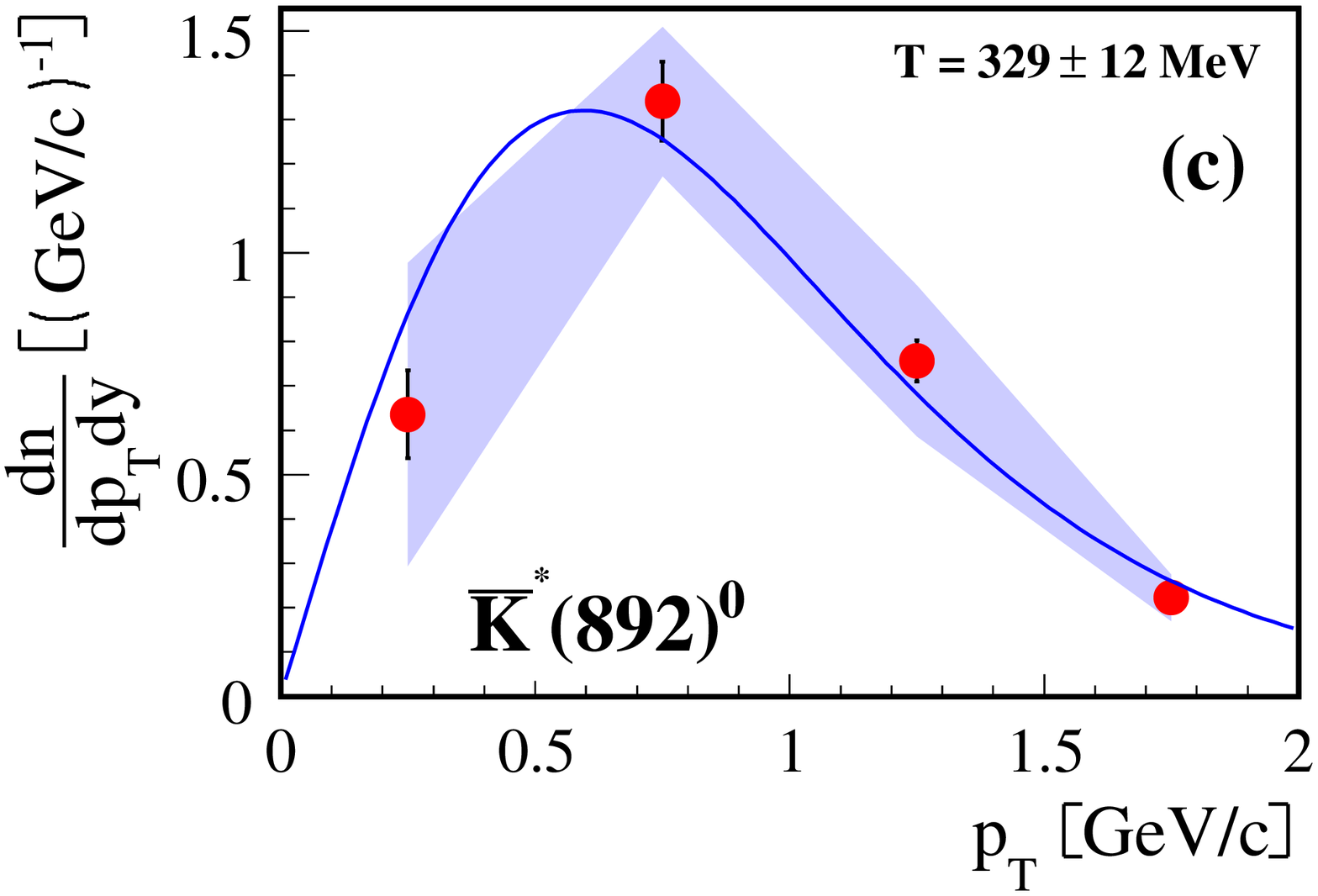}
\includegraphics[width=0.40\linewidth]{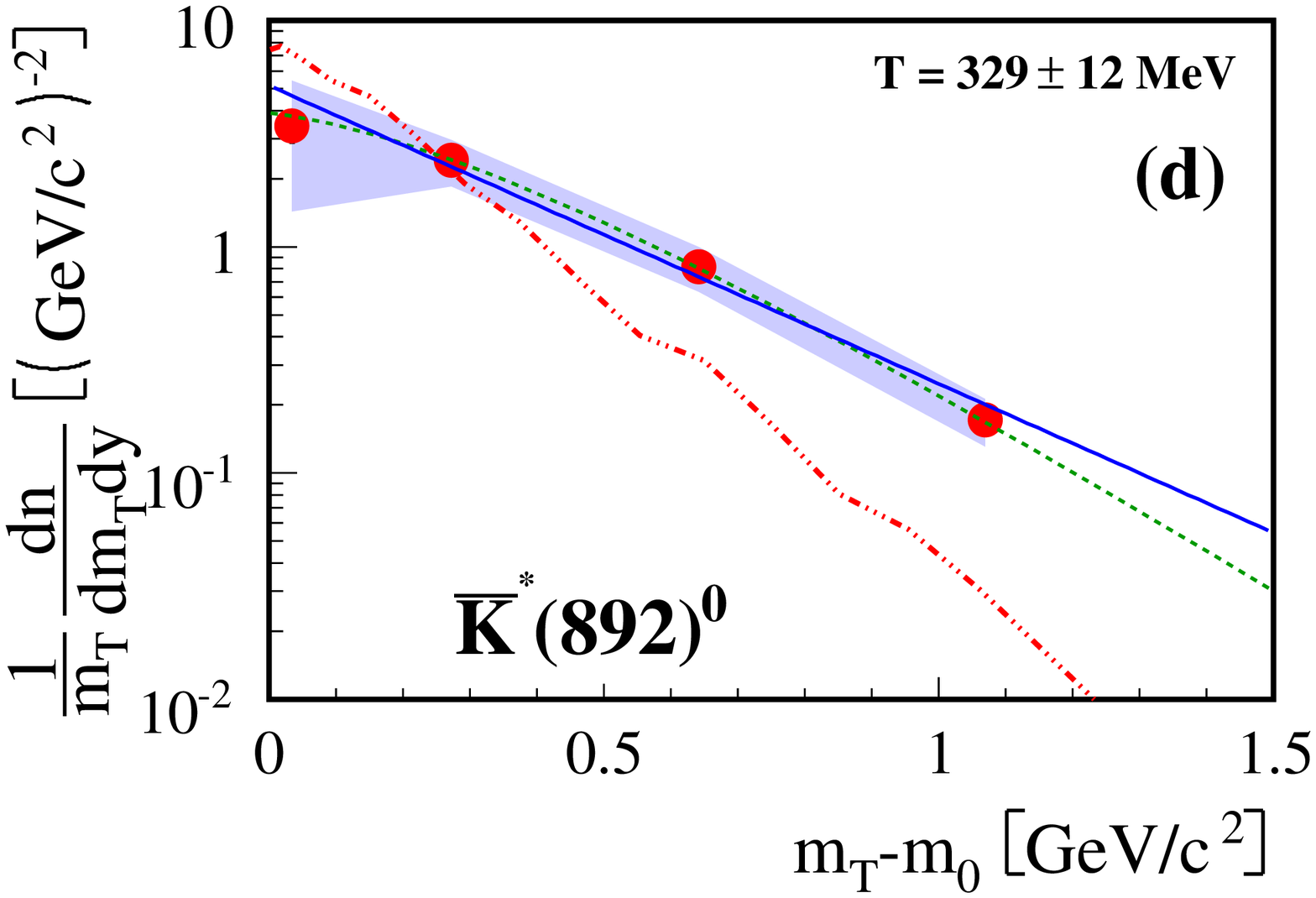}
}
\caption{\label{fig7} (Color online)
   Differential yields in central Pb+Pb collisions as function of transverse momentum 
   $p_T$ (a,c) and transverse mass $m_T$ (b,d) for $K^{\ast}(892)^0$ (a,b) and
   $\overline{K}^{\ast}(892)^0$ (c,d) in the rapidity interval $0.43 < y < 1.78$.
   The bars show statistical errors, the bands indicate 
   systematic errors. Solid curves show results of exponential 
   fits~(\Eq{equmt}, see text). Dotted curves result from blast-wave calculations using
   the parameters fitted in ref.~\cite{becorrna49}. The dashed-dotted curves depict
   predictions of the UrQMD model~\cite{urqmd_vogel}.
  }
\end{figure}
%---------------------------------------------------------------
%

For obtaining distributions in transverse momentum yields were extracted in
four $p_T$ bins for the rapidity range $0.43 < y < 1.78$. The results
for $\frac{dn}{dp_Tdy}$ are plotted in panels (a,c) of \Fi{fig7} and listed in \Ta{tab:results_dndpt}.
From these measurements the yield as function of transverse mass $m_T = \sqrt{p_T^2 + m_0^2}$
(where $m_0$ is the $K^*$ mass)
was calculated. The obtained values of $\frac{1}{m_T}\frac{dn}{dm_Tdy}$ are shown in 
panels (b,d) of \Fi{fig7} and are listed in \Ta{tab:results_dndmt}. An 
exponential function:
\begin{eqnarray}\label{equmt}
\frac{1}{m_T} \frac{dn}{dm_Tdy} = A \cdot e^{-\frac{m_T}{T}} ,
\end{eqnarray}
was fitted to these measurements, 
where $T$ is the inverse slope parameter and $A$ a normalisation constant. The resulting values 
of $T$~=~339~$\pm$~9~MeV for $K^{\ast}(892)^0$ and $T$~=~329~$\pm$~12~MeV for $\overline{K}^{\ast}(892)^0$
are much larger than for kaons, but closer to that for the higher mass $\phi$ meson \cite{phina49}.
Moreover, $\frac{1}{m_T}\frac{dn}{dm_Tdy}$ seems to exhibit a convex shape in the logarithmic representation
of \Fi{fig7}. This behaviour may be due to the participation of the $K^{\ast}$ in the 
strong radial flow~\cite{becorrna49}. Indeed, a blast-wave calculation, also shown in 
\Fi{fig7}~(b,d) by dotted curves, provides a good description of the $K^{\ast}$ spectra
using parameters fitted to pion, kaon, proton and anti-proton spectra 
($T = 93$~MeV, $\rho_0 = 0.91$~\cite{becorrna49}). Alternatively, the convex shape
could be due to the attenuation of the $K^{\ast}$ in the fireball which is expected
to be strongest for low values of $p_T$.

\begin{table}[tbh]
\caption{\label{tab:results_dndpt}
Yields $\frac{dn}{dp_Tdy}$ of $K^{\ast}(892)^0$ and $\overline{K}^{\ast}(892)^0$ per event 
in central Pb+Pb collisions as a function of
transverse momentum $p_T$ in the rapidity interval $0.43 < y < 1.78$.
Both statistical (first) and systematic (second) errors are shown for
the differential yields. The inverse slope parameters $T$ of exponential fits 
according to \Eq{equmt} are listed with their statistical errors.
}
\begin{center}
\begin{tabular}{c|cc}
\hline
\hline
$p_T$ [\gevc]   & \hspace{0.5cm} $\frac{dn}{dp_Tdy}(K^{\ast}(892)^0)$ [(\gevc)$^{-1}$] \hspace{0.5cm} & \hspace{0.5cm} $\frac{dn}{dp_Tdy}(\overline{K}^{\ast}(892)^0)$ [(\gevc)$^{-1}$] \hspace{0.5cm} \\
\hline
 0.0 - 0.5   &      1.49 $\pm$ 0.12 $\pm$ 0.66   &      0.64 $\pm$ 0.10 $\pm$ 0.34   \\
 0.5 - 1.0   &      3.04 $\pm$ 0.13 $\pm$ 0.38   &      1.34 $\pm$ 0.09 $\pm$ 0.17   \\
 1.0 - 1.5   &      1.59 $\pm$ 0.07 $\pm$ 0.31   &      0.76 $\pm$ 0.04 $\pm$ 0.17   \\
 1.5 - 2.0   &      0.55 $\pm$ 0.04 $\pm$ 0.09   &      0.22 $\pm$ 0.02 $\pm$ 0.05   \\
\hline
$T$ [GeV]      &     0.339 $\pm$ 0.009   &     0.329 $\pm$ 0.012   \\
\hline
\hline
\end{tabular}
\end{center}
\end{table}
%------------------------------------------------------------------
%

%
\begin{table}[tbh]
\caption{\label{tab:results_dndmt}
Yields $\frac{1}{m_T}\frac{dn}{dm_Tdy}$ of $K^{\ast}(892)^0$ and $\overline{K}^{\ast}(892)^0$ per event 
in central Pb+Pb collisions as a function of
transverse mass $m_T - m_0$ in the rapidity interval $0.43 < y < 1.78$.
Both statistical (first) and systematic (second) errors are listed.
}
\begin{center}
\begin{tabular}{c|cc}
\hline
\hline
$m_T - m_0$ [GeV/$c^2$] & \hspace{0.2cm} $\frac{1}{m_T}\frac{dn}{dm_Tdy}(K^{\ast}(892)^0)$ [(GeV/$c^2$)$^{-2}$] \hspace{0.2cm} & \hspace{0.2cm} $\frac{1}{m_T}\frac{dn}{dm_Tdy}(\overline{K}^{\ast}(892)^0)$ [(GeV/$c^2$)$^{-2}$] \hspace{0.2cm}  \\
\hline
 0.034   &      5.95 $\pm$ 0.48 $\pm$ 2.48   &      2.54 $\pm$ 0.39 $\pm$ 1.37   \\
 0.272   &      4.05 $\pm$ 0.16 $\pm$ 0.50   &      1.79 $\pm$ 0.12 $\pm$ 0.22   \\
 0.642   &      1.27 $\pm$ 0.06 $\pm$ 0.24   &      0.61 $\pm$ 0.04 $\pm$ 0.13   \\
 1.070   &      0.32 $\pm$ 0.02 $\pm$ 0.05   &      0.13 $\pm$ 0.01 $\pm$ 0.03   \\
\hline
\hline
\end{tabular}
\end{center}
\end{table}
%------------------------------------------------------------------
%

%--------------------------------------------------------------
\begin{figure}[htb]
\includegraphics[width=0.45\linewidth]{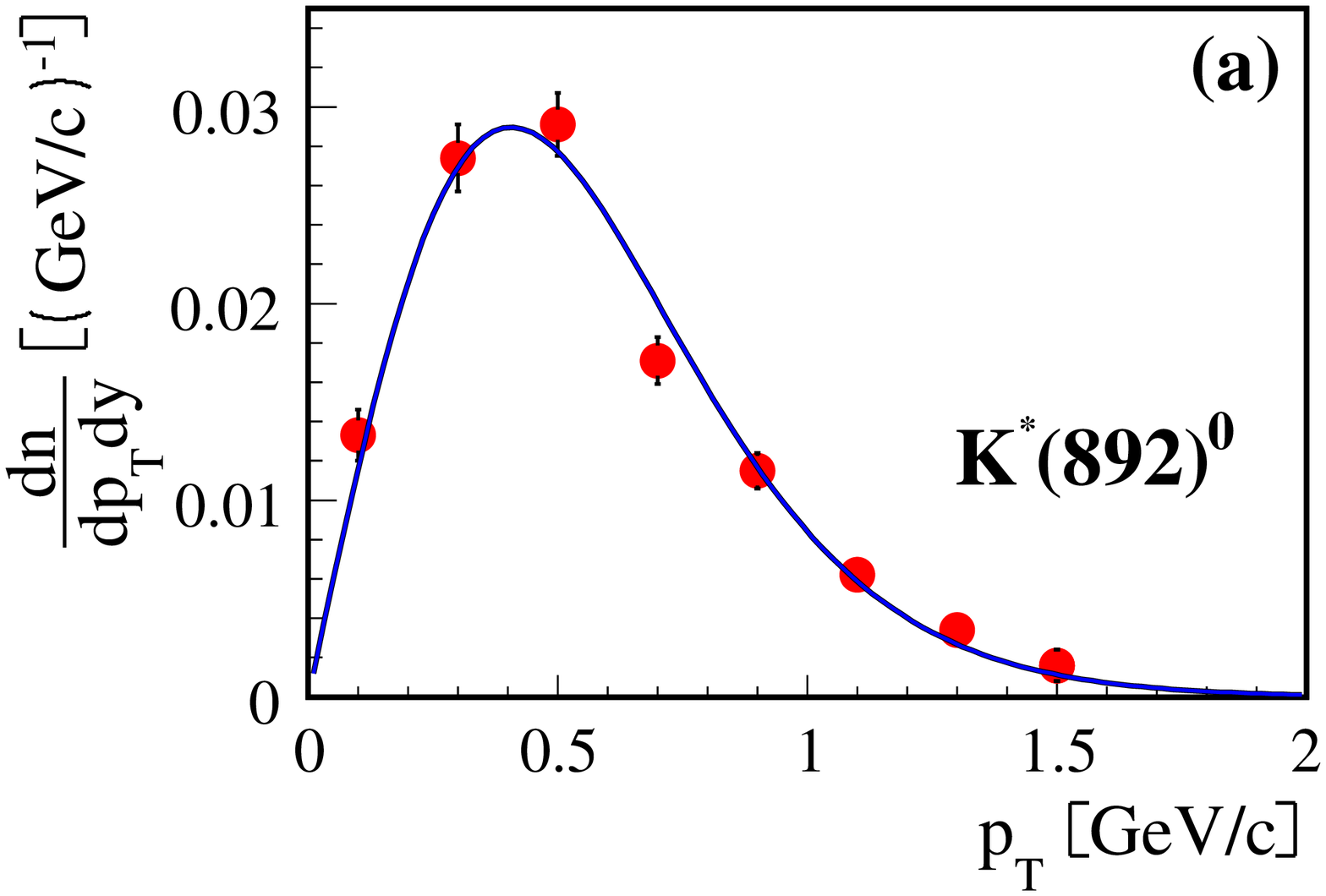}
\includegraphics[width=0.45\linewidth]{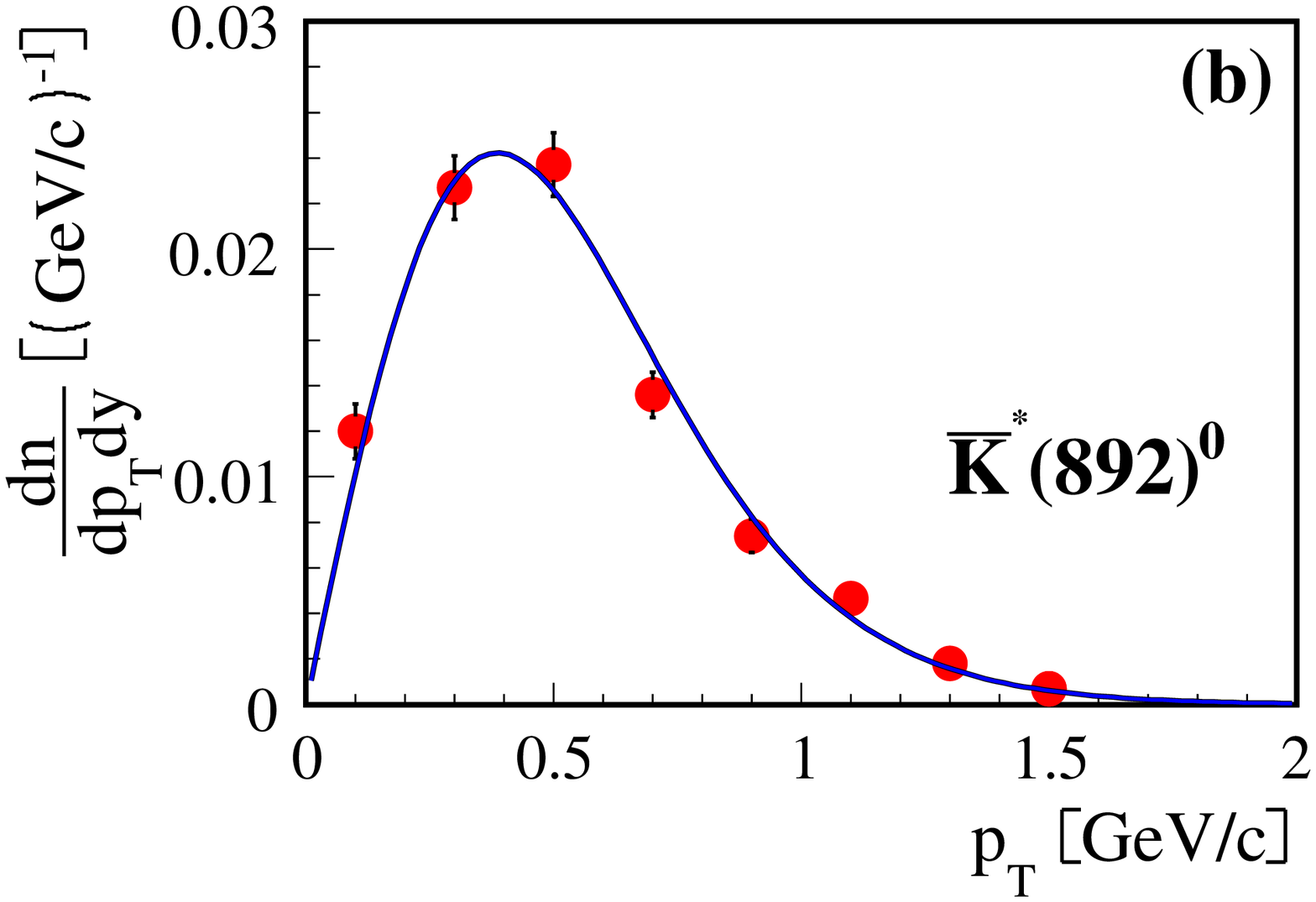}
\caption{\label{kstar_pt_pp}
   Transverse momentum distribution of $K^{*}(892)^{0}$~(a) 
   and $\overline{K}^{*}(892)^{0}$~(b) for the rapidity interval
   $0.2 < y < 0.7$ in inelastic p+p collisions at 158~\gevc~incident momentum. 
   Only statistical errors are shown; the overall systematic error of
   normalisation is 9~\%.
   Curves show results of fits with the exponential function \Eq{equmt}.
  }
\end{figure}
%---------------------------------------------------------------

%--------------------------------------------------------------
\begin{figure}[htb]
\includegraphics[width=0.45\linewidth]{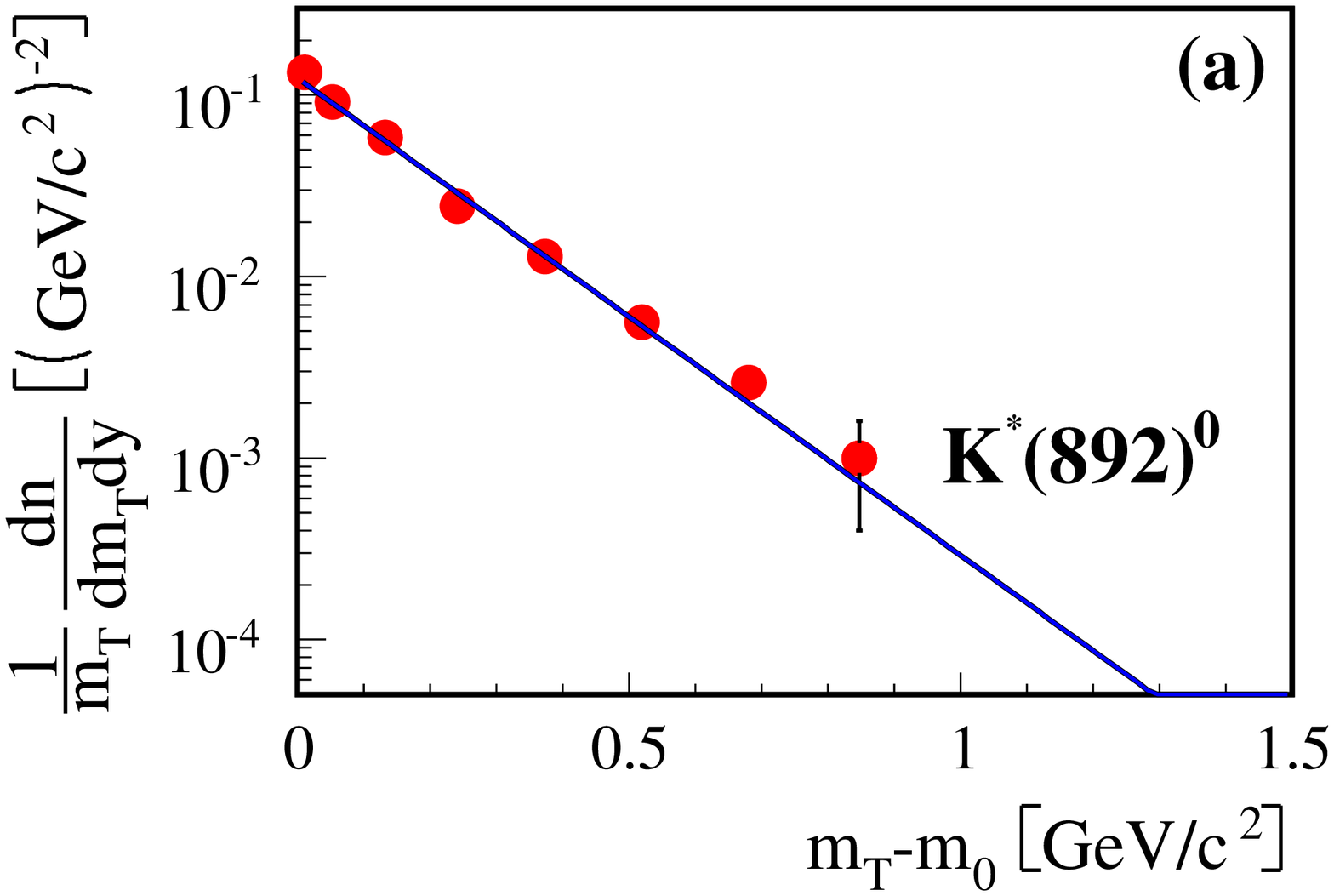}
\includegraphics[width=0.45\linewidth]{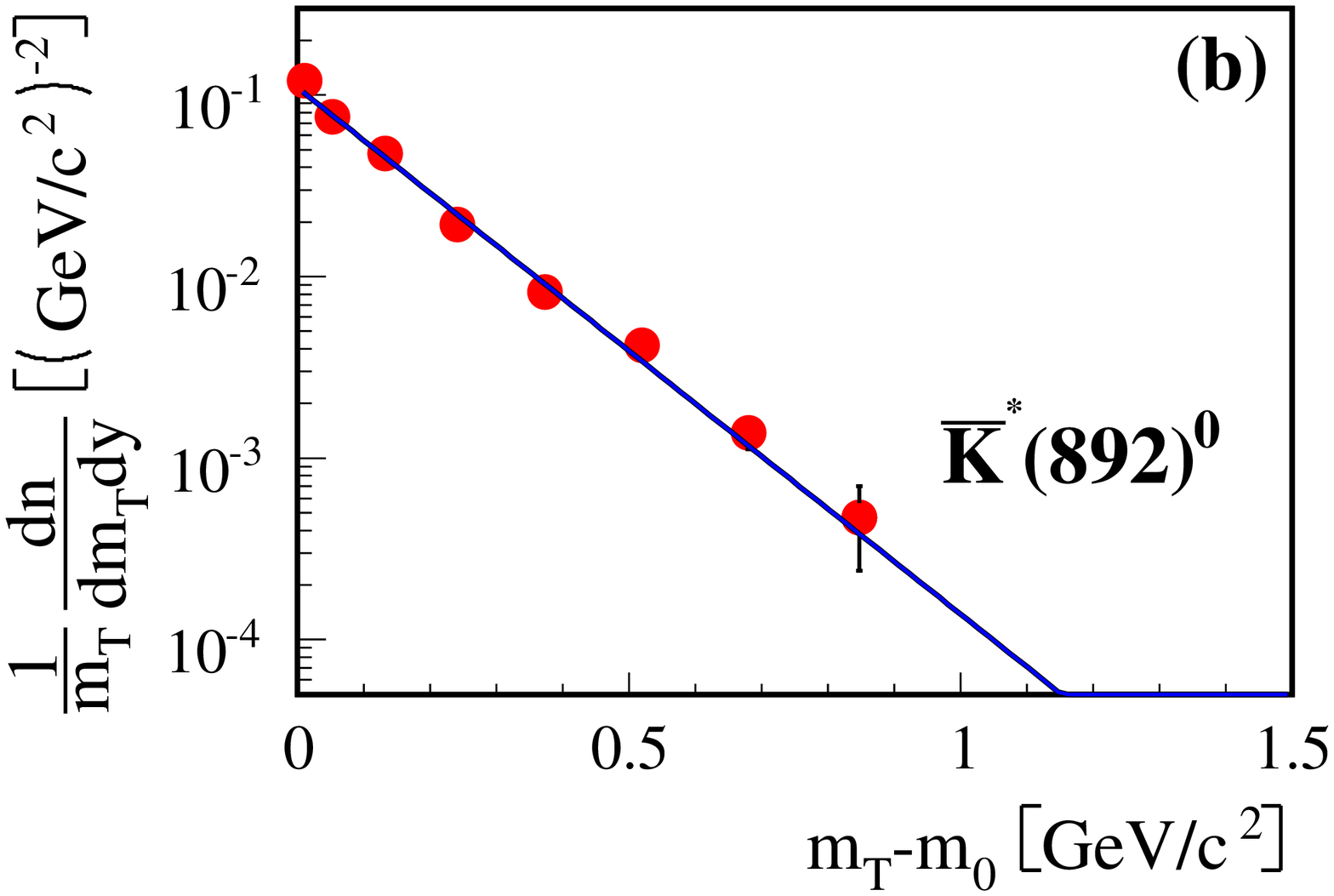}
\caption{\label{kstar_mt_pp}
   Transverse mass distribution of $K^{*}(892)^{0}$~(a) 
   and $\overline{K}^{*}(892)^{0}$~(b) for the rapidity interval
   $0.2 < y < 0.7$ in inelastic p+p collisions at 158~\gevc~incident momentum. 
   Only statistical errors are shown; the overall systematic error of
   normalisation is 9~\%.
   Curves show results of fits with the exponential function \Eq{equmt}.
  }
\end{figure}
%---------------------------------------------------------------

%--------------------------------------------------------------
\begin{figure}[htb]
\includegraphics[width=0.45\linewidth]{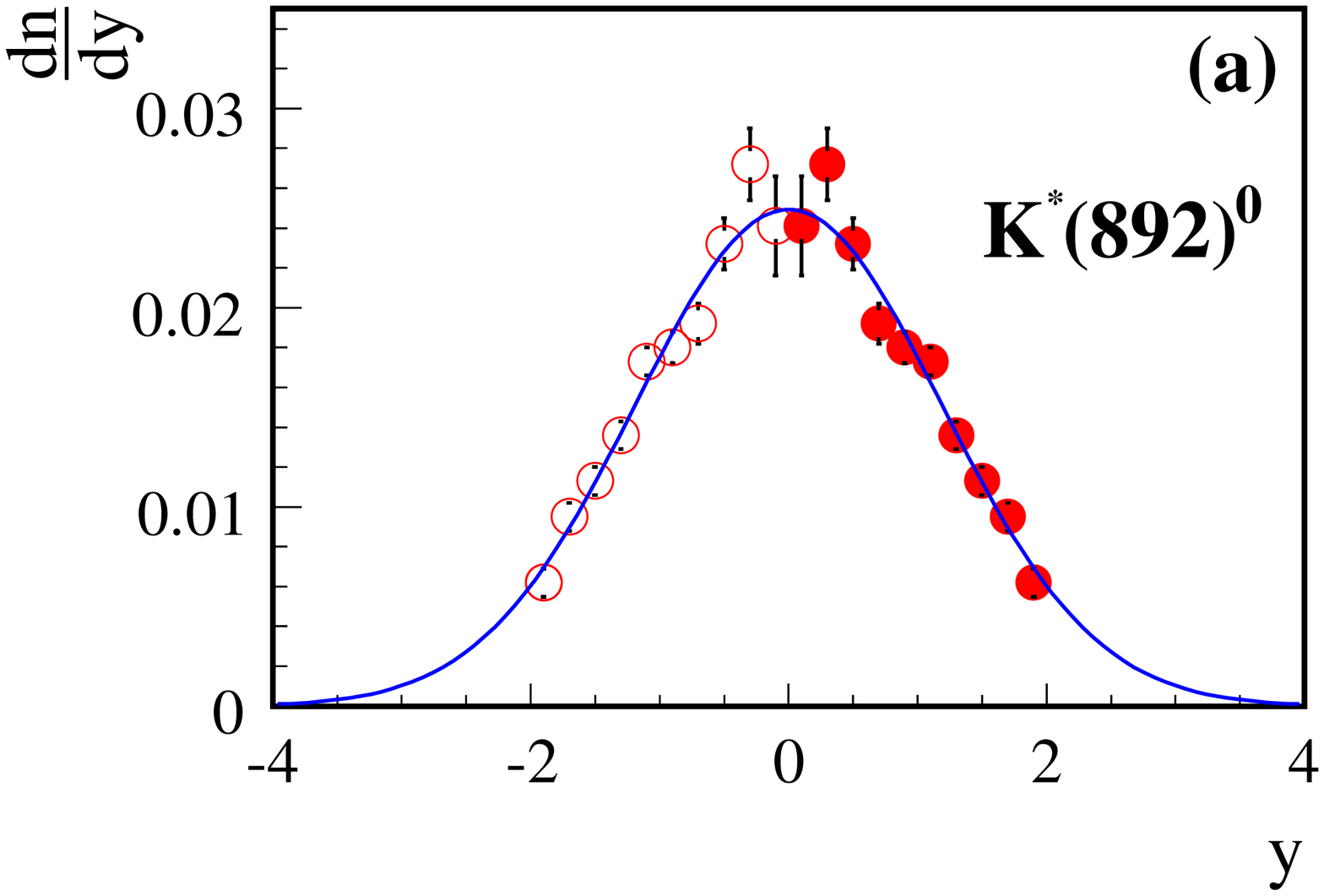}
\includegraphics[width=0.45\linewidth]{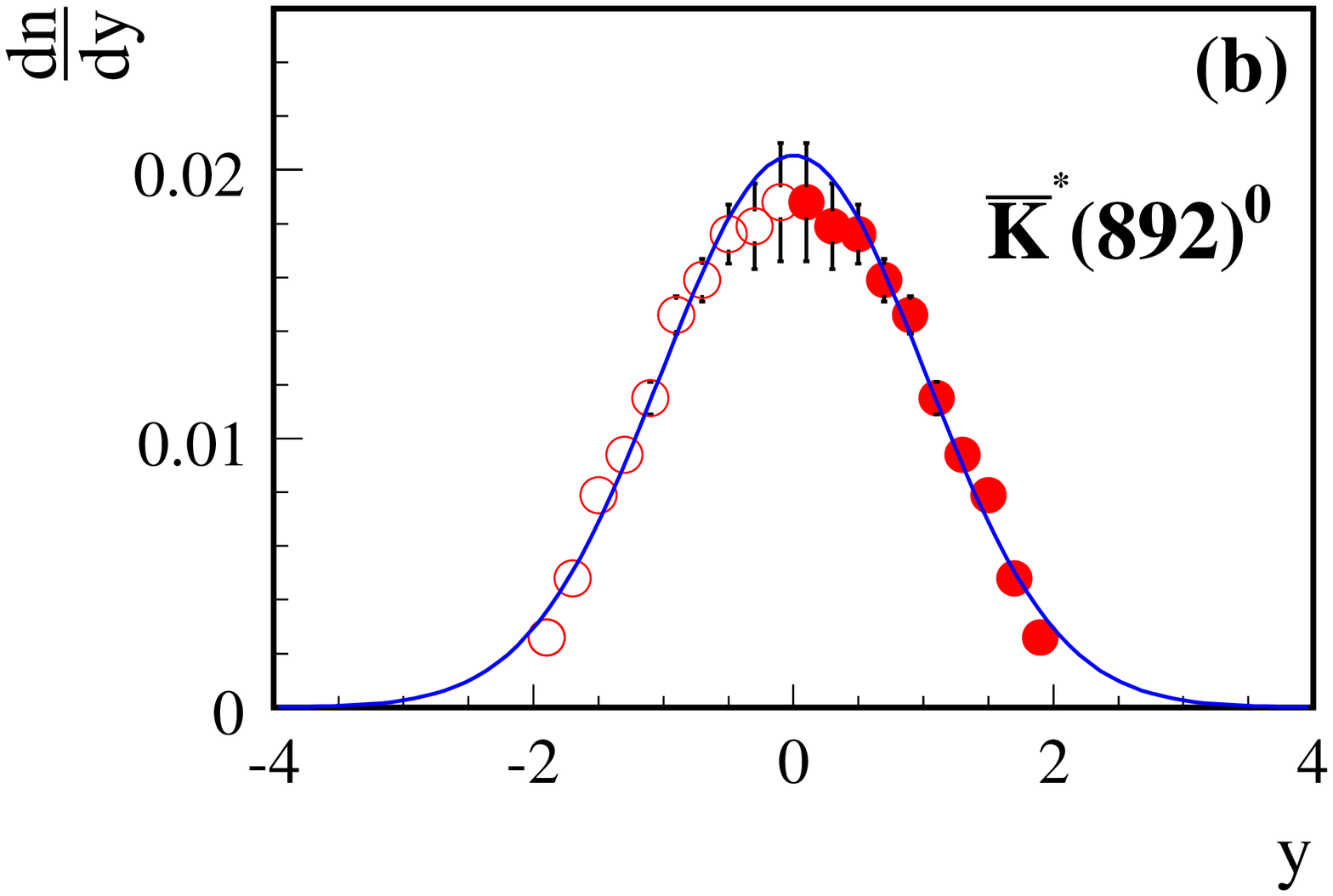}
\caption{\label{kstar_y_pp}
   Rapidity distributions for $K^{*}(892)^{0}$~(a) and $\overline{K}^{*}(892)^{0}$~(b)
   integrated over $p_T < 1.5$~GeV/$c$ in inelastic p+p collisions at 158~\gevc~incident momentum. 
   Only statistical errors are shown; the overall systematic error of
   normalisation is 9~\%.
   Curves show fits with a Gaussian function centered at mid-rapidity $y = 0$.
  }
\end{figure}
%---------------------------------------------------------------

\subsection{p+p collisions}

%---------------------------------------------------------------
\begin{table}[htb]
 {\caption{\label{tab:pp_dndy}
   Yields of $K^{\ast}(892)^0$ and $\overline{K}^{\ast}(892)^0$ per event with
   statistical errors in inelastic p+p collisions as a function of 
   rapidity $y$ and integrated over $p_T < 1.5$~GeV/$c$. The systematic error of normalisation (not shown) is 9~\%.
   Also listed are the widths $\sigma_{y}$,
   the central rapidity yields $\frac{dn}{dy}\mid_{y=0}$ 
   and the total yields obtained from
   Gaussian fits with statistical and systematic errors.
 }}
 \begin{center}
  \begin{tabular}{c | c c}
  \hline
  \hline
  $y$ &  \hspace{1.5cm} $\frac{dn}{dy}$($K^{\ast}(892)^0$) \hspace{1.5cm} & \hspace{1.5cm} $\frac{dn}{dy}$ ($\overline{K}^{\ast}(892)^0$) \hspace{1.5cm} \\
  \hline
  0.0 - 0.2 &  0.0241 $\pm$ 0.0025 & 0.0188 $\pm$ 0.0022 \\
  0.2 - 0.4 &  0.0272 $\pm$ 0.0018 & 0.0179 $\pm$ 0.0016 \\
  0.4 - 0.6 &  0.0232 $\pm$ 0.0013 & 0.0176 $\pm$ 0.0011 \\
  0.6 - 0.8 &  0.0192 $\pm$ 0.0010 & 0.0159 $\pm$ 0.0008 \\
  0.8 - 1.0 &  0.0180 $\pm$ 0.0008 & 0.0146 $\pm$ 0.0007 \\
  1.0 - 1.2 &  0.0173 $\pm$ 0.0007 & 0.0115 $\pm$ 0.0006 \\
  1.2 - 1.4 &  0.0136 $\pm$ 0.0007 & 0.0094 $\pm$ 0.0005 \\
  1.4 - 1.6 &  0.0113 $\pm$ 0.0007 & 0.0079 $\pm$ 0.0005 \\
  1.6 - 1.8 &  0.0095 $\pm$ 0.0007 & 0.0048 $\pm$ 0.0004 \\
  1.8 - 2.0 &  0.0062 $\pm$ 0.0007 & 0.0026 $\pm$ 0.0004 \\
 \hline
  $\sigma_{y}$ & 1.17 $\pm$ 0.03 $\pm$ 0.07 & 1.01 $\pm$ 0.02 $\pm$ 0.06\\
  $\frac{dn}{dy}\mid_{y=0}$ & 0.0257 $\pm$ 0.0031 $\pm$ 0.0023 & 0.0183
      $\pm$ 0.0027 $\pm$ 0.0016\\
  total yield & 0.0741 $\pm$ 0.0015 $\pm$ 0.0067 & 0.0523 $\pm$ 0.0010 $\pm$ 0.0047\\
  \hline
  \hline
  \end{tabular}
 \end{center}
\end{table}
%---------------------------------------------------------------

%---------------------------------------------------------------
\begin{table}[htb]
 {\caption{\label{tab:pp_dndpt}
   Yields $\frac{dn}{dp_{T}dy}$ of $K^{\ast}(892)^0$ and $\overline{K}^{\ast}(892)^0$ per event
   in inelastic p+p collisions for the rapidity interval $0.2 < y < 0.7$ as a 
   function of transverse momentum $p_T$ with statistical errors. The systematic
   error of normalisation (not shown) is 9~\%. 
   The inverse slope parameters $T$ of exponential fits according to \Eq{equmt} 
   are also listed with statistical and systematic errors.
 }}
 \begin{center}
  \begin{tabular}{c | c c }
  \hline
  \hline
  $p_T$ [GeV/$c$] & \hspace{1.5cm} $\frac{dn}{dp_{T}dy}$ ($K^{*}(892)^{0}$) \hspace{1.5cm} & \hspace{1.5cm} $\frac{dn}{dp_{T}dy}$ ($\overline{K}^{\ast}(892)^{0}$) \hspace{1.5cm} \\
  \hline
   0.0 - 0.2 & 0.0133 $\pm$ 0.0013 & 0.0120 $\pm$ 0.0012 \\
   0.2 - 0.4 & 0.0274 $\pm$ 0.0017 & 0.0227 $\pm$ 0.0014 \\
   0.4 - 0.6 & 0.0291 $\pm$ 0.0016 & 0.0237 $\pm$ 0.0014 \\
   0.6 - 0.8 & 0.0171 $\pm$ 0.0012 & 0.0136 $\pm$ 0.0010 \\
   0.8 - 1.0 & 0.0115 $\pm$ 0.0009 & 0.00741 $\pm$ 0.00073 \\
   1.0 - 1.2 & 0.0062 $\pm$ 0.0007 & 0.00466 $\pm$ 0.00050 \\
   1.2 - 1.4 & 0.0034 $\pm$ 0.0005 & 0.00180 $\pm$ 0.00034 \\
   1.4 - 1.6 & 0.0016 $\pm$ 0.0008 & 0.00070 $\pm$ 0.00035 \\
  \hline
   $T$ [GeV] & 0.166 $\pm$ 0.011 $\pm$ 0.010 & 0.150 $\pm$ 0.010 $\pm$ 0.010   \\
  \hline
  \hline
  \end{tabular}
 \end{center}
\end{table}

\begin{table}[htb]
 {\caption{\label{tab:pp_dndmt}
   Yields $\frac{1}{m_T}\frac{dn}{dm_Tdy}$ of $K^{\ast}(892)^0$ and $\overline{K}^{\ast}(892)^0$ 
   per event in inelastic p+p collisions for the rapidity interval $0.2 < y < 0.7$ 
   as a function of $m_T-m_0$. Only statistical errors are shown.  The systematic
   error of normalisation (not shown) is 9~\%.
 }}
 \begin{center}
  \begin{tabular}{c | c c }
  \hline
  \hline
   $\langle m_T \rangle - m_0$ [GeV/$c^2$] & \hspace{0.5cm} $\frac{1}{m_T}\frac{dn}{dm_Tdy}$ ($K^{*}(892)^{0}$) \hspace{0.5cm} & \hspace{0.5cm} $\frac{1}{m_T}\frac{dn}{dm_Tdy}$ ($\overline{K}^{\ast}(892)^{0}$) \hspace{0.5cm} \\
      \hline
     0.011 & 0.1331 $\pm$ 0.0132& 0.1200 $\pm$  0.0117 \\
     0.053 & 0.0914 $\pm$ 0.0054& 0.0757 $\pm$  0.0048 \\
     0.132 & 0.0583 $\pm$ 0.0032& 0.0477 $\pm$  0.0027 \\
     0.241 & 0.0244 $\pm$ 0.0018& 0.0194 $\pm$  0.0015 \\
     0.373 & 0.0129 $\pm$ 0.0010& 0.0082 $\pm$  0.0008 \\
     0.520 & 0.0056 $\pm$ 0.0006& 0.0042 $\pm$  0.0005\\
     0.680 & 0.0026 $\pm$ 0.0004& 0.00138 $\pm$  0.00026 \\
     0.847 & 0.0010 $\pm$ 0.0006& 0.00047 $\pm$  0.00023 \\
  \hline
  \hline
  \end{tabular}
 \end{center}
\end{table}
%---------------------------------------------------------------

For p+p reactions results are presented as yields per inelastic reaction.
Transverse momentum spectra (see Fig.~\ref{kstar_pt_pp} and \Ta{tab:pp_dndpt}) 
were extracted in a rapidity range of
$0.2 < y < 0.7$. The range was chosen as close to mid-rapidity as the 
acceptance for the $K^{*}(892)$ allows.
The $m_{T}$-spectra (see Fig.~\ref{kstar_mt_pp} and \Ta{tab:pp_dndmt}) show a thermal shape 
which can be well described by an exponential function (\Eq{equmt}) with an
inverse slope parameter $T = 166 \pm 15$ MeV for $K^{*}(892)^{0}$ and $150 \pm 14$~MeV 
for $\overline{K}^{*}(892)^{0}$, respectively. These values are consistent with
those found for other mesons in p+p collisions \cite{hoehne2003} indicating the absence of
radial flow in these reactions. The
$p_{T}$-integrated rapidity spectrum (see Fig.~\ref{kstar_y_pp} and \Ta{tab:pp_dndy}) was
extracted in the range of $p_T < 1.5$~\gevc, except for the last
rapidity bin $1.8 < y < 2.0$ which had a reduced range of $p_T < 1.2$~\gevc~
because of the upper momentum limit imposed by the $dE/dx$ identification procedure. The 
range $p_T < 1.5$~\gevc~contains 99.1\% of all $K^{*}(892)$ (for $T = 160$ MeV). The total 
yields (listed in \Ta{tab:yields_na49}) were extracted by fitting a Gaussian distribution centered 
at $y = 0$ to the rapidity distribution. The resulting widths of the rapidity
distributions and the mid-rapidity yields are listed in \Ta{tab:pp_dndy}. 
The extracted width of the rapidity distribution is consistent with the one for charged kaons
\cite{na49_pp-kaon-paper}. The extracted yields fit well into the trend of results
from p+p collisions at higher and lower energies (see \cite{hoehne2003}
for a more detailed comparison).

%---------------------------------------------------------------
%\begin{table}[h]
%\caption{\label{tab:kstar_table_pp}
%     Summary of results for $K^{*}$(892)$^{0}$ and
%     $\overline{K}^{*}(892)^{0}$ in minimum-bias p+p
%     collisions. The first error quoted is statistical, the second systematic.
%     Midrapidity $dN/dy$ values are averaged over the interval $0 < y < 0.4$.
%}
%\begin{center}
%\begin{tabular}{c c c}
%\hline
%& K$^{*}$(892)$^{0}$ & $\overline{\text{K}^{*}}(892)^{0}$ \\
%\hline
%$m_{0}$ & (892 $\pm$ 5) MeV & (892 $\pm$ 5) MeV\\
%$\Gamma_{0}$ & 50.5 MeV (fixed) & 50.5 MeV (fixed)\\
%$\sigma_{m}$ & 0.87 MeV (fixed) & 0.87 MeV (fixed)\\
%\hline
%$T$ & (166 $\pm$ 11 $\pm$ 10) MeV & (150 $\pm$ 10 $\pm$ 10) MeV\\
%$\sigma_{y}$ & 1.17 $\pm$ 0.03 $\pm$ 0.07 & 1.01 $\pm$ 0.02 $\pm$ 0.06\\
%\hline
%
%
%$\langle \text{N} \rangle$ & 0.0792 $\pm$ 0.0016 $\pm$ 0.0063 &
%0.0559 $\pm$ 0.0011 $\pm$ 0.0045\\
%$\frac{dN}{dy}\mid_{y=0}$ & 0.0275 $\pm$ 0.0033 $\pm$ 0.003 & 0.0196
%
%
%$\pm$ 0.0029 $\pm$ 0.002\\
%\hline
%\end{tabular}
%\end{center}
%\end{table}
%---------------------------------------------------------------

\subsection{C+C and Si+Si collisions}

Due to the limited number of recorded events only total yields per event could be estimated
(see Sect. III.C). The results with their statistical and sytematic uncertainties
are listed in \Ta{tab:yields_na49}.

\section{Discussion of results}

Both $K^{\ast}(892)^0$ ($\overline{K}^{\ast}(892)^0$) and $K^+$ ($K^-$) contain an anti-strange (strange)
valence quark in addition to a light quark and should therefore be similarly sensitive
to the strangeness content of the produced matter. The ratio of total yields
$\langle K^{\ast}(892)^0 \rangle$/$\langle \overline{K}^{\ast}(892)^0 \rangle$ is about 2 
in C+C, Si+Si and Pb+Pb collisions (see \Ta{tab:yields_na49}) and is similar to the ratio 
$\langle K^+ \rangle$/$\langle K^- \rangle$ $\approx$~2.0~\cite{syssz_2005,deconf} as expected. 
The yields per wounded nucleon are compared graphically in \Fi{fig13}~(a).
This quantity seems to increase from p+p to C+C and Si+Si collisions
and then to decrease to central Pb+Pb collisions. 
This behaviour may result from an interplay between strangeness enhancement in nucleus-nucleus
collisions and the interaction of the $K^{\ast}(892)$ and its decay products
in the produced fireball.

Because kaons and $K^{\ast}(892)$ both contain the same valence quarks, the system size
dependence of the ratios
$\langle K^{\ast} \rangle$/$\langle K^+ \rangle$ and 
$\langle \overline{K}^{\ast} \rangle$/$\langle K^-\rangle$ is expected to be
sensitive mostly to the interactions in the surrounding medium
while the effect of strangeness enhancement should approximately cancel. As can be seen
from \Fi{fig13}~(b), the ratios decrease by about a factor 2 from C+C and Si+Si reactions
to central Pb+Pb collisions and about a factor of 3 when taking p+p reactions as the reference. 
Thus $K^{\ast}(892)^0$ yields seem to be strongly affected by interactions in the produced fireball
with destruction dominating regeneration.
Published measurements from the STAR collaboration at RHIC energies~\cite{adams05} 
show a weaker suppression of the
$\langle K^{\ast} \rangle$/$\langle K^+ \rangle$ ratio for central Cu+Cu and Au+Au collisions
compared to inelastic p+p reactions of only about 30~\%. 
%
%--------------------------------------------------------------
\begin{table}[htb]
\caption[]{\label{tab:yields_na49}Total yields of $K^{\ast}(892)^0$ and
$\overline{K}^{\ast}(892)^0$ in inelastic p+p and in central C+C, Si+Si and 
Pb+Pb collisions at 158\agev~beam energy. Statistical and systematic errors
were added in quadrature.  In addition, model predictions are listed
from HGM (fit A of \cite{Be:05}) and UrQMD~1.3~\cite{urqmd_vogel} 
(predictions are for the $K^+\pi^-$ and $K^-\pi^+$ decay channels, respectively, and
were scaled by 3/2 to account for the branching ratio).
}
\begin{center}
\begin{tabular}{ll|cccc}
\hline
\hline
reaction             & & \bf{p+p} & \bf{C+C} & \bf{Si+Si} & \bf{Pb+Pb}\\
\hline
centrality           & & min. bias & 15.3\% & 12.2\% & 23.5\% \\
$\left<N_{W}\right>$ & & 2 &$ 14\pm 2$ & $37 \pm 3$ & $ 262 \pm 6$ \\
\hline
$\left<K^{\ast}\left(892\right)^0\right>$& this analysis & $0.0741  \pm 0.0069 $ & $0.8 \pm
0.24$ & $2.2 \pm 0.66$ & $10.3 \pm 2.5$ \\
    & HGM   & 0.074 &  0.964 & 2.76 & 25.1 \\
    & UrQMD & 0.076 & 0.74 & 2.25 & 22.2 \\
\hline
$\left<\overline{K}^{\ast}\left(892\right)^0\right>$ & this analysis & $0.0523  \pm 0.0048 $ & $0.43
\pm 0.14$ & $1.3 \pm 0.4$ & $5.2 \pm 1.7$ \\
    & HGM   & 0.041 & 0.455 & 1.33 & 12.5 \\
    & UrQMD & 0.043 & 0.41 & 1.19 & 9.5 \\
\hline
\hline
\end{tabular}
\end{center}
\end{table}
%--------------------------------------------------------------
%

%
%--------------------------------------------------------------
\begin{figure}[htb]
\includegraphics[width=0.45\linewidth]{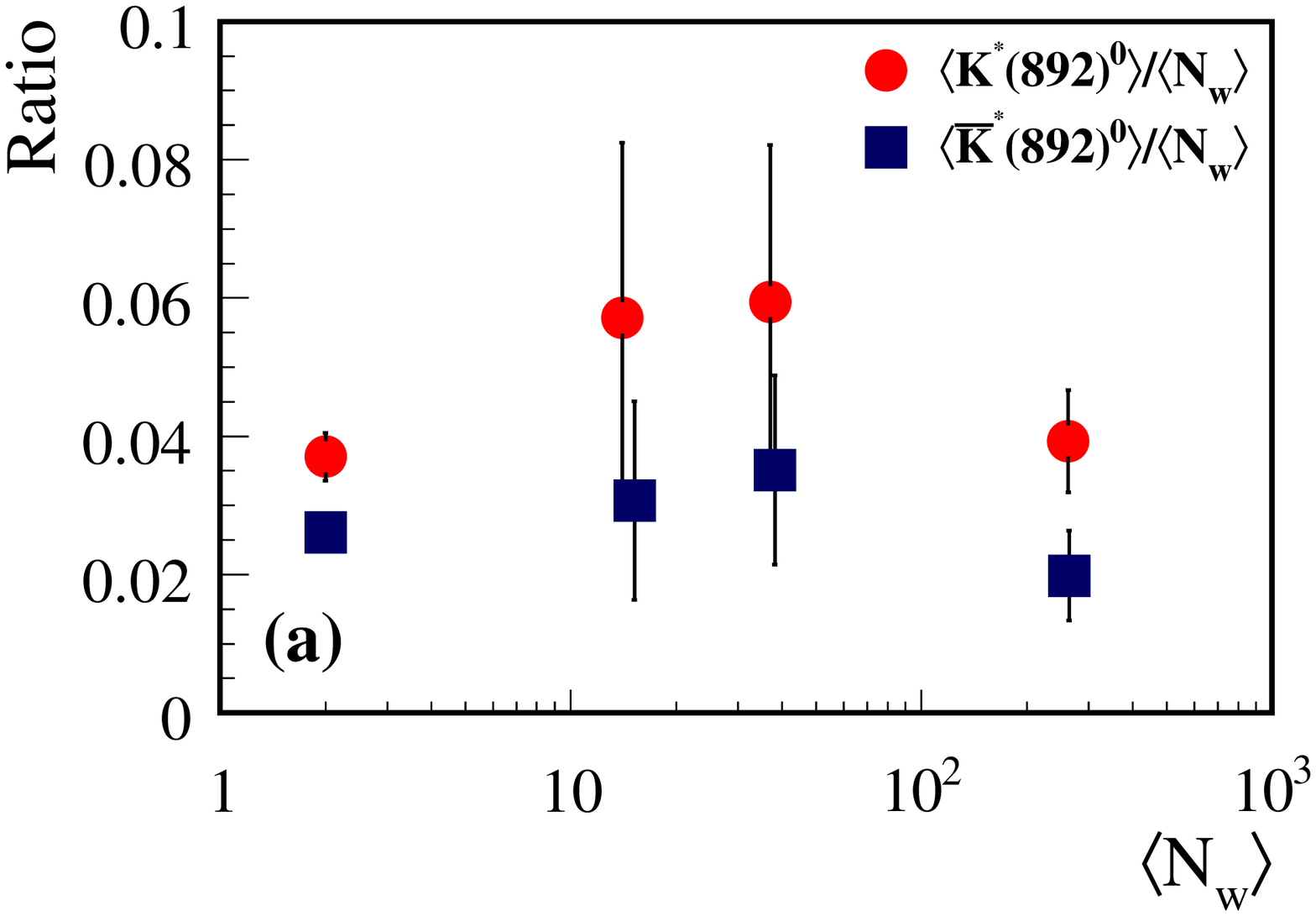}
\includegraphics[width=0.45\linewidth]{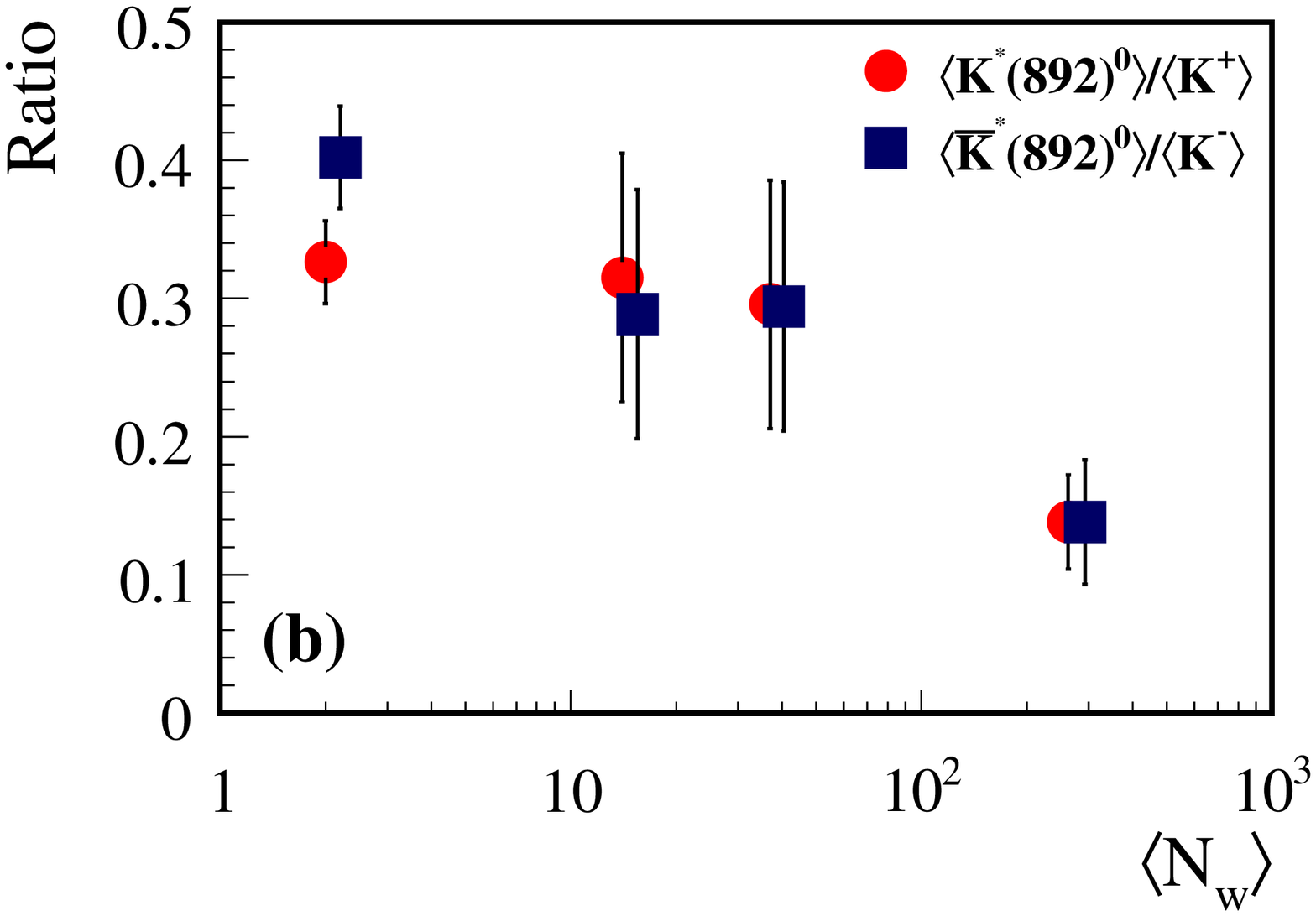}
\caption{\label{fig13} (Color online)
   (a) Yields per wounded nucleon $K^{\ast}(892)^0$/$\langle N_W \rangle$ (dots) and 
   $\overline{K}^{\ast}(892)^0$/$\langle N_W \rangle$ (squares) versus size of the collision system.
   (b) Ratios $\langle K^{\ast}(892)^0 \rangle$/$\langle K^+ \rangle$ (dots)
   and $\langle \overline{K}^{\ast}(892)^0 \rangle$/$\langle K^-\rangle$ (squares)
   versus size of the collision system (p+p, C+C, Si+Si and Pb+Pb collisions). 
   Total kaon yields were taken from refs.~\cite{deconf,syssz_2005,na49_pp-kaon-paper}
   and appropriately scaled by $\langle N_W \rangle$. For evaluating error bars the
   quadratic sums of statistical and systematic errors were used.
  }
\end{figure}
%  pp data (a) 0.0371 +- 0.0035     0.0262 +- 0.0024
%          (b) 0.326  +- 0.030      0.402  +- 0.037
%---------------------------------------------------------------
%

Microscopic models of hadron production in nucleus-nucleus collisions have been
used to study the modification of resonance yields during the space-time
evolution of the fireball. In the UrQMD model~\cite{urqmd_bass} particle
production proceeds via string excitation and decay at high energies, and evolves further by
interactions and coalescence in the produced matter. During the model
simulations track is kept of the full history for each particle thus allowing
to study the phenomena of destruction and regeneration of resonance 
states~\cite{urqmd_vogel,urqmd_aichelin} which was first considered in~\cite{torr01}. 
For comparison with our measurements we extracted the predicted yields in the
$K^+\pi^-$ and $K^-\pi^+$ decay channels, respectively, and scaled them by 3/2 to take
into account the decay branching ratio as was done for the data. The 
measured rapidity spectra of $K^{\ast}(892)^0$ and $\overline{K}^{\ast}(892)^0$
in central Pb+Pb collisions at 158\agev~
are compared to results from UrQMD model calculations  in \Fi{fig6} (dashed-dotted curves).
While one observes good agreement for the shape of the rapidity distributions, yields are 
overpredicted by roughly a factor of two. Moreover, the predicted transverse mass 
distributions are steeper than those of the data (see dashed-dotted curves in \Fi{fig7}~(b,d)). 

Inspection of the particle histories in the simulated UrQMD events indicates that 
of the originally produced $K^{\ast}(892)^0$ ($\overline{K}^{\ast}(892)^0$) about 
2~\% (2~\%) in p+p,  12~\% (13~\%) in central C+C, 23~\% (27~\%) 
in central Si+Si and 44~\% (62~\%) in central Pb+Pb collisions are lost owing to 
in-medium interactions and decay (rescattering of the decay products).
The contributions of the various mechanisms are illustrated by the curves shown in \Fi{fig14}.
The first reduction is due to reinteractions of the $K^{\ast}$ in the fireball medium
and the second reduction is the effect of the decay branching ratio into $K^+\pi^-$
and $K^-\pi^+$, respectively.
The last reduction is caused by the scattering of the $K^{\ast}$ decay daughters in the medium. 
The model calculations thus suggest a sizable duration
of the hadronic phase of the fireball to allow for such reinteractions.

Predictions of the UrQMD model for total yields are listed in \Ta{tab:yields_na49}.
The agreement with the measurements for p+p, C+C and Si+Si reactions suggests 
that UrQMD reproduces the absorption effects in these smaller systems. In contrast,
predicted total $K^{\ast}(892)^0$ and $\overline{K}^{\ast}(892)^0$ yields 
for central Pb+Pb collisions exceed the experimental
results by roughly a factor of 2. This might imply that the lifetime of the hadronic
phase is larger than suggested by the model calculation.

%
%--------------------------------------------------------------
\begin{figure}[htb]
\mbox{
\includegraphics[width=0.45\linewidth]{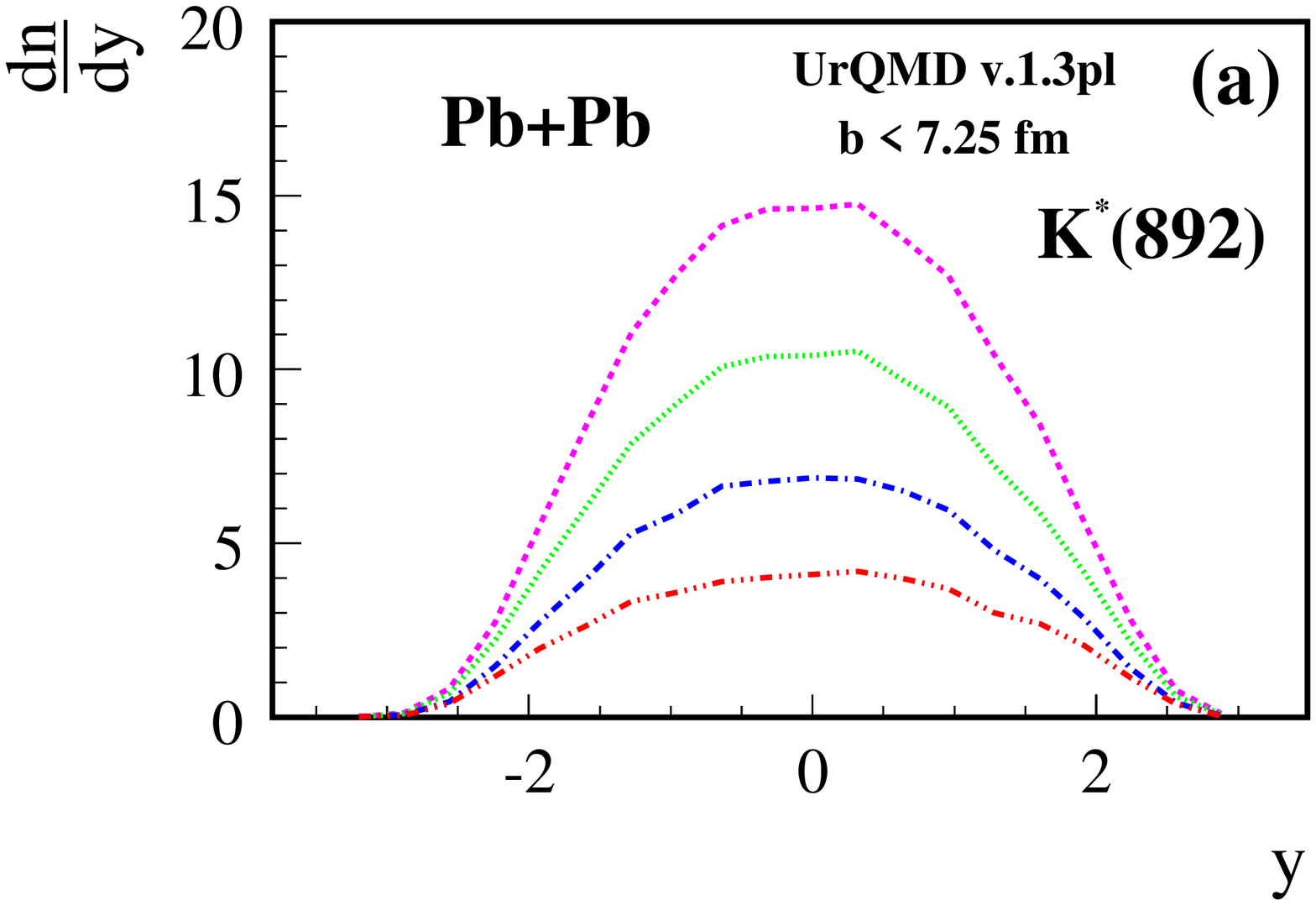}
\includegraphics[width=0.45\linewidth]{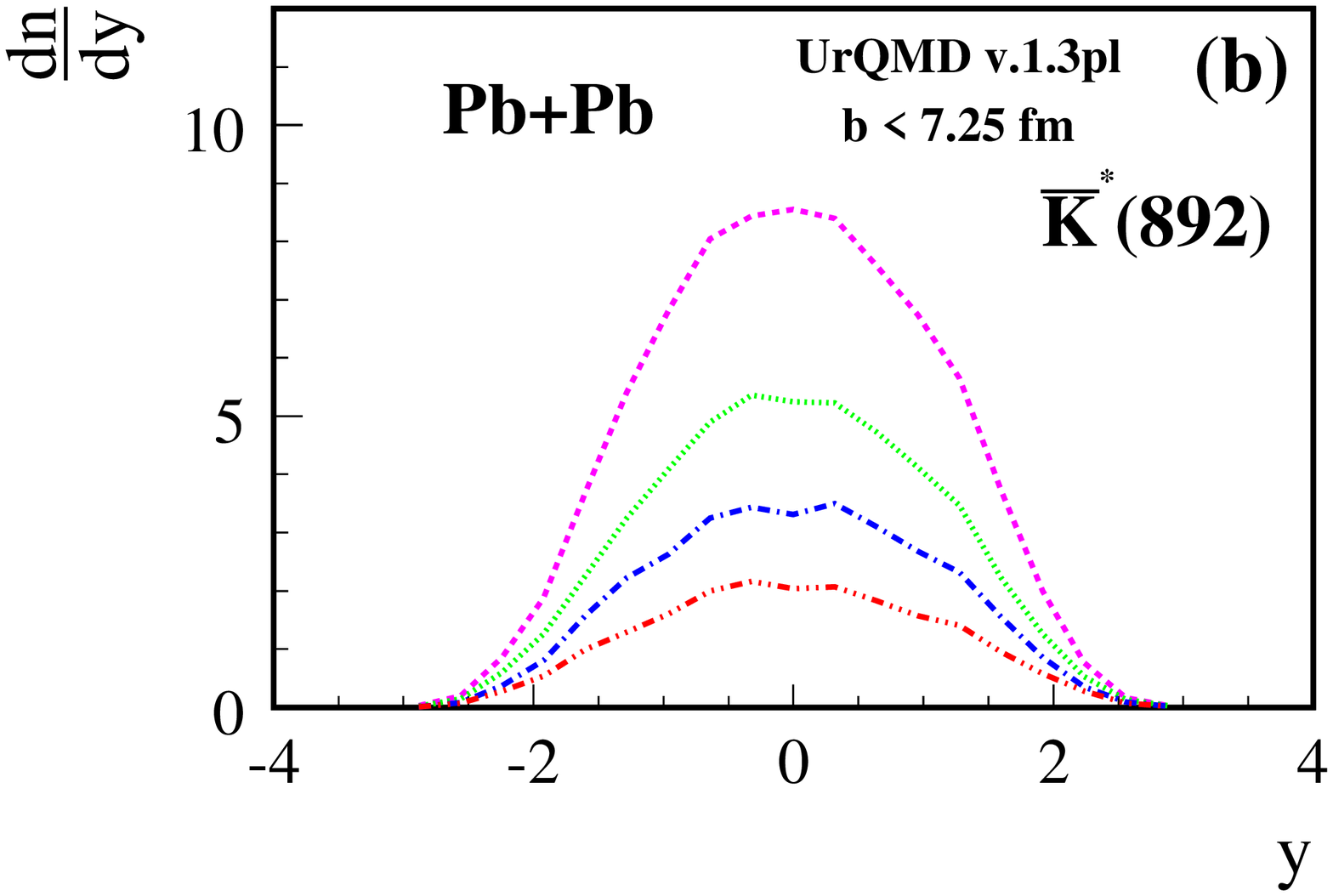}
}
\mbox{
\includegraphics[width=0.45\linewidth]{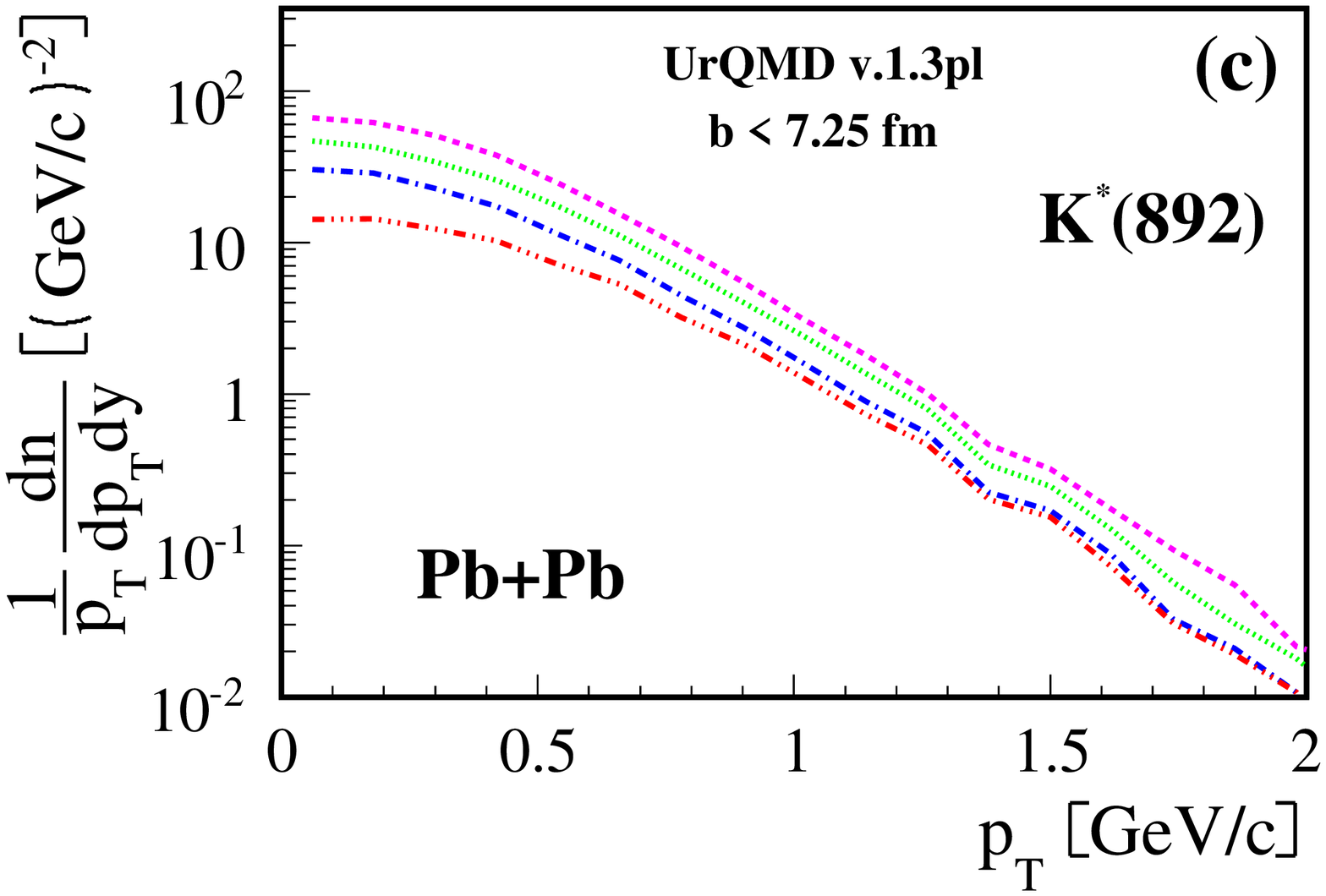}
\includegraphics[width=0.45\linewidth]{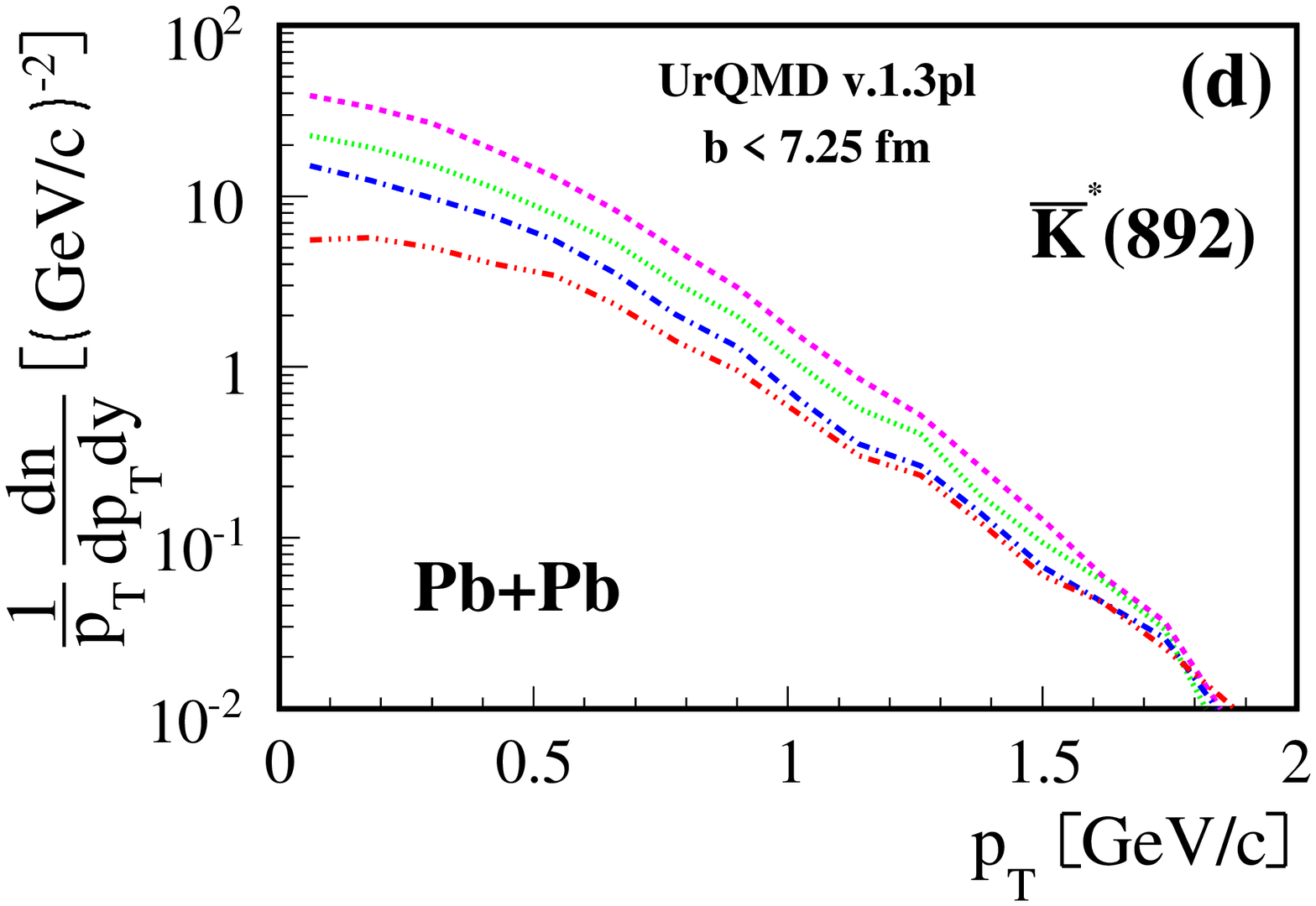}
}
\caption{\label{fig14} (Color online)
   Predictions of the UrQMD model version 1.3p1 ~\cite{urqmd_vogel} for
   the 23.5\% most central Pb+Pb collisions at 158\agev:
   distributions of rapidity (a,b) and transverse momentum $p_T$ (c,d)
   in the rapidity interval $0.43 < y < 1.78$
   of $K^{\ast}(892)^0$ (a,c) and $\overline{K}^{\ast}(892)^0$ (b,d).
   The curves show the successive reduction of yields owing to various
   interaction mechanisms of the $K^{\ast}(892)^0$ and $\overline{K}^{\ast}(892)^0$
   and their decay daughters in the fireball (see text).
   Note that the final results show the yield predictions for the $K^+\pi^-$ and
   $K^-\pi^+$ decay channels respectively. The curves in \Fi{fig6} and \Fi{fig7}
   were obtained by scaling these by 3/2.
  }
\end{figure}
%---------------------------------------------------------------
%

The statistical hadron gas model (HGM) was found to provide a good fit to total
yields of stable hadrons produced in elementary $e^+ + e^-$, $p + p$ and
nucleus+nucleus collisions using as adjustable parameters the hadronisation
temperature $T_{chem}$, the baryo-chemical potential $\mu_B$ and a strangeness
saturation parameter $\gamma_s$ \cite{Be:03,Be:05}. The predictions for 
$K^{\ast}(892)$ yields (which were not included in the fit of the model
parameters) in p+p and nucleus+nucleus collisions are compared
to the measurements in \Fi{fig15}~(a). One finds that the HGM predictions are
consistent with the measurements for p+p and light nuclei collisions, 
but exceed by more than a factor of 2 the observed yields in central Pb+Pb reactions.

%
%--------------------------------------------------------------
\begin{figure}[htb]
\includegraphics[width=0.45\linewidth]{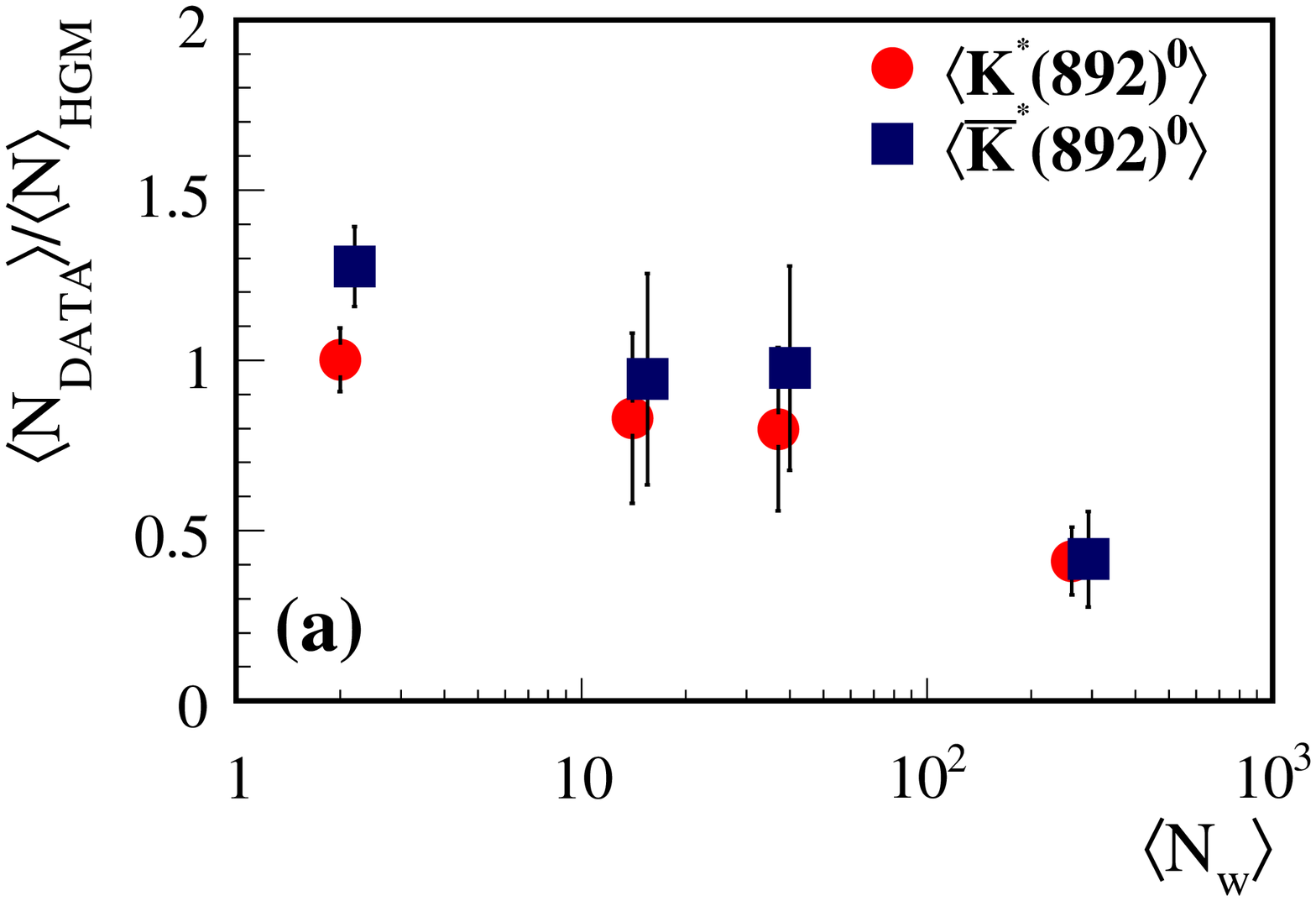}
\includegraphics[width=0.45\linewidth]{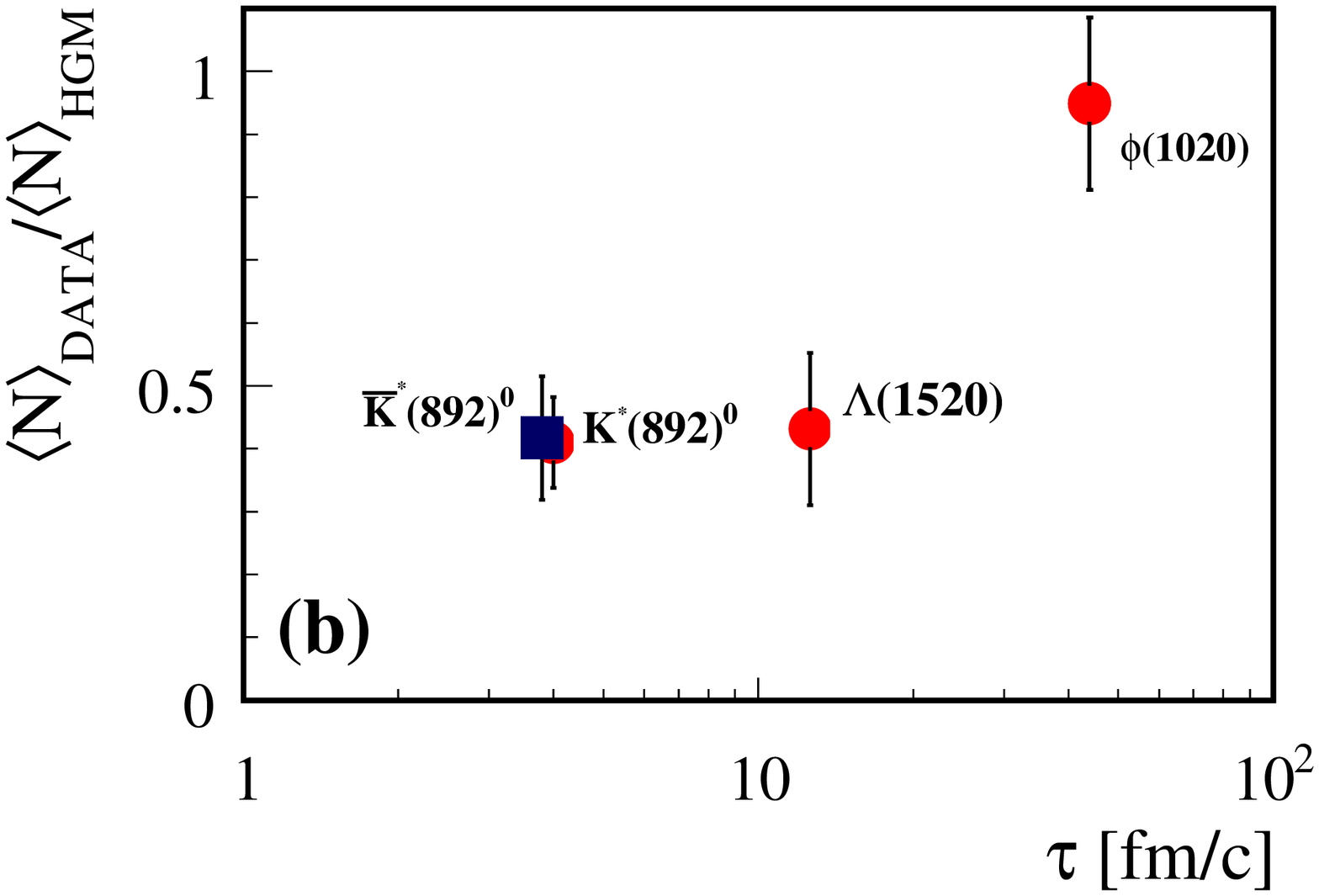}
\caption{\label{fig15} (Color online)
   (a) Ratio of measured total yields of $K^{\ast}(892)^0$ (dots) and
   $\overline{K}^{\ast}(892)^0$ (squares) to statistical hadron gas model (HGM)
   predictions \cite{Be:05} versus the size of the collision system
   (p+p, C+C, Si+Si and Pb+Pb collisions). (b) Ratio of measured
   yield in central Pb+Pb collisions to the statistical hadron-gas model 
   prediction for $K^{\ast}(892)^0$ (dot),
   $\overline{K}^{\ast}(892)^0$ (square), $\phi$ meson~\cite{phina49} and
   preliminary measurement of $\Lambda(1520)$~\cite{lam1520} versus the lifetime 
   $\tau$ of the resonance state. For evaluating error bars the
   quadratic sums of statistical and systematic errors were used.
  }
\end{figure}
%   pp data  (a)   1.001 +- 0.093     1.276 +- 0.117
%---------------------------------------------------------------
%

Yields of several resonance states were investigated by the NA49 collaboration,
namely of $K^{\ast}(892)^0$, $\Lambda(1520)$ \cite{lam1520} and the $\phi$ meson \cite{phina49}.
The ratios between the measured yields and the predictions of the HGM model are
plotted in \Fi{fig15}~(b) versus the respective lifetimes (3.91, 12.7 and 46.5 fm/$c$).
The suppression with respect to the HGM predictions seems to get stronger with
decreasing lifetime of the resonance. This suggests that a large part of the reduction
of the $K^{\ast}(892)^0$ yield may be caused by rescattering of its decay daughters during
the hadronic stage of the fireball and implies that this stage lasts for a time
(estimated about 6~fm/$c$ in Ref.~\cite{becorrna49}) at least comparable to the 
lifetime of the resonance. 
 
Alternatively, one may reconsider the assumption of simultaneous chemical freeze-out of the hadrons
from the fireball at a unique temperature. In the statistical hadron gas model the yield ratio of
two hadrons with identical strangeness, isospin
and baryon number allows to estimate the freeze-out temperature $T_{fo}$ 
(in the Boltzmann approximation, and neglecting feed-down from resonances) as:
\begin{eqnarray}\label{equSM}
    T_{fo} = ( m_2 - m_1 ) / \ln \left( 
      \left( \frac{2J_1+1}{2J_2+1} \right) 
      \left( \frac{m_1}{m_2} \right)^{\frac{3}{2}}
      \left( \frac{N_2}{N_1} \right) \right)
\end{eqnarray}
where $m_1$, $m_2$ are the masses, $J_1$, $J_2$ the spins and $N_1$, $N_2$ the 
produced multiplicities. Inserting the numbers for the pairs $\Lambda(1520)$, $\Lambda$
and $K^{\ast}(892)^0$, $K^{\pm}$ 
one obtains apparent freeze-out temperatures of 90 MeV for $\Lambda(1520)$ and 100 MeV 
for $K^{\ast}(892)^0$ respectively. The chemical freeze-out temperature fitted to the yields
of stable hadrons is $T_{chem} = 155$~MeV~\cite{Be:05}.
This would lead to the conclusion that short-lived resonances freeze out at the end of the fireball
evolution when the temperature has fallen below that for stable hadrons.

\section{Summary}

Production of the $K^{\ast}(892)^0$ and $\overline{K}^{\ast}(892)^0$ resonances was studied via their
$K^+ \pi^-$ and $K^- \pi^+$ decay modes in central
Pb+Pb, Si+Si, C+C and inelastic p+p collisions at 158\agev~($\sqrt{s_{NN}} = 17.3$~GeV) 
with the NA49 detector at the CERN SPS.
Transverse momentum and rapidity distributions were obtained and total yields were estimated.
The following conclusions were reached:

\begin{itemize}

\item
The yield of $K^{\ast}(892)^0$ exceeds that of $\overline{K}^{\ast}(892)^0$ by about a factor of two.
This observation can be understood from the similar ratio of the 
$K^+$ and $K^-$ yields and the valence quark composition of these mesons.

\item
The yield of $K^{\ast}(892)^0$ and $\overline{K}^{\ast}(892)^0$ per wounded nucleon appears to 
increase from p+p to C+C and Si+Si collisions and then tends to decrease to central Pb+Pb reactions.
This behaviour seems to reflect an interplay between strangeness enhancement in nucleus-nucleus
collisions and attenuation of $K^{\ast}(892)^0$ and $\overline{K}^{\ast}(892)^0$ in the produced fireball.

\item
The ratios $\langle K^{\ast}(892)^0 \rangle$/$\langle K^+ \rangle$ and
$\langle \overline{K}^{\ast}(892)^0 \rangle$/$\langle K^-\rangle$ decrease strongly with increasing
size of the colliding nuclei. These ratios are expected to be mostly sensitive to 
interactions of the $K^{\ast}(892)^0$ and its decay daughters with the produced dense matter. 
The decrease of the ratios suggests a substantial duration of the hadronic stage
of the fireball.

\item
The UrQMD model, although including rescattering of $K^{\ast}(892)^0$ and $\overline{K}^{\ast}(892)^0$ and their
decay daughters in the hadronic phase, is not able to provide a quantitative description of 
$K^{\ast}(892)^0$ production in nucleus-nucleus collisions at SPS energies.

\item
Yields of $K^{\ast}(892)^0$ mesons in central Pb+Pb collisions are about a factor of 2.5 below 
the predictions of the statistical
hadron gas model using parameters fitted to the yields of stable hadrons.

\end{itemize}

In summary, the predicted suppression of $K^{\ast}(892)^0$ yields~\cite{torr01} was observed 
in central Pb+Pb collisions at the SPS. It was found to be stronger at SPS than at RHIC energies.
More comprehensive studies of the energy and system-size dependence of the 
suppression of hadron resonance production  will help to better understand
the hadronisation process and the evolution of the high-density matter droplet created in
nucleus-nucleus collisions.

%============================================================================

\section*{Acknowledgments}

Acknowledgements: This work was supported by
the US Department of Energy Grant DE-FG03-97ER41020/A000,
the Bundesministerium fur Bildung und Forschung, Germany,
the German Research Foundation (grant GA 1480/2-1),
the Polish Ministry of Science and Higher Education (1~P03B~006~30, 1~P03B~127~30, 0297/B/H03/2007
/33, N~N202~078735, N~N202~204638),
the Hungarian Scientific Research Foundation (T068506),
the Bulgarian National Science Fund (Ph-09/05),
the Croatian Ministry of Science, Education and Sport (Project 098-0982887-2878)
and
Stichting FOM, the Netherlands.

%============================================================================


\newpage

\begin{thebibliography}{9} 
\section*{References}

\bibitem{Hz_Jac_2000} U.~Heinz and M.~Jacob, preprint nucl-th/0002042 (2000).

\bibitem{deconf} C.~Alt {\it et al.}, Phys.~Rev.~C~{\bf 77}, 024903 (2008).

\bibitem{Ga_Go_Se_2011} M.~Ga\'zdzicki, M.~Gorenstein and P.~Seyboth, 
                       Acta~Phys.~Polon.~B~{\bf 42}, 2705 (2011).

\bibitem{stenh} J.~Rafelski and B.~M\"uller, Phys.~Rev.~Lett.~{\bf 48}, 1066 (1982).

\bibitem{gazdz} M.~Ga\'zdzicki and M.~Gorenstein, Acta~Phys.~Polon.~B~{\bf 30}, 2705 (1999).

\bibitem{rapp00} R.~Rapp and J.~Wambach, Adv.~Nucl.~Phys. {\bf 25}, 1 (2000).

\bibitem{torr01} G.~Torrieri and J.~Rafelski, Phys.~Lett.~B~{\bf 509}, 239 (2001);
   Z.~Xu~(STAR collaboration) Nucl.~Phys~B~{\bf 698}, 607c (2002).

\bibitem{adler02} C.~Adler {\it et al.}, Phys.~Rev.~C~{\bf 66}, 061901R (2002).

\bibitem{adams05} J.~Adams {\it et al.}, Phys.~Rev.~C~{\bf 71}, 064902 (2005);
   M.~Aggarwal {\it et al.}, preprint arXiv:1006.1961 (2010);
   S.~Dash (STAR collaboration), J.~Phys.~G~{\bf 35}, 104057 (2008).

\bibitem{qm08} P.~Seyboth, J.~Phys.~G~{\bf 35}, 104008 (2008):
               R.~Barton et al., J.~Phys.~G~{\bf 27}, 367 (2001).

\bibitem{urqmd_bass} S.~Bass {\it et al.}, Prog.~Part.~Nucl.~Phys.~{\bf 41}, 255 (1998);
  M.~Bleicher {\it et al.}, J.~Phys.~G~{\bf 25}, 1859 (1999).

\bibitem{Be:05}
  F.~Becattini, J.~Manninen and M.~Ga\'zdzicki, Phys.~Rev.~C~{\bf 73}, 044905 (2006);
   results of fit A were used for comparisons.

\bibitem{na49_nim} S.~Afanasiev {\it et al.}, Nucl.~Instrum.~Meth.~A~{\bf 430}, 210 (1999).

\bibitem{na49_pions_in_pp} C.~Alt {\it et al.}, Eur.~Phys.~J.~C~{\bf 45}, 343 (2006).

\bibitem{bialas}
A.~Bia{\l}as, M.~B{\l}eszy\'nski and W.~Czy\.z, Nucl.~Phys.~B~{\bf 111}, 461 (1976). 
(Note that $N_W$ is often referred to less precisely as the number of participants.
$N_W$ does not include nucleons participating only in secondary interactions)

\bibitem{Laszlo_cent} A.~Laszlo, CERN EDMS Id 885329.

\bibitem{venus} K.~Werner, Phys.~Rep.~{\bf 232}, 87 (1993).

\bibitem{syssz_2005} C.~Alt {\it et al.}, Phys.~Rev.~Lett.~{\bf 94}, 052301 (2005).

\bibitem{pdg2010} K.~Nakamura {\it et al.} (Particle Data Group), J.~Phys.~G~{\bf 37}, 075021 (2010).

\bibitem{minvres} C.~Alt {\it et al.}, Phys.~Rev.~C~{\bf 73}, 034910 (2006).

\bibitem{geant} Geant Detector Description and Simulation Tool, CERN Program Library
                Long Writeup W5013.

\bibitem{thesis_slod} M.~S{\l}odkowski, PhD Thesis, Warsaw University of Technology (2008),
                CERN EDMS Id 999736.

\bibitem{na49_pp-kaon-paper} T.~Anticic {\it et al.},  Eur.~Phys.~J.~C~{\bf 68}, 1 (2010).

\bibitem{event_mixing} D.~Drijard, H.~Fischer, and T.~Nakada,
  Nucl.~Instr.~Meth.~{\bf 225}, 367 (1984).
 
\bibitem{ehs86} T.~Aziz et al., Z.~Phys.~C~{\bf 30}, 381 (1986).

\bibitem{hoehne2003} C.~H\"ohne, PhD Thesis, Marburg University (2003), CERN EDMS Id 816035.
   Note that the corrections for inelastic events not accepted by the trigger or having
   no tracks in the TPCs were revised.

\bibitem{phina49} C.~Alt {\it et al.}, Phys.~Rev.~C~{\bf 78}, 044907 (2008).

\bibitem{becorrna49} C.~Alt {\it et al.}, Phys.~Rev.~C~{\bf 77}, 064908 (2008).

%\bibitem{ehs91} M.~Aguilar-Benitez et al., Z.~Phys.~C~{\bf 50}, 405 (1991).

\bibitem{urqmd_vogel}  
            S.~Vogel and M.~Bleicher, preprint nucl-th/0505027 (2005);
            S.~Vogel, private communication (2008). Version 1.3p1 was used to 
            calculate the predictions.

\bibitem{urqmd_aichelin} M.~Bleicher and J.~Aichelin, Phys.~Lett.~B~{\bf 530}, 81 (2002).

\bibitem{Be:03}
  F.~Becattini and U.~Heinz, Z.~Phys.~C~{\bf 76}, 269 (1997).

\bibitem{lam1520} V.~Friese~(NA49 collaboration), Nucl.~Phys.~A~{\bf 698}, 487 (2002);
  C.~Markert, PhD Thesis, Frankfurt University (2001), CERN EDMS Id 816027.

\end{thebibliography}

%============================================================================

\end{document}